\documentclass[a4paper,fleqn,usenatbib]{mnras}

\usepackage{newtxtext,newtxmath}
\usepackage[T1]{fontenc}
\usepackage{ae,aecompl}

\usepackage{graphicx}	% Including figure files
\usepackage{amsmath}	% Advanced maths commands
\usepackage{amssymb}	% Extra maths symbols

\newcommand{\txn}{\textnormal}

\newcommand{\us}{\,} %space between number and corresponding unit 
\newcommand{\diff}{\txn{d}}
\newcommand{\code}{\nolinkurl} %plays nice with underscores etc.
\newcommand{\specwizard}{\textsc{specwizard}}
\newcommand{\eagle}{EAGLE}

\title[O~VII and O~VIII in EAGLE]{The abundance and physical properties of O~{\sc vii} and O~{\sc viii} X-ray absorption systems in the EAGLE simulations}

\author[N.\ A.\ Wijers et al.]{
Nastasha A.\ Wijers,$^{1}$\thanks{E-mail: wijers@strw.leidenuniv.nl}
Joop Schaye,$^{1}$
Benjamin D.\ Oppenheimer,$^{2,3}$
Robert A.\ Crain,$^{4}$ \newauthor
Fabrizio Nicastro,$^{3,5}$
\\
$^{1}$Leiden Observatory, Leiden University, PO Box 9513, NL-2300 RA Leiden, the Netherlands\\
$^{2}$CASA, Department of Astrophysical and Planetary Sciences, University of Colorado, 389 UCB, Boulder, CO 80309, USA\\
$^{3}$Harvard-Smithsonian Center for Astrophysics, 60 Garden St., Cambridge,
MA 02138, USA\\
$^{4}$Astrophysics Research Institute, Liverpool John Moores University, 146 Brownlow Hill, Liverpool L3 5RF, UK\\
$^{5}$INAF Osservatorio Astronomico di Roma, Via Frascati, Monte Porzio Catone (RM), Italy
}

\date{Accepted XXX. Received YYY; in original form ZZZ}

\pubyear{2019}

\begin{document}
\label{firstpage}
\pagerange{\pageref{firstpage}--\pageref{lastpage}}
\maketitle

\begin{abstract} % 253 words
%This is a simple template for authors to write new MNRAS papers.
%The abstract should briefly describe the aims, methods, and main results of the paper.
%It should be a single paragraph not more than 250 words (200 words for Letters).
%No references should appear in the abstract.
We use the {\eagle} cosmological, hydrodynamical simulations to predict the column density and equivalent width distributions of intergalactic \ion{O}{vii} ($E=574 \us \txn{eV}$) and \ion{O}{viii}  ($E=654 \us \txn{eV}$) absorbers at low redshift. These two ions are predicted to account for $40 \us \%$ of the gas-phase oxygen, which implies that they are key tracers of cosmic metals.
%, and to investigate the physical conditions probed by these absorption systems. 
We find that their column density distributions evolve little at observable column densities from redshift~$1$ to~$0$, and that they are sensitive to AGN feedback, which strongly reduces the number of strong (column density $N \gtrsim 10^{16} \us \txn{cm}^{-2}$) absorbers. The distributions have a break at $ N \sim 10^{16}\us \txn{cm}^{-2}$, corresponding to overdensities of $\sim 10^2$, likely caused by the transition from sheet/filament to halo gas. Absorption systems with $N \gtrsim 10^{16} \us \txn{cm}^{-2}$ are dominated by collisionally ionized \ion{O}{vii} and \ion{O}{viii}, while the ionization state of oxygen at lower column densities is also influenced by photoionization. At these high column densities, \ion{O}{vii} and \ion{O}{viii} arising in the same structures probe systematically different gas temperatures, meaning their line ratio does not translate into a simple estimate of temperature.
%Finally, we investigate correlations between these two ions and UV line absorption. 
While \ion{O}{vii} and \ion{O}{viii} column densities and covering fractions correlate poorly with the \ion{H}{i} column density at $\txn{N}_{\txn{H}\,\txn{I}} \gtrsim 10^{15} \us \txn{cm}^{-2}$, \ion{O}{vii} and \ion{O}{viii} column densities are higher in this regime than at the more common, lower \ion{H}{i} column densities.
%While neutral hydrogen column densities $\gtrsim 10^{15} \us \txn{cm}^{-2}$ strongly increase the probability of the presence high \ion{O}{vii} and \ion{O}{viii} column densities, the column densities and covering fractions of \ion{O}{vii} and \ion{O}{viii} correlate poorly with $\txn{N}_{\txn{H}\,\txn{I}}$ in this regime. 
The column densities of \ion{O}{vi} and especially \ion{Ne}{viii}, which have strong absorption lines in the UV, are good predictors of the strengths of \ion{O}{vii} and \ion{O}{viii} absorption and can hence aid in the detection of the X-ray lines. 
\end{abstract}


%%%%%%%%%%%%%%%%%%%%%%%%%%%%%%%%%%%%%%%%%%%%%%%%%%

%%%%%%%%%%%%%%%%% BODY OF PAPER %%%%%%%%%%%%%%%%%%

%\section*{Writing: nota bene}
%\begin{itemize}
%\item{Put in `column density distribution function (CDDF)' at first occurrence, IGM, CIE, PIE, EW. }
%\item{Check for I/we/passive consistency (default `we')}
%\end{itemize}
%%# O6: $1032\,\mathrm{\AA}$    O7: $21.60\,\mathrm{\AA}$   O8: $18.97\,\mathrm{\AA}$ Fe17: $17.05\,\mathrm{\AA}$
%% CHIANTI weighted oscillator strengths: 	5.53e-01 for 18.9671 A, 2.77e-01 for 18.9726 A
%% chianti file in cloudy file seems to contain osc. strengths from the comments, but I don't know which line and column to use...
%
%
%\section*{todo: research}
%\begin{itemize}
%\item{{\todo Some sample spectrum plots with $\rho$, $T$? Illustrate velocity structure and compare to the weighted phase diagram plots for validation of that description. (Seems like a waste not to use this stuff more extensively, and these validations address the approximation of the CDDF used throughout the paper.)}}
%\item{Phase diagram plots, fixed-z CDDF comparison for $z=1.0$: get at reason for the evolution found (more investigation in halo connection paper?)}
%\item{check impact of multiple absorption systems along the line of sight on EW distributions}
%\item{compare column densities in $n_{\mathrm{H}}$--$T$ space to predictions of shock heating model (
%On the Physical Origin of OVI Absorption-Line Systems
%T.M. Heckman, C.A. Norman, D.K. Strickland, K. R. Sembach) }
%\end{itemize}

\section{Introduction}

%This is a simple template for authors to write new MNRAS papers.
%See \texttt{mnras\_sample.tex} for a more complex example, and \texttt{mnras\_guide.tex}
%for a full user guide.

%All papers should start with an Introduction section, which sets the work
%in context, cites relevant earlier studies in the field by \citet{Others2013},
%and describes the problem the authors aim to solve \citep[e.g.][]{Author2012}.

Within extragalactic astronomy, the missing baryon problem is well-established. We know from cosmic microwave background measurements \citep[e.g.][]{planck_2013} and big bang nucleosynthesis \citep[see the review by][]{cyburt_fields_etal_2016} how many baryons there were in the very early universe. However, at low redshift, we cannot account for all of these baryons by adding up observed populations: stars, gas observed in the interstellar, circumgalactic, and intra-cluster media, and photoionized absorbers in the intergalactic medium (IGM). According to a census presented by \citet{shull_smith_danforth_2012}, about $30\%$ of baryons are observationally missing at low redshift. 

Cosmological simulations predict that missing baryons reside in shock-heated regions of the IGM. \citet{cen_ostriker_1999} already found that the missing gas is mainly hot and diffuse. They called this gas the warm-hot intergalactic medium (WHIM), defined as gas with temperatures in the range $10^5$--$10^7 \us \mathrm{K}$. Some of this gas has already been detected. The cooler component is traced largely by \ion{O}{vi} absorption: $T \sim 10^{5.5} \us \txn{K}$ in collisional ionization equilibrium (CIE), which applies to high-density gas. \ion{O}{vi} absorption has been studied observationally by many groups \citep[e.g.,][]{tumlinson_etal_2011, johnson_chen_mulchaey_2013}, often using the Cosmic Origins Spectrograph (COS) on the Hubble Space Telescope (HST). Others have investigated \ion{O}{vi} absorption in galaxy and cosmological simulations \citep[e.g.,][]{cen_fang_2006, tepper-garcia_richter_etal_2011, cen_2012, shull_smith_danforth_2012, rahmati_etal_2016, oppenheimer_etal_2016, oppenheimer_2018_fossilAGN_cos, Nelson_etal_2018_hiO}.  
Predictions for the hotter part of the WHIM have been made with simulations \citep[e.g.,][]{cen_fang_2006, branchini_ursino_etal_2009, bertone_schaye_etal_2010, cen_2012, Nelson_etal_2018_hiO}, typically, but not exclusively, focussing on the \ion{O}{vii} and \ion{O}{viii} X-ray lines ($T \sim 10^{5.4}$--$10^{6.5}$ and $\sim 10^{6.1}$--$10^{6.8} \us \txn{K}$ in CIE, respectively). 
Others have made predictions for the WHIM gas using analytical models \citep{perna_loeb_1998}, or using a combination of analytical models and simulations \citep{fang_bryan_canizares_2002, furlanetto_phillips_kamionkowski_2005}. 
%\citet{bertone_etal_2008} give a review of numerical work on the WHIM, while \citet{bregman_2007_review} gives a more observationally-focussed review. 

Besides being a major baryon reservoir, the WHIM also provides an important way to understand accretion and feedback processes in galaxy formation. As gas collapses onto a galaxy, some of it forms stars or is accreted by the central supermassive black hole. This creates feedback, where these stars (through, for example, supernova explosions) and active galactic nuclei (AGN) inject energy and momentum into their surrounding gas, slowing, stopping, or preventing further gas accretion. Current-generation cosmological simulations like {\eagle} \citep{eagle_paper}, IllustrisTNG \citep{pillepich_springel_etal_2018}, and Horizon-AGN \citep{dubois_pichon_etal_2014} include star formation and AGN feedback, and metal enrichment from stars. The feedback in {\eagle} and IllustrisTNG, and the AGN feedback in Horizon-AGN, is calibrated to reproduce galaxy (stellar and black hole) properties, because individual stars, supernova explosions, and black hole accretion disks are too small to be resolved in these simulations. That means that these simulations require a model of the effect of this feedback on scales they can resolve. 

The effect of this feedback on the circumgalactic medium (CGM) is not as well-constrained. For example, feedback in BAHAMAS \citep{mccarthy_schaye_etal_2017} and IllustrisTNG is calibrated to halo gas fractions, but this only applies to high-mass haloes where  observations are available ($M_{500c} > 10^{13} \us \txn{M}_{\sun}$). One way to constrain the effects of feedback on the CGM, is to study how far metals, which are created in galaxies, spread outside their haloes. In other words, we can investigate what fraction of the metals ends up in the intergalactic medium, which impacts how many metal absorbers we expect to find in quasar sightlines. This is not the only possible effect, as feedback also impacts the temperature, density, and kinematics of gas. 

The main way we expect to find the hot WHIM gas, is through absorption in e.g.\ quasar spectra \citep[e.g.,][]{brenneman_smith_etal_2016, nicastro_etal_2018, kovacs_orsolya_etal_2019}. This is because this WHIM gas typically has low densities: since emission scales with the density squared and absorption with the density along the line of sight, this makes absorption more readily detectable than emission in most of the WHIM. Observationally, absorption from these ions has been found around the Milky Way, as described by e.g. \citet{bregman_2007_review}, but claims of extragalactic \ion{O}{vii} and \ion{O}{viii} are rare \citep[e.g.,][]{nicastro_mathur_etal_2005} and often disputed \citep[e.g.,][]{kaastra_werner_etal_2006} or uncertain \citep[e.g.,][]{bonamente_nevalainen_etal_2016}. \citet{nicastro_krongold_etal_2017} review these WHIM searches in absorption. \citet{nicastro_etal_2018} recently found two extragalactic \ion{O}{vii} absorbers, using very long observations with the XMM-Newton RGS of the spectrum of the brightest X-ray blazar. These are consistent with current predictions, though the small number of absorbers means uncertainties on the total absorber budget are still large.

Besides these oxygen ions, other ions are also useful for studying the WHIM. 
For example, \ion{Ne}{VIII} traces WHIM gas that is somewhat hotter than \ion{O}{vi} traces ($T \sim 10^{5.8} \us \txn{K}$ in CIE). Theoretically, \citet{tepper-garcia_richter_etal_2013} investigated its absorption in the OWLS cosmological, hydrodynamical simulations \citep{schaye_dalla-vecchia_etal_2010_owls} and \citet{rahmati_etal_2016} did so with {\eagle}.
Observationally, e.g., \citet{meiring_tripp_etal_2013} studied \ion{Ne}{viii} absorption, and recently, \citet{burchett_tripp_etal_2018} found \ion{Ne}{viii} absorption associated with the circumgalactic medium (CGM). 
A different tracer of the WHIM is broad Ly$\alpha$ absorption ($T \gtrsim 10^{5} \us \txn{K}$), which traces the cooler component of the WHIM, similar to \ion{O}{vi} but without the need for metals \citep[e.g.,][]{tepper-garcia_richter_etal_2012, shull_smith_danforth_2012}. 
These ions are useful because they trace the WHIM themselves, but they can also be used to find \ion{O}{vii} and \ion{O}{viii} absorbers. Recently, \citet{kovacs_orsolya_etal_2019} did this: they found extragalactic \ion{O}{vii} absorption by stacking quasar spectra using the known redshifts of 17 Ly$\alpha$ absorbers associated with massive galaxies.  

\citet{mroczkowski_nagai_etal_2018} give an overview of how the Sunyaev-Zeldovich (SZ) effect has been used to search for the WHIM. This effect probes line-of-sight integrated pressure (thermal SZ) or free electron bulk motion (kinetic SZ), which both have the same density dependence as absorption and are therefore less biased towards high-density gas than emission is. Attempts to detect the IGM by this method have focussed on SZ measurements of cluster pairs to detect filaments between them, and cross-correlations with other tracers of large-scale structure \citep[and references therein]{mroczkowski_nagai_etal_2018}. For example, \citet{de-graaff_cai_etal_2017} used stacked massive galaxy pairs (mean stellar mass $10^{11.3} \, \mathrm{M}_{\sun}$, virial mass $\sim 10^{13} \us \mathrm{M}_{\sun} \, h^{-1}$) to detect a filamentary thermal SZ signal. \citet{tanimura_hinshaw_etal_2019} similarly stacked luminous red galaxy pairs with stellar masses $> 10^{11.3} \, \mathrm{M}_{\sun}$, and found a filamentary thermal SZ signal larger than, but consistent with, predictions from the BAHAMAS simulations \citep{mccarthy_schaye_etal_2017}. Using the {\eagle} simulations, \citet{davies_crain_etal_2019} made theoretical predictions for the thermal SZ effect from the CGM at different halo masses. 

Another way to find hotter gas is through X-ray emission, but this scales with the density squared, and is therefore generally best for studying dense gas. \citet{bregman_2007_review} reviews some X-ray observations of extragalactic gas, though this is limited to the denser parts of the intracluster medium (ICM) and CGM. Following earlier work on WHIM X-ray emission in simulations by e.g., \citet{pierre_bryan_gastaud_2000} and \citet{yoshikawa_yamasaki_etal_2003}, \citet{bertone_schaye_etal_2010} studied soft X-ray line emission, and found that this emission should indeed be tracing mostly denser and metal-rich gas, i.e., ICM and CGM. \citet{tumlinson_peeples_werk_2017_cgmreview} discuss some more recent results on X-ray line emission from the CGM. Instead of lines, \citet{davies_crain_etal_2019} investigated broad-band soft X-ray emission from the CGM in {\eagle}, and found that it can be used as a proxy for the CGM gas fraction.

%{\todo Some mention of multiphase CGM gas (recent stuff form the journal club)? Or is that material for the discussion?}

One way to search for the missing baryons in absorption is by doing blind surveys, where observers look for absorbers in the lines of sight to suitable background sources. There are a number of upcoming and proposed X-ray telescopes for which there are plans to carry out such surveys, including Athena \citep{athena_ifu_2016}, Arcus \citep{brenneman_smith_etal_2016, smith_abraham_etal_2016_arcus}, and Lynx \citep{lynx_2018_08}. The \ion{O}{viii} and especially \ion{O}{vii} ions tend to be the focus of such plans, since oxygen is a relatively abundant element \citep{allendeprieto_lambert_asplund_2001} and these ions have lines with large oscillator strengths \citep{verner_verner_ferland_1996}, making these lines more readily detectable in the hot WHIM.

In simulations, \ion{O}{vii} and \ion{O}{viii} absorption has been studied: e.g.\ \citet{branchini_ursino_etal_2009} and \citet{cen_fang_2006} predicted the equivalent width distributions for blind surveys using an earlier generation of simulations and mock spectra generated from these. More recently, \citet{Nelson_etal_2018_hiO} studied absorption by these ions in the IllustrisTNG simulations, but they did not study equivalent widths. 

Here, we will study the column density and equivalent width distributions of \ion{O}{vii} ($21.60 \us${\AA}) and \ion{O}{viii} ($18.967, 18.973  \us${\AA}) in the {\eagle} simulations \citep{eagle_paper, eagle_calibration}, to predict which absorption systems may be detected by future missions, and to establish the physical conditions these absorption systems probe. {\eagle} has been used to predict column density distributions that agree reasonably well with observations at a range of redshifts, for a variety of ions \citep[e.g.,][]{eagle_paper, rahmati_etal_2016}. However, these studies all focussed on ions with ionization energies lower than \ion{O}{vii} ($739 \us \txn{eV}$) and \ion{O}{viii} ($871 \us \txn{eV}$) \citep{crc_handbook}.

This paper is organised as follows.
In Section~\ref{sec:methods}, we discuss our methods: the {\eagle} simulations themselves (Section~\ref{sec:eagle}), and how we extract column densities (Section~\ref{sec:cddf_methods}) and equivalent widths (Section~\ref{sec:ew_methods}) from them. 
We present the column density distributions we find in Section~\ref{sec:cddfs}, and discuss our mock spectra (Section~\ref{sec:spectra}) and the equivalent width distributions we infer from these (Section~\ref{sec:EW}). We discuss the origin of the shape of the column density distribution (Section~\ref{sec:break}) and how it probes AGN feedback (Section~\ref{sec:agn}).
We discuss the physical properties of \ion{O}{vii} and \ion{O}{viii} absorption systems in Section~\ref{sec:pds}. In Section~\ref{sec:ioncorr}, we discuss how absorption by these two ions correlates with three ions with UV lines: \ion{H}{i}, \ion{O}{vi}, and \ion{Ne}{viii}. In Section~\ref{sec:ioncomp}, we compare the gas traced by different ions along the same sightlines. 
In Section~\ref{sec:discussion}, we outline what our results predict for three planned or proposed missions in Section~\ref{sec:missions}, and what some limitations of our work may be (Section~\ref{sec:caveats}). 
Finally, we summarise our results in Section~\ref{sec:conclusion}.

%In the appendices, we look into the convergence of our column density distributions (appendix~\ref{app:conv}) and the impact of some technical choices in how to obtain equivalent width distributions from those column density distributions (appendix~\ref{app:projchoice}). In appendix~\ref{app:dXvar}, we show that spread in the measured column density distribution in blind surveys follows a Poisson distribution: large-scale structure does not have a significant impact on this, at least on the scales we measure.
%In this paper, we will not investigate the spatial distribution of the absorption systems in any detail. We will study this in a later paper. 
%Throughout this paper, we will use the prefix `c' for length units to denote comoving units, and a `p' to denote proper units; centimetres are always proper units. We will give velocities and equivalent widths in rest-frame units, unless we explicitly indicate otherwise.  

\section{Methods}
\label{sec:methods}

We study predictions for \ion{O}{vii} and \ion{O}{viii} absorption in the {\eagle} simulations \citep{eagle_paper, eagle_calibration}.
We use tabulated ion fractions as a function of temperature, density, and redshift, as well as the element abundances from the simulation, to calculate the number density of ions at different positions in the simulation. By projecting the number of ions in thick slices through these simulations onto 2-dimensional column density maps, we obtain column density distributions from the simulations. To calculate equivalent widths for some of these columns, we generated synthetic absorption spectra at sightlines through their centres.

% specifically in those used by \citet{bertone_schaye_etal_2010}
Oxygen abundances in this paper are given in solar units. Here, the solar oxygen mass fraction is $0.00586$ \citep{allendeprieto_lambert_asplund_2001}. This is simply used as a unit, and should not be updated to more recent solar abundance measurements in further work. 
 Length units include  p (`proper') or c (`comoving'). The exception is the centimetre we use in (column) densities: centimetres are always proper units.
% 0.005862311051135351

\subsection{The EAGLE simulations}
\label{sec:eagle}

In this section, we provide a short summary of the {\eagle} (Evolution and Assembly of GaLaxies and their Environments) simulations. More details can be found in \citet{eagle_paper}, the paper presenting the simulations calibrated to observations, \citet{eagle_calibration}, which describes the calibration of these simulations, and \citet{mcalpine_helly_etal_2016}, which describes the data release of the {\eagle} galaxy and halo data. 
%The {\eagle} simulations are smoothed particle hydrodynamics (SPH) simulations of a cosmological volume, aimed at understanding the formation and evolution of galaxies, and their interactions with their environments. 

The code used is a modified version of \code{Gadget3}, last described by \citet{springel_2005}, using the TREE-PM gravitational force calculation. The modifications include an implementation of smoothed particle hydrodynamics (SPH) known as \textsc{anarchy} (\citeauthor{eagle_paper} \citeyear{eagle_paper}, appendix~A; \citeauthor{anarchy_effect} \citeyear{anarchy_effect}). A $\Lambda$CDM cosmogony is used, with the parameters $(\Omega_m,\Omega_\Lambda,\Omega_b, h, \sigma_8, n_s, Y) = (0.307, 0.693, 0.04825, 0.6777, 0.8288, 0.9611, 0.248)$ \citep{planck_2013}. 
%Here, $\sigma_8$ determines the normalisation of the initial density perturbation power spectrum and $n_s$ is its slope . The value $Y$ is the primordial helium abundance (mass fraction), $h$ is the Hubble parameter $H_0/ 100 \us \txn{km} \,\txn{s}^{-1}\txn{Mpc}^{-1}$, and $\Omega_i$ is the average density of component $i$ (matter, cosmological constant, and baryons, respectively), relative to the critical density. 

Gas cooling is implemented following \citet{wiersma_schaye_smith_2009}, using the tracked abundances of 11 elements and cooling rates for each. Collisional and photoionization equilibrium is assumed, with photoionization from the \cite{HM01} UV/X-ray background model. With those assumptions, and the element abundances as tracked in the simulation, we calculate the number of ions in a column through the simulated box. To do this, we use ion fraction tables from \citet{bertone_schaye_etal_2010, bertone_schaye_etal_2010_uv}, who investigated line emission with tables computed under the same assumptions as the {\eagle} gas cooling. The ion fractions were computed using Cloudy, version~c07.02.00\footnote{This older version is consistent with the {\eagle} cooling rate calculations.} \citep{cloudy}. These tables give the ion fraction (fraction of nuclei of a given element that are part of a particular ion) as a function of $\log_{10}$ hydrogen number density, $\log_{10}$ temperature, and redshift. We interpolated these tables linearly, in log space for temperature and density, to obtain ion balances for each SPH particle. 
%Calculations were made for hydrogen number densities of $10^{-9}$--$10^{3} \us \mathrm{cm}^{-3}$ at $0.25 \us \mathrm{dex}$ intervals, and temperatures of $10^{1}$--$10^{9} \us \mathrm{K}$ at $0.1 \us \mathrm{dex}$ intervals. For the redshift range we investigate here ($0$--$1$), the redshifts the tables were calculated for are spaced by $\Delta z \approx 0.05$--$0.1$.

These assumptions mean we ignore both local ionization sources, such as stars and AGN, and non-equilibrium ionization. This also means we ignore the effect of flickering AGN. These could boost the abundances of highly ionized species, such as those we are interested in, even if the AGN is not `active' when we observe it \citep{oppenheimer_schaye_2013, segers_oppenheimer_etal_2017, oppenheimer_2018_fossilAGN_cos}. This boost is caused by species being ionized by the AGN radiation, and then not having time to recombine before observations are made. For typical AGN duty cycles and ion recombination times, the ions can be out of ionization equilibrium like this for a large fraction of systems. We discuss the impact of these assumptions in more detail in Section~\ref{sec:caveats}.   

Since the resolution of the simulations is too low to resolve individual events of star formation and feedback (stellar winds and supernovae), or accretion disks in AGN, the star formation and stellar and AGN feedback on simulated scales are implemented using subgrid models, with model feedback parameters calibrated to reproduce the $z=0.1$ galaxy stellar mass function, the relation between black hole mass and galaxy mass, and reasonable galaxy disc sizes. Star formation occurs where the local gas density is high enough, with an additional metallicity dependence following \citet{schaye_2004}. The rate itself depends on the local pressure, in a way that reproduces the  Kennicutt-Schmidt relation  \citep{schaye_dalla-vecchia_2008}. Stars lose mass to surrounding gas particles as they evolve, enriching them with metals. This is modelled according to \citet{wiersema_etal_2009_insim}.

A major problem in galaxy formation simulations at {\eagle}-like resolution is that if reasonable amounts of stellar and AGN feedback energy are injected into gas surrounding a single-age population of stars or a black hole at each time step, the energy is radiated away before it can do any work. This causes gas in galaxies to form too many stars. Following \citet{booth_schaye_2009} and \citet{dalla-vecchia_schaye_2012}, the solution for this problem used in {\eagle} is to statistically `save up' the energy released by these particles, until it is enough to heat neighbouring particles by a fixed temperature increment of $10^{7.5} \us \txn{K}$ (stellar feedback) or $10^{8.5} \us \txn{K}$ (AGN feedback). Which particles are heated and when is determined stochastically. The expectation value depends on the local gas density and metallicity \citep{eagle_calibration}. 
%such that the injected energy is, on average, what is predicted by the calibrated stellar and AGN feedback models. The temperature difference for AGN feedback was one of the calibration parameters. 

Table~\ref{tab:sims} gives an overview of the different {\eagle} simulations we use in this work. The reference feedback model was calibrated for the standard {\eagle} resolution, as used in e.g., \code{Ref-L100N1504}. 
%In the \code{AGNdT9} model, AGN feedback heats particles by $10^{9.0} \us \txn{K}$, together with some other feedback parameter changes \citep{eagle_paper}. This corresponds to the \code{S15_AGNdT9} model in the public data release. It is a different successful calibration at the standard {\eagle} resolution. In this feedback model, the AGN feedback is more explosive than in \code{Ref}: the total energy injection is the same, but it is spread out over fewer energy injctions. In Section~\ref{sec:agn} , we examine the differences between CDDFs from these two calibrated {\eagle} models. 
Simulation \code{Recal-L025N0752} was calibrated in the same way as the reference model, but at an eight times higher mass resolution. This is used to test the `weak convergence' of the simulations, in the language of \cite{eagle_paper}, compared to strong convergence tests, which use the same feedback parameters at different resolutions (appendix~\ref{app:conv}). The idea behind this is that the parameters for subgrid feedback will generally depend on which scale is considered subgrid, so the subgrid model parameters are expected to be resolution-dependent.  Finally, \code{NoAGN-L050N0752} is a variation of \code{Ref-L050N0752} without AGN feedback, which was not described in \citet{eagle_paper} or \citet{eagle_calibration}. 
Except for appendix~\ref{app:conv}, where we test resolution and box size convergence, we will only use the  \code{Ref-L100N1504} (reference), 
\code{Ref-L050N0752} ($50\us\txn{cMpc}$ reference) and \code{NoAGN-L050N0752} (no AGN) simulations.

\begin{table}
	\centering
	\caption{The simulations used in this work. The names consist of three parts, in the format \code{<name>-L<size>N<particles>}. The name is the name or abbreviation for the stellar and AGN feedback model, as explained in the text. The size is the size of the simulation box in comoving Mpc, and the last part is the cube root of the number of dark matter particles (equal to the initial number of gas particles) used in the simulation. The table lists the dark matter particle mass ($m_{\txn{DM}}$), the initial gas particle mass $m_{\txn{gas,\, init}}$, and the Plummer-equivalent gravitational softening length at low redshift ($l_{\txn{soft}}$).}
	\label{tab:sims}
	\begin{tabular}{rccc} % four columns, alignment for each
		\hline
		simulation name & $m_{\txn{DM}}$ & $m_{\txn{gas,\, init}}$ & $l_{\txn{soft}}$ \\
		     & ($\txn{M}_{\sun}$)    & ($\txn{M}_{\sun}$)          & (pkpc) \\
		\hline
		\code{Ref-L100N1504} &    $9.70 \times 10^6 $ & $1.81 \times 10^6 $ & $0.70$ \\
		\code{Ref-L050N0752} &    $9.70 \times 10^6 $ & $1.81 \times 10^6 $ & $0.70$ \\
		\code{Ref-L025N0376} &    $9.70 \times 10^6 $ & $1.81 \times 10^6 $ & $0.70$ \\
		\code{Ref-L025N0752} &    $1.21 \times 10^6 $ & $2.26 \times 10^5 $ & $0.35$ \\
		\code{Recal-L025N0752} &  $1.21 \times 10^6 $ & $2.26 \times 10^5 $ & $0.35$ \\
%		\code{AGNdT9-L050N0752} & $9.70 \times 10^6 $ & $1.81 \times 10^6 $ & $0.70$ \\
		\code{NoAGN-L050N0752} &	$9.70 \times 10^6 $ & $1.81 \times 10^6 $ & $0.70$ \\
		\hline
	\end{tabular}
\end{table}
% This is taken from \citet[table~2]{eagle_paper}.

\subsection{Column density calculation}
\label{sec:cddf_methods}
%{\todo Methodolgy: basic projection code, kernels, reference to appendix for checks. Assumptions in calculating the ion balances and what effect these have. (Need to add in tests of different backgrounds?)
%Getting CDDF from this is easy enough; need a formula for dX etc.\ }

%Observationally, a column density is derived from the amount of absorption along a sightline, which is essentially infinitesimally thin. 
To obtain column densities from the {\eagle} simulations, we calculate the number of ions in columns (elongated rectangular boxes) of finite area and fixed length. 
We `slice' the simulation along the $Z$-axis, then divide each slice along the $X$- and $Y$-directions into narrow 3-dimensional columns, which, when projected, become the pixels of a (2-dimensional) column density map. We make such a column density map for each slice along the $Z$-direction.
We use $32,000$ columns along both $100 \us \txn{cMpc}$ sides of the simulation box, meaning each column has an area of $3.125^2 \us \txn{ckpc}^2$ (comoving kpc) and each map has $1.024 \times 10^{9}$~pixels. We use 16 slices, each $6.25 \us \txn{cMpc}$ thick. In appendix~\ref{app:conv}, we verify that the column density statistics are converged at this pixel size, and examine the effect of slice thickness. These columns are an approximation of what is done observationally, where absorbers are defined by regions of statistically significant absorption in spectra of nearly point-like sources (typically quasars), and column densities are obtained by fitting Voigt profiles to these regions.

To project the ions of each SPH particle onto a grid, we need to know the shape of the gas distribution that individual particles model. There is no unique function for this, since anything smaller than a single SPH particle is, by definition, unresolved. However, there is a sensible choice. In SPH, a similar assumption about the gas distribution must be made, which is used to evolve the gas particles' motions and thermodynamic properties. The function describing this is known as a kernel. 
We use this same function in projection: the C2-kernel\footnote{In the simulation, the function depends on the 3D distance to the particle position; for the projection, we input the 2D distance (impact parameter) instead.} \citep{wendland_1995}. Along the $Z$-axis, we simply place each particle in the slice that contains its centre. We have verified that the results are insensitive to the chosen kernel: the difference in the column density distribution function (CDDF, the main statistic we are interested in) compared to using a different kernel (the {\textsc Gadget} kernel) is $\lesssim 0.05\us\txn{dex}$ for $10^{11}\us \txn{cm}^{-2} < N_{\mathrm{O\, VII, VIII}} < 10^{16.5} \us\txn{cm}^{-2}$, which covers the column densities we are interested in. We discuss the kernels in more details in Appendix~\ref{app:conv}.

An issue that arises in {\eagle} is that for cold, dense gas (star-forming gas), the temperature is limited by a fixed equation of state, used to prevent artificial fragmentation during the simulation \citep{schaye_dalla-vecchia_2008}. This means that the temperature of this gas does not represent the thermodynamic state we would actually expect the gas to be in, and ion balance calculations using this temperature are unreliable. Following e.g.\ \citet{rahmati_etal_2016},  we therefore fix the temperature of all star-forming gas to $10^4 \us \txn{K}$, typical for the warm phase of the ISM. However, we found that the treatment of this gas has virtually no impact, as \ion{O}{vii} and \ion{O}{viii} are high-energy ions, and therefore mainly exist in hot and/or low-density gas (see Section~\ref{sec:pds}). The difference between CDDFs calculated using our standard method, using the temperatures in the simulation output, and excluding star-forming gas altogether is $<0.01 \us \txn{dex}$ at $10^{11}\us \txn{cm}^{-2} < N < 10^{16.5} \us\txn{cm}^{-2}$.

The column density distribution function (CDDF) is defined as
\begin{equation}\label{eq:cddf}
f(N, z) \equiv \frac{\partial^2 n}{ \partial N \partial X},
\end{equation} 
where $X$ is the dimensionless absorption length, $n$ is the number of absorption systems, and $N$ is the ion column density. 
%This quantity $f$ is also what we will plot as column density distributions. (In figures, we will show the right-hand or left-hand side of this equation.) Here, $\diff X$ is the total absorption length, added over all the columns we use to roughly define absorption systems. They give 
Here,
\begin{equation}\label{eq:dX}
\diff X = \diff z \, \left[ H_0 / H(z) \right] (1+z)^2,
\end{equation}  
%This follows from the absorption length definition of :
%\begin{equation}\label{eq:Xabs}
%X(z) = \int_0^z (1+z)^2 \left[ H_0 / H(z) \right] dz.
%\end{equation}
which means that at $z=0$, $\diff X = \diff z$ \citep{bahcall_peebles_1969}. At higher redshifts, using $\diff X$ accounts for changes in the CDDF due to the uniform expansion of the Universe for absorption systems of fixed proper size.
% Therefore, $f(N,z)$ will not change with $z$ if all that happens over cosmic time is that absorbers retain their proper sizes in an expanding universe. Any remaining changes in the CDDF over time are then due to e.g.\ structure formation and enrichment of gas by metals.
We calculate $\diff z$ from the thickness of the spatial region we project (i.e., the length along the projection axis), through the Hubble flow, using the redshift of the snapshot and the cosmological parameters used in the simulation.

\subsection{Spectra and equivalent widths}
\label{sec:ew_methods}

We calculated the equivalent widths along EAGLE sightlines using mock spectra obtained using a program called {\specwizard}. 
%\citet{theuns_efstathiou_etal_1998} give a quick overview of the calculation this program uses in their appendix~A4.
It was developed by Joop Schaye, Craig M. Booth and Tom Theuns, and is described in \citet[section~3.1]{tepper-garcia_richter_etal_2011}.
 A given line of sight is divided into (1-dimensional) pixels in position space. First, the number of ions is calculated for each SPH particle intersecting the line of sight. Then the column density in each pixel is calculated by integrating the particles' assumed ion distribution (defined by the SPH kernel) over the extent of the pixel. This guarantees that the column density is correct at any resolution. For the ion distribution, we assume a 3-dimensional Gaussian, since this is easy to integrate. We have verified that the difference with column densities and equivalent widths obtained with a different kernel is negligible. The ion-number-weighted temperature and peculiar velocity along the line of sight are also calculated in each pixel.

Once this real space column density spectrum has been calculated, it is used to obtain the absorption spectrum in velocity space. For each pixel in the column density spectrum, the optical depth distribution in velocity space is calculated as for a single absorption line. The absorption is centred based on the position of the pixel and the Hubble flow at the simulation output redshift, together with the peculiar velocity of the pixel. The thermal line width ($b$ parameter) is calculated from the temperature. We have not modelled any subgrid/unresolved turbulence, and neglect Lorentz broadening in our calculations. We also use these `ideal' spectra directly, and do not model continuum estimation, noise, line detection, or a detecting instrument in our equivalent width calculations.     

We calculate rest frame equivalent widths, $EW$, from these mock spectra by integrating the entire spectrum:  
\begin{equation} \label{eq:EW}
EW = \left(1 - \sum_{i=0}^{N-1} \frac{F_{i}}{N} \right) \lambda_{\mathrm{rest}} \frac{H(z)\, l_{\mathrm{sl}, \mathrm{com}}}{c\, (1+z)},
\end{equation}     
where $N$ is the number of pixels, $F_{i}$ is the flux in pixel $i$, normalised to the continuum, $\lambda_{\mathrm{rest}}$ is the rest-frame wavelength of the absorption line, $z$ is the redshift, $H(z)$ is the local Hubble flow, $c$ is the speed of light, and $l_{\mathrm{sl}, \mathrm{com}}$ is the comoving box size. The (observer-frame) velocity difference across the full $100 \us \mathrm{cMpc}$ box is $7113 \us \mathrm{km}\,\mathrm{s}^{-1}$ ($\Delta z  = 0.02373$) at redshift $0.1$, and $12026 \us \mathrm{km}\,\mathrm{s}^{-1}$ ($\Delta z  = 0.04011$) at redshift $1$ with the {\eagle} cosmological parameters. The spectra we generate inherit the periodic boundary conditions of the simulation box, and therefore probe velocity differences of at most half these values.

We use \citet{verner_verner_ferland_1996} oscillator strengths and wavelengths for the $18.97\us${\AA} \ion{O}{viii} doublet: $f_{\txn{osc}} = 0.277, 0.139$ and  $\lambda = 18.9671, 18.9725 \us${\AA}, respectively. For the \ion{O}{vii} resonant line, we use values consistent with theirs: $\lambda = 21.6019\us${\AA} and $f_{\txn{osc}} = 0.696$, but ours come from a data compilation by \citet{kaastra_2018_pc}. The other \ion{O}{vii} He-like lines (forbidden and intercombination) have wavelengths sufficiently far from the resonant line \citep[e.g.,][]{bertone_schaye_etal_2010} that these should be clearly separated from each other at the resolutions achieved by instruments aboard  Arcus \citep{smith_abraham_etal_2016_arcus, brenneman_smith_etal_2016}, Athena  \citep{Athena_2017_11, Athena_2018_07}, and Lynx \citep{lynx_2018_08} (see Sections~\ref{sec:spectra} and~\ref{sec:missions} for further discussion of the telescopes and instruments). Therefore, we only discuss the resonant line here, and note that for lines at the detection limits of these instruments, the lines are not blended. 
Also, these lines have much weaker oscillator strengths, by a factor of at least $\sim 6000$, so although they can be important in emission studies, the forbidden and intercombination lines will not be important in absorption. 
%(NIST database, I think, but I'll have to wait till the shutdown ends to check).}   

Since we compute the equivalent width by integrating the entire spectrum, if there is more than one absorption system along the line of sight, we will only recover the total equivalent width. However, the equivalent widths we are interested in are the potentially detectable ones, i.e., the rarer, larger values, corresponding to the highest column densities. These will not generally be `hidden' in projection by even larger EW absorption systems, though we find that multiple systems of similar strength along such lines of sight are not uncommon. We will discuss this issue further in Section~\ref{sec:EW}, where we examine some spectra and the conversion between column density and equivalent width. 

%For selected pixels in the column density maps (using $100 \us \txn{cMpc}$ columns), we calculated mock spectra at the pixel centres. This allowed us to find the relation between the (projected) column density and equivalent width in this simulation. By extrapolating this relation to sightlines that were not selected, we can gauge the equivalent width distribution in \code{Ref-L100N1504} up to higher equivalent widths than with a random sample, since computational costs excludes the possibility of computing $32000^2$ mock spectra from this simulation.   

To determine the relation between the (projected) column density and equivalent width, we obtained mock spectra along 16384 lines of sight through the reference simulation (\code{Ref-L100N1504}) output at redshift $0.1$.
% and calculated equivalent widths for these sightlines. 
Since we are mainly interested in the systems with large equivalent widths, we did not choose random lines of sight.
%, since the CDDFs indicate that the highest column-density systems would be rare in such samples. 
Instead, we selected sightlines using the column density maps for \ion{O}{vi}, \ion{O}{vii}, and \ion{O}{viii}: for each ion, we selected samples randomly from evenly spaced log column density bins. 
%For each ion, the highest column density bins had fewer sightlines available than an even sampling would require. The sightlines `left over' this way were selected from the next-highest column density bins, with the requirement that the number of sightlines in these bins was at most double that initially assigned. In this initial selection, the same line of sight was sometimes selected for more than one ion. The selection process was then repeated until the desired total sample size was reached. 
We used column densities over the full $100\us\mathrm{cMpc}$ box depth, along the $Z$-direction, for this selection.      

The matching of equivalent widths to projected column densities is the reason we calculate the equivalent widths along $100\us\mathrm{cMpc}$ sightlines, despite the possibility of combining two absorption systems this way. The peculiar velocities of the absorbers are often $\sim 500 \us \mathrm{km}/\mathrm{s}$, and can reach $1000 \us \mathrm{km}/\mathrm{s}$ or more. Compared to a Hubble flow across the redshift~$0.1$ box of $\sim 7000 \us \mathrm{km}/\mathrm{s}$, this means that matching absorption in the spectrum to one of 16 slices along the line of sight would not be a straightforward process. A mistake here could have major implications for the reconstructed equivalent width distribution, since the highest column density absorption systems are orders of magnitude more rare than more typical ones. This means that mismatches of absorption systems with large equivalent widths to lower column densities could lead to a significant overestimation of the abundance of these rare absorption systems. For $100\us\mathrm{cMpc}$, there is no possibility of mismatches along the line of sight.

With this sample of column densities and equivalent widths, we make a 2-dimensional histogram of column density and equivalent width. By normalising it by the number of absorbers at each column density, we obtain a matrix from this histogram that we use to convert column density distributions to equivalent width distributions. Though we make these matrices for $100\us \txn{cMpc}$ sightlines, we will also apply them to column density distributions obtained using $6.25 \us \txn{cMpc}$ slices of the simulation. That is to say, we assume that the relation between column density and equivalent width does not depend significantly on the slice thickness (sightline length).

\section{Results}
\label{sec:res}

\subsection{Column density distributions}
\label{sec:cddfs}

\begin{figure}
\includegraphics[width=\columnwidth]{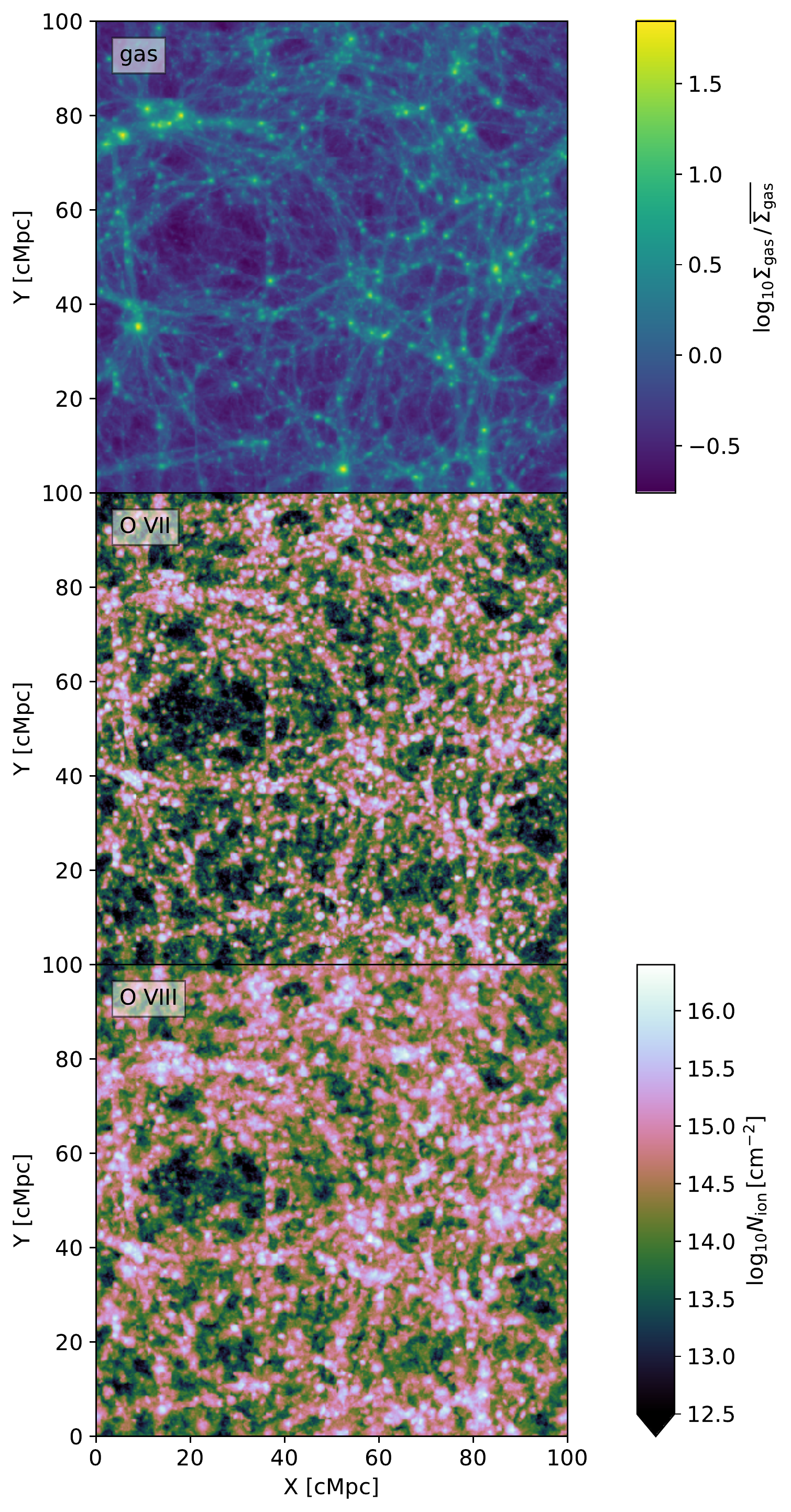}
\caption{A map of the \ion{O}{vii} (middle) and \ion{O}{viii} (bottom) column densities in the $100 \us \txn{cMpc}$ {\eagle} \code{Ref-L100N1504} simulation at $z=0.1$, at a resolution of $400^2$ pixels and with a column depth of $100 \us \txn{cMpc}$. The full column density range is not shown. The top panel shows the corresponding gas surface overdensities. \ion{O}{vii} and \ion{O}{viii} trace large-scale structures.}
\label{fig:coldensmaps}
\end{figure}

The starting points for our column density distributions are column density maps.
Fig.~\ref{fig:coldensmaps} shows a map of column densities calculated as we described in Section~\ref{sec:cddf_methods}. These column densities are for columns through the full depth of the $100 \us \txn{cMpc}$ box, using $400^2$ pixels, and therefore do not produce the converged CDDFs we use in the rest of the paper. They do demonstrate that \ion{O}{vii} and \ion{O}{viii} trace the large-scale structure in the box, as indicated by the total gas surface density map. We will investigate the spatial distribution of these ions (column densities around haloes of different masses) in more detail in an upcoming paper.

\begin{figure}
\centering
\includegraphics[width=0.5\textwidth]{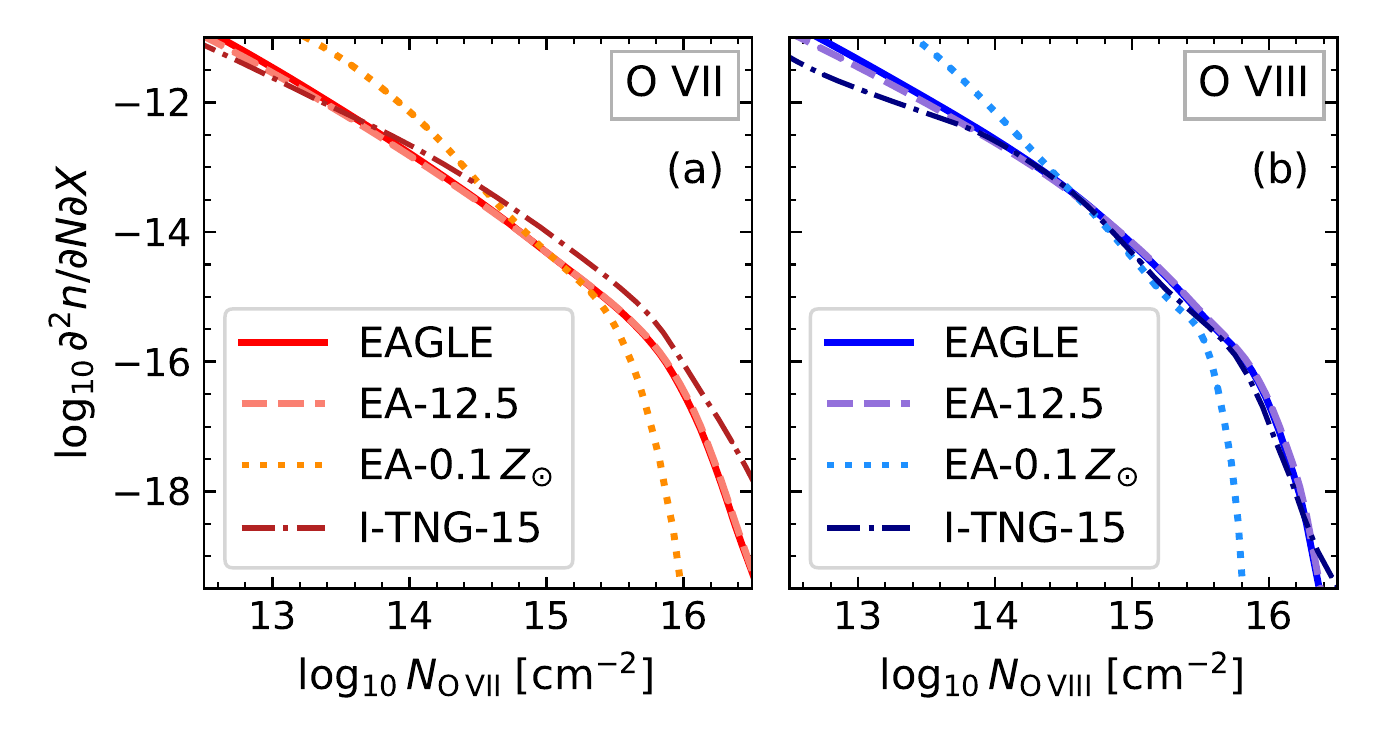}
\caption{The column density distribution function (CDDF), as described in equation~\ref{eq:cddf}, of \ion{O}{vii} (a) and \ion{O}{viii} (b) in the $100 \us \txn{cMpc}$ {\eagle} reference simulation at $z=0.0$. The solid lines (EAGLE) show our standard CDDF: it uses the simulation oxygen abundances, and column densities are measured in $6.25\us\txn{cMpc}$ long columns. The dotted lines (EA-$0.1Z_{\sun}$) use $0.1$ times the solar oxygen abundance for all gas instead. All the {\eagle} column densities are calculated from $32000^2$ columns in each slice along the line of sight direction; each column has an area of $3.125^2 \us \txn{ckpc}^2$. We also compare to the IllustrisTNG~100-1 CDDF  \citep[I-TNG-15]{Nelson_etal_2018_hiO}, shown with dot-dashed lines. Since these column densities were measured in $15\us\txn{cMpc}$ columns, we also show the CDDF we obtain using $12.5\us\txn{cMpc}$ columns for comparison (dashed lines, EA-$12.5$). This shows that {\eagle} and IllutrisTNG-100 predict similar column density distributions for these ions, and that using realistic metallicities is crucial in determining the column density distribution.}
\label{fig:cddf_z0p1}
\end{figure}

Fig.~\ref{fig:cddf_z0p1} shows the \ion{O}{vii} and \ion{O}{viii} CDDFs (solid lines) we obtained from the reference simulation (\code{Ref-L100N1504}). The main feature visible here is the `knee' or break in the distributions for both ions at column densities around $10^{16} \us \txn{cm}^{-2}$. We will investigate this feature in detail in Section~\ref{sec:break}, where we find that it marks the transition from absorption arising in extrahalo and intrahalo gas.
 %where we find that it coincides with where absorption mainly comes from gas at overdensities $\sim 100$, likely associated with haloes. 
We show in appendix~\ref{app:conv} that these distributions are converged with pixel size in the column density range we shown here (up to $10^{16.5}\us\txn{cm}^{-2}$), and reasonably converged with the size and resolution of the simulations. We have verified that the effects of some technical choices in the calculation of the column densities are small. 

The dotted lines indicate the distribution we would get if all gas had an oxygen abundance  $0.1$ times the solar value. This demonstrates the impact of non-uniform metal enrichment: it causes the CDDF to be much shallower, extending to larger maximum column densities. We have verified that hot or high-density absorbers are most enriched with metals, while the less dense IGM (especially the cooler part) is less enriched. This means the \ion{O}{vii} and \ion{O}{viii} CDDFs are sensitive to the distribution of oxygen.

From a different perspective, $40 \us \%$ of gas-phase oxygen and $27 \us \%$ of total oxygen\footnote{ The total amount of oxygen here is the amount of oxygen released by stars, and does not count oxygen that remains in stellar remnants (a significant quantity) or is swallowed by black holes (a relatively insignificant quantity). Compared to gas-phase oxygen, the total also includes oxygen that was in gas, but is now in stars that formed from that gas.} is in these two ions at $z=0.1$ in the reference simulation (\code{Ref-L100N1504}). That means that absorption from these ions is also important in determining where the bulk of the metals produced in galaxies go.

We also compare our \ion{O}{vii} and \ion{O}{viii} CDDFs to those of \citet{Nelson_etal_2018_hiO}. They obtained column density distributions for these ions from the IllustrisTNG~100-1 simulation, using a similar method to ours. \citet{Nelson_etal_2018_hiO} use 
% \citep{pillepich_springel_etal_2018, weinberger_springel_etal_2017}
%tracked element abundances and 
metallicity-dependent ion balances to calculate their column densities, but remark that the dependence of ion fractions on metallicity is minimal. 
%They measure column densities by calculating the number of ions in rectangular boxes in their simulations, like we do. 
They use $15 \us \txn{cMpc}$ long columns to calculate their column densities, but we find that this makes almost no difference for the CDDF in the column density range shown here. They also use a different UV/X-ray background than adopted here. 

The IllustrisTNG~100-1 simulation uses a similar cosmology, volume, and resolution as {\eagle}, but a different hydrodynamics solver (moving mesh instead of SPH). While it also includes star formation and stellar and AGN feedback, it models these processes using different subgrid prescriptions.  
Fig.~\ref{fig:cddf_z0p1} shows that the CDDFs from the {\eagle} reference simulation (\code{Ref-L100N1504}) and IllustrisTNG-100-1 agree remarkably well; the differences are small compared to the effect of assuming a constant metallicity.
Note that the {\eagle} and IllustrisTNG-100-1 absorber numbers differ by a factor $\approx 2$ at column densities above the break for \ion{O}{vii}, though the differences are small compared to the dynamical range shown. A comparison at fixed metallicity (not shown) did not decrease the differences, meaning they are not dominated by different metal distributions in the two simulations. 
%The breaks in the CDDFs seem to be at similar column densities. 
%\citet{Nelson_etal_2018_hiO} do not show data at column densities above $10^{16}\us\txn{cm}^{-2}$, and they use somewhat larger pixels than we do to calculate column densities (which means the distribution is converged up to a lower maximum column density), so we do not consider the diverging CDDFs at the highest column densities shown to be a major issue.
%However, this transition seems to happen at slightly higher column densities in {\eagle} \code{Ref-L100N1504} than in IllutrisTNG~100-1, and the slight curves in the roughly power law behaviour of the \ion{O}{viii} CDDF at lower column densities seem to be stronger in  IllutrisTNG~100-1 than in {\eagle} \code{Ref-L100N1504} at both redshifts.

\begin{figure}
\includegraphics[width=0.5\textwidth]{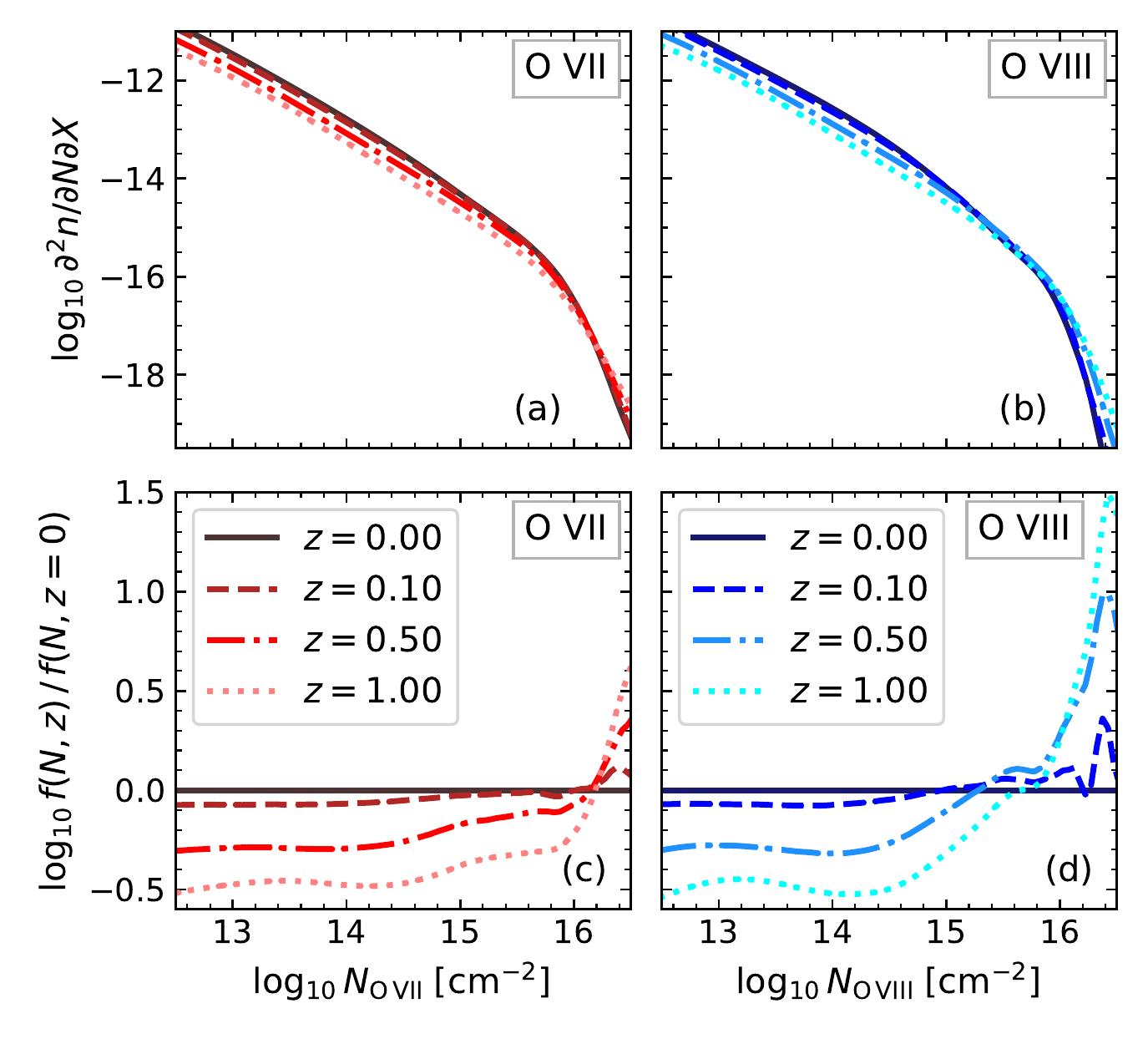}
\caption{The evolution of the column density distribution function (CDDF) of \ion{O}{vii} (a, c) and \ion{O}{viii} (b, d) in the $100 \us \txn{cMpc}$ {\eagle} reference simulation from $z=1$ to $0$.  The top panels show the CDDFs themselves, while the bottom panels show the CDDFs relative to the $z=0$ CDDF. This shows that the CDDF evolves only mildly between redshifts~$0$ and~$1$.}
\label{fig:cddf_zev}
\end{figure}

Fig.~\ref{fig:cddf_zev} shows that the CDDF evolves little between redshifts~$0$ and~$1$. 
The shape of the CDDF does not change much, but there is some evolution: the incidence of high column densities decreases with time, while that of low column densities increases. The distribution evolves the least around its break. We have verified that these changes are not simply the result of the changing pixel area and slice thickness in physical units at the fixed comoving sizes we used, by comparing these changes to the effects of changing column dimensions described in appendix~\ref{app:conv}. 
The evolution we find in {\eagle} is similar to that shown in Fig.~5 of \citet{Nelson_etal_2018_hiO}. 
%Their plot shows that for both \ion{O}{vii} and \ion{O}{viii}, the incidence rate of absorption systems at low column densities increases by $\sim 0.5 \us \mathrm{dex}$, while they remain roughly equal at column densities $\sim 10^{16} \us \mathrm{cm}^{-2}$. In both simulations, the transition occurs at higher column densities for \ion{O}{vii} than \ion{O}{viii}. 

\subsection{Absorption spectra}
\label{sec:spectra}

\begin{figure*}
\begin{center}
\includegraphics[width=2\columnwidth]{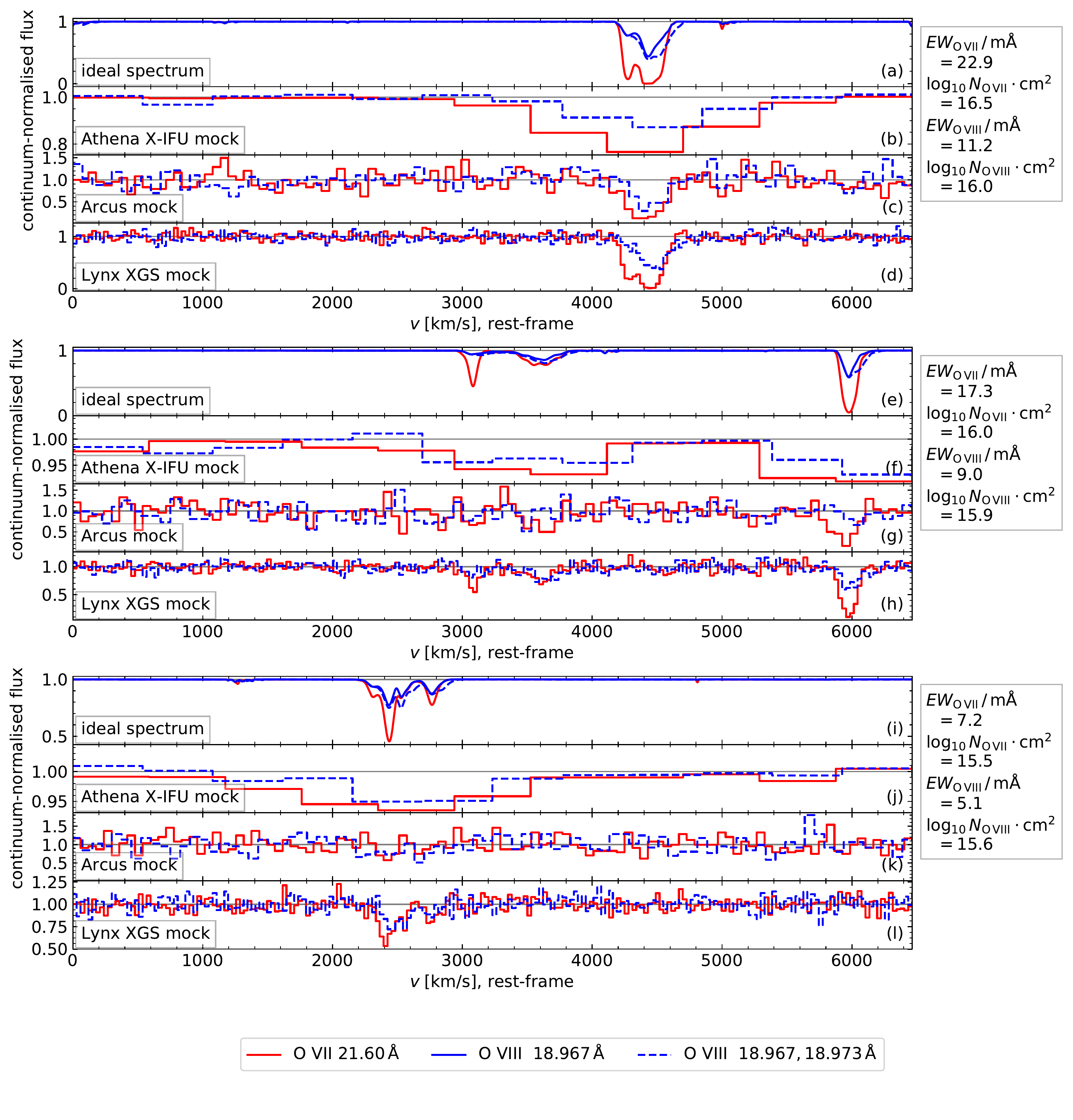}
\end{center}
\caption{Example mock spectra from the {\eagle} reference simulation (\code{Ref-L100N1504}) at $z=0.1$. On the x-axis, we show rest-frame velocity, on the y-axis, we show the flux, normalised to the continuum. The (a), (e), and (i) panels show the ideal spectra the other panels are derived from. These have a resolution of $2 \us\mathrm{km}\,\mathrm{s}^{-1}$ and are noise-free. In the other panels, we show mock spectra for the Athena~X-IFU (b, f, j), Arcus (c, g, k), and Lynx~XGS (d, h, l) effective areas and resolutions. These spectra have Poisson noise added to them, but no other error or uncertainty sources. We show the spectra for the \ion{O}{vii} resonant line and the \ion{O}{viii} doublet. In the ideal spectra, we also show the spectrum using only the stronger of the two \ion{O}{viii} doublet lines to gauge the effect of the weaker one. The three panel groups (a, b, c, d), (e, f, g, h), and (i, j, k, l) represent three different sightlines, selected to illustrate some general trends. We chose sightlines with \ion{O}{vii} column densities of $10^{16.5}$, $10^{16.0}$, and $10^{15.5}\us\mathrm{cm^{-2}}$. The column densities and EWs measured over the whole sightline, for the ideal spectra, are shown to the right of the spectra. For Athena~X-IFU mock spectra, we use a (full-width half-maximum) resolution of $2.1 \us \mathrm{eV}$ and an effective area (specified for soft X-rays) of $1.05 \us \mathrm{m}^2$  \citep{Athena_2018_07, Athena_2017_11}. For Arcus, we used a spectral resolving power $E / \Delta E$ (where $\Delta E$ is the full-width half-maximum and $E$ is the observed energy) of 2000 below $21.60$~{\AA} (observed wavelength), and 2500 at longer wavelengths, with an effective area (average at $16$--$28\us${\AA}) of $250 \us \mathrm{cm}^{2}$ \citep{smith_abraham_etal_2016_arcus}. We used $E / \Delta E = 5000$ and an effective area (specified for $0.3$--$0.7 \us \txn{keV}$) of $4000 \us \txn{cm}^2$ for the Lynx XGS \citep{lynx_2018_08}. The background (blazar) source flux was $1 \times 10^{-11} \us \mathrm{erg}\, \mathrm{cm}^{-2} \mathrm{s}^{-1}$ between $2$ and $10\us \mathrm{keV}$ with a photon spectral index of $\Gamma = 1.8$  as specified for the Athena $5 \sigma$ weak line sensitivity limit by \citet{Athena_2017_11}, and an observing time of $100 \us \mathrm{ks}$. The spectra, like the simulations they come from, are periodic. We bin the instrument mock spectra to two pixels per full-width half-maximum.}
\label{fig:mockspectra}
\end{figure*}

We move on to an examination of a few of the spectra we will obtain equivalent widths (EWs) from. Three example spectra are shown in Fig.~\ref{fig:mockspectra}. We show our ideal spectra (resolution of $2\us\mathrm{km}\,\mathrm{s}^{-1}$, no noise), and mock spectra with spectral resolutions and effective areas corresponding to the Athena~X-IFU, Arcus\footnote{The effective collecting area for Arcus is $500 \us \txn{cm}^{-2}$, but the spectrometer only uses half of this to improve spectral resolution (subaperturing).}, and Lynx~XGS instruments for a $100 \us \mathrm{ks}$ observation of a blazar sightline. 
For the blazar, we use a source flux of $1 \times 10^{-11} \us \mathrm{erg}\, \mathrm{cm}^{-2} \mathrm{s}^{-1}$ between $2$ and $10\us \mathrm{keV}$ with a photon spectral index of $\Gamma = 1.8$,  as specified for the Athena $5 \sigma$ weak line sensitivity limit by \citet{Athena_2017_11}.
%We have checked that the equivalent widths in our total sample are converged at this resolution. 
We model the Poisson noise, but no other sources of error/uncertainty, and determine the unabsorbed number of photons per bin from the blazar spectrum and the redshifted line energy. We do not account for the slope of the blazar spectrum across our sightline. The spectra are periodic, like the simulations we derived them from, which is why we see absorption at low line of sight velocities in panel~(f). We selected these spectra to be roughly representative of a larger sample we examined: ten spectra at column densities of $10^{14}$--$10^{17} \us \mathrm{cm}^{-2}$, in $0.5 \us \txn{dex}$ intervals, for each ion.  

The ideal spectra illustrate that most absorption systems have more than one component. Single-component systems do occur, though. Athena is more sensitive than Arcus, but it cannot distinguish close components as easily. 
%For Athena, any absorption system will appear as one absorber. This justifies the use of our estimated absorption system equivalent width distribution when making predictions for these instruments in Section~\ref{sec:missions}. It also means that 
Line widths, in Athena~X-IFU and even in Arcus spectra, may not provide very stringent limits on absorber temperatures, since these widths are inferred mostly from the component structure. With Lynx, it may be more feasible to separate the different components.

We note that it is not uncommon to find multiple absorption systems along a single line of sight. This should not be a major problem for our $6.25 \us \txn{cMpc}$ CDDFs, but it could affect the relation between column density and EW, since the column densities and equivalent widths of the different systems simply add up, even when the stronger system's absorption is saturated. However, we show in appendix~\ref{app:projchoice} that the EW distributions obtained using different conversions between column density and equivalent width are quite similar around the expected detection thresholds of Athena, Arcus, and Lynx, so we do not expect this to have a significant impact on our survey predictions.
%, especially compared to the differences between the predictions we get for different sightline lengths. 

We see clear differences between the spectra obtained with the simulated instruments: Arcus has much higher resolution than Athena, enabling clearer measurements of line shapes and detections of different components, while Athena's much higher effective area enables it to detect weaker lines in the same observing time. Note that the science requirements for these two instruments specify different observing times for the proposed blind WHIM surveys. The Lynx~XGS is planned to have the highest spectral resolution of the three and a large effective area, meaning it recovers these lines best. We will discuss these instruments in more detail in Section~\ref{sec:missions}. 

Finally, the ideal spectra confirm that we need to model the \ion{O}{viii} doublet as two blended lines. They are intrinsically blended, but the offset between the lines is large enough that it might affect the line saturation (panel~a), so modelling the two lines as one may also be inadequate. The EW of the detected line will determined by both doublet components together.

\subsection{Equivalent widths}
\label{sec:EW}
%{\todo Main question: what do we need to convert a CDDF to an EW distribution? What are the detection prospects for O7 and O8 with e.g.\ Athena, Arcus, Lynx?}

As described in Section~\ref{sec:ew_methods}, we calculated the rest-frame equivalent width (EW) distribution in the reference simulation (\code{Ref-L100N1504}) by first  obtaining the relation between the column density from the projection and the total EW along the same line of sight for a subset of sightlines with relatively high column densities. This relation includes the scatter in EW at fixed column density. We then used this relation to obtain the EW distribution from the column density distribution. This approximation is necessary because computing absorption spectra corresponding to the full $32000^2$ grid of pixels is too computationally expensive. We did not fit any models to this relation, but simply binned the distribution of EWs as a function of column density, and used it as a conversion matrix. Below the minimum column density of $10^{13} \us \mathrm{cm}^{-2}$ for which we selected sightlines, we used the linear curve of growth to make the conversion. We discuss the calculation of the EW distributions and the effect of including scatter in more detail in Appendix~\ref{app:projchoice}.
%, or the best-fit $b$ parameter; there were no significant differences between the two. 

% http://www.physics.sfsu.edu/~lea/courses/grad/cog.PDF, checked that they have the same b_thermal definition
\begin{figure}
\includegraphics[width=0.5\textwidth]{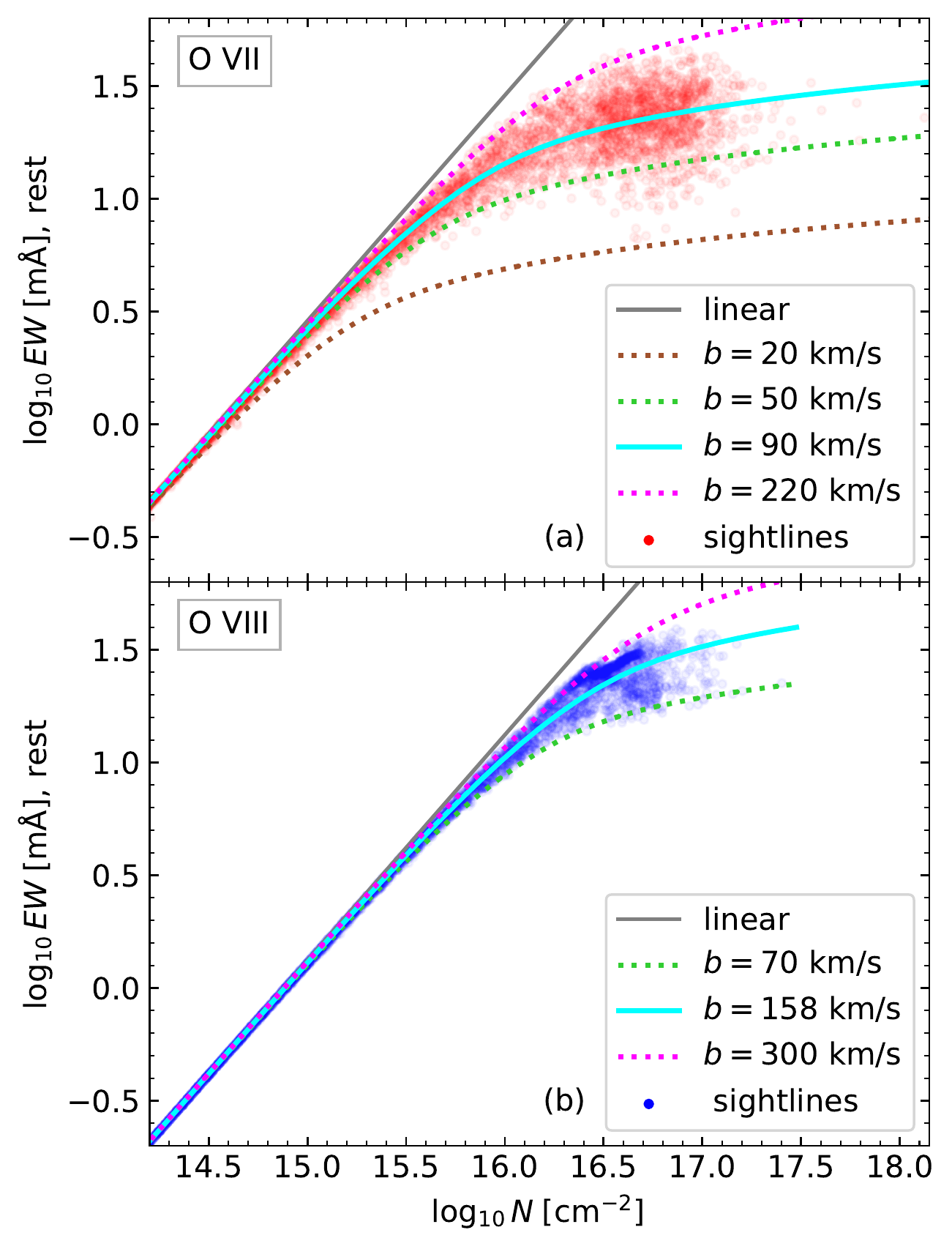}
\caption{The curve of growth for the \ion{O}{vii} resonant line (a) and the \ion{O}{viii} doublet (b) in the reference (\code{Ref-L100N1504}) simulation at redshift~$0.1$. We plot the rest frame equivalent width calculated over the full $100 \us \mathrm{cMpc}$ sightlines against the column density along those same sightlines. The solid cyan lines are for the best-fit $b$ parameters to the log equivalent width, using only the sightlines selected for the ion in question. The solid gray lines indicate the linear column density equivalent width relation, which applies to optically thin gas. The dotted lines roughly indicate the upper and lower range of $b$ parameters, with the brown line indicating the bare minimum for \ion{O}{vii} while the green lines indicate values below which absorbers are rare. This demonstrates our sampling of the column density equivalent width relation, and the range of single-absorber-equivalent widths associated with it.}
\label{fig:cog}
\end{figure} 

Fig.~\ref{fig:cog} shows the rest-frame equivalent width (EW) as obtained by integrating the absorption spectra, as a function of the column density computed with the same code ({\specwizard}) used to generate the absorption spectra\footnote{Here, for each ion, we only show the sightlines selected uniformly in column density for that particular ion, while we use the very similar distribution of all sightlines we have spectra for when converting column densities to EWs. 
%If we show all the sightlines for which we obtained mock spectra, the distribution of points in the plot looks very similar. One small difference is that for the \ion{O}{vii} resonant line, equivalent widths consistent with $b$ parameters down to $20 \us \txn{km}\,\txn{s}^{-1}$ are present at all column densities, though they are rare. Another difference is that the distributions are smoother, likely due to the larger size of the dataset. The best-fit $b$-parameters obtained from fits to the subsamples and full datasets differ by $\leq 5 \us \txn{km}\,\txn{s}^{-1}$. 
If we plot the EWs against the column densities from our column density maps, there is slightly more scatter in the relation, which is clear at lower column densities. This is due to mismatches between the column densities calculated using the two different methods. These differences are generally small. At the highest column densities, they are due to non-convergence of the highest column densities with projection resolution (see appendix~\ref{app:conv}). More generally, the spectra and 2d projections assume slightly different gas distributions for a single SPH particle, and the 2d projections deal with SPH particles smaller than the projection column size in ways that the spectra do not. We account for these small differences when we calculate the EW distributions: we use the relation between column densities from our grid projections and EWs calculated at matching sightlines.}. 
This shows that for the highest column densities and EWs ($\gtrsim 10^{15} \us \txn{cm}^{-2}$ for \ion{O}{vii}, $\gtrsim 10^{15.5} \us \txn{cm}^{-2}$ for \ion{O}{viii}) , the relation between the two is no longer linear. Indeed, inspection of the mock spectra (examples in Fig.~\ref{fig:mockspectra}) shows the lines at these column densities are typically saturated. Furthermore, the scatter in the relation at the highest column densities is large, meaning a single curve of growth cannot be used to accurately convert between column density and EW. We discuss the effect of scatter on the EW distributions in Appendix~\ref{app:projchoice}.
 
We characterise this relation using a $b$-parameter dependent column density-EW relation. This $b$ is a measure of the line width: a model Gaussian absorption line profile has a shape $ 1- \exp(-\tau(\Delta v))$, with $\tau(\Delta v) \propto N b^{-1} \exp(-(\Delta v\, b^{-1})^2 )$. Here $\Delta v$ is the distance from the line centre in rest-frame velocity units, $\tau$ is the optical depth, and $N$ is the column density. The constant of proportionality depends on the atomic physics of the transition producing the absorption. 
The solid and dotted lines in Fig.~\ref{fig:cog} show curves of growth for different $b$ parameters. There is no clear trend of $b$ parameter with column density.

\begin{figure}
\includegraphics[width=0.5\textwidth]{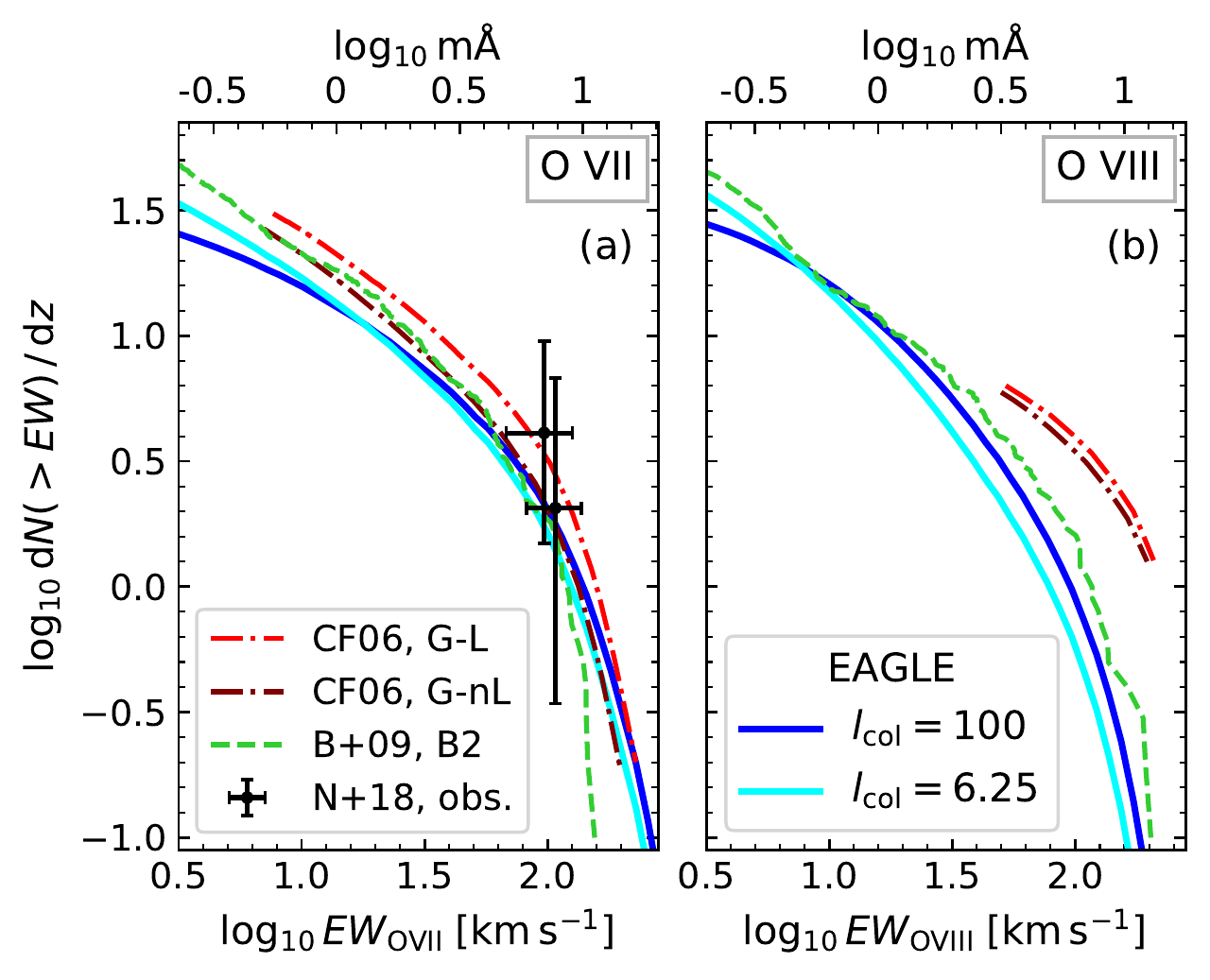}
\caption{Comparison between the \ion{O}{vii} (a) and \ion{O}{viii} (b) rest-frame equivalent width distributions in the {\eagle} reference simulation (\code{Ref-L100N1504}) at $z=0.1$, the GSW model of \citet{cen_fang_2006} (CF06) at $z\sim 0$ with (G-L) and without (G-nL) assuming ionization equilibrium, the \citet{branchini_ursino_etal_2009} (B+09) B2 model at $z \leq 1$, and observations by \citet{nicastro_etal_2018} with the update of \citet{nicastro_2018} (N+18). Note that the units differ from those in Fig.~\ref{fig:cog}: the equivalent widths are in km/s (still rest-frame), and the number of absorption systems per unit redshift $z$ is counted, rather than the number per unit absorption length $X$. The EAGLE distributions were computed using the $100 \us \txn{cMpc}$ and $6.25 \us \txn{cMpc}$ CDDFs and the $100 \us \txn{cMpc}$ column density equivalent width relation calibrated to all the sightlines in our sample. This shows the rough agreement between the equivalent width distribution we predict for \ion{O}{vii} and that predicted in other work, and an agreement with the available observations. The \ion{O}{viii} distributions have larger differences, and there are no extant observational constraints we are aware of.}
\label{fig:ew_litcomp}
\end{figure}

In Fig.~\ref{fig:ew_litcomp}, we show the EW distributions we obtain from the CDDFs measured over $6.25\us\txn{cMpc}$ and $100\us\txn{cMpc}$ sightlines, and compare them to simulation predictions from other groups and a recent measurement. First, comparing our two distributions, the differences are as expected: when we group together multiple absorption systems along a longer sightline, we measure a higher number density of high EWs, while lower-EW absorption systems are `swallowed up' and therefore less common. The differences are larger for \ion{O}{viii}, matching what we see in the CDDF when we vary the slice thickness in appendix~\ref{app:conv}.

We also compare our distributions to predictions made by \citet{cen_fang_2006} and \citet{branchini_ursino_etal_2009}, and to recent observations by \citet{nicastro_etal_2018}, including the update for the equivalent width of the more distant absorber given by \citet{nicastro_2018}. 
\citet{nicastro_2018} measured two \ion{O}{vii} resonant line absorption systems at redshifts~$0.43$ and~$0.36$ in the spectrum of a very bright blazar, and used that to calculate the EW distribution for such absorbers. 
These measurements are consistent with all the predictions made so far, in part because the distribution measured from two systems is still quite uncertain\footnote{The $68\us \% $ error bars in the figure account for this. Considering the overlapping error bars, we do not interpret these two data points as pointing to a steep slope in the distribution function.}.
Still, this agreement is encouraging. Fig.~3 of \citet{nicastro_etal_2018} shows a similar agreement. 
The {\eagle} EW distribution shown there was obtained from the column density distribution using a fixed $b$-parameter, and the $100\us\txn{cMpc}$ sightline CDDF.

The \citet{cen_fang_2006} predictions we compare to are based on the \citet{cen_ostriker_2006} simulations in a $123 \us \mathrm{cMpc}$ box with $1024^3$ cells, and a dark matter particle mass of $3.9 \times 10^{8} \us \txn{M}_{\sun}/h$. They use somewhat different cosmological parameters than {\eagle}. Since these simulations solve the hydrodynamics equations on a fixed Eulerian grid, galaxies and the effects of their feedback are much less well-resolved in these simulations than in {\eagle}.
%, though they have higher resolution in low-density regions. 
%The \ion{O}{iv} through \ion{O}{ix} oxygen ions are followed in the simulation, and evolved without assuming ionization equilibrium, although the total oxygen abundance was rescaled after the simulation to better match abundances in the intra-cluster medium. 
The results we show here are from their simulations with galactic feedback (`GSW'). The `G-nL' curves are derived from simulations tracking ion abundances without assuming ionization equilibrium, while the `G-L' curves were obtained by using equilibrium ion abundances instead. For the photoionization in this model, they use the $z=0$ radiation field from the simulation, which is consistent with observations. They measured the EW distribution by making mock spectra with a resolution of $19\us \txn{km}\,\txn{s}^{-1}$ \citep{fang_bryan_canizares_2002} for random sightlines. They then identified and counted absorption lines in those. Their predictions are specified for redshift~$0$.

In Fig.~\ref{fig:ew_litcomp}, we also compare our EW distributions to the B2 model of \citet[Fig.~4]{branchini_ursino_etal_2009}. In this work, they present three models, of which they consider B2, the model shown here, to be the most realistic. It is also the most optimistic about what we can detect. The B2 model uses simulations by \citet{borgani_murante_etal_2004}. This is an SPH simulation using $2 \times 480^3$ particles (gas + dark matter), in a $192 \us \mathrm{cMpc}/h$ box and a different set of cosmological parameters from \citet{eagle_paper} or \citet{cen_ostriker_2006}. The mass resolution is three orders of magnitude lower than for {\eagle}. It includes star formation, supernova feedback, and radiative cooling/heating (for primordial gas). \citet{branchini_ursino_etal_2009} impose a density-metallicity relation to get the oxygen abundance for each SPH particle. In model~B2, the relation from the simulations of  \citet{cen_ostriker_1999} is imposed, including scatter. They calculate ion abundances with \code{Cloudy}, assuming ionization equilibrium. They measure the EW distribution from these sightlines by constructing lightcones, from which they obtain mock spectra at a resolution of $9\us \txn{km}\,\txn{s}^{-1}$, and then identify lines in these. \citet{branchini_ursino_etal_2009} plot observed equivalent widths in their figure, for absorbers at redshifts $0$--$0.5$. We show the rest-frame distribution obtained by assuming all their absorbers are at redshift~$0.25$. 
For \ion{O}{vii}, both the \citet{branchini_ursino_etal_2009} and the \citet{cen_fang_2006} models agree reasonably well with our EW distribution, especially at higher column densities. 
%We match the non-equilibrium model of \citet{cen_fang_2006} better than their non-equilibrium model, although the differences are generally small. Where our predictions differ most from the others, Fig.~\ref{fig:EWdist} shows that the conversion between column-density and equivalent width is quite certain, so our rough conversion method will not explain the differences. 
For \ion{O}{viii}, the differences are much larger. 
The \citet{branchini_ursino_etal_2009} B2~model is roughly consistent with {\eagle} in this respect, while the latter predicts many fewer absorption systems at high column densities than \citet{cen_fang_2006}.  

\subsection{The physical origin of the break}
\label{sec:break}

We now focus on the main feature of the \ion{O}{vii} and \ion{O}{viii} column density distributions: the break (`knee') at large column densities (just below $10^{16} \us \txn{cm}^{-2}$, see Fig.~\ref{fig:cddf_z0p1}). We look into the temperatures, densities, and metallicities of absorption systems at different column densities to investigate this.

\begin{figure}
\includegraphics[width=\columnwidth]{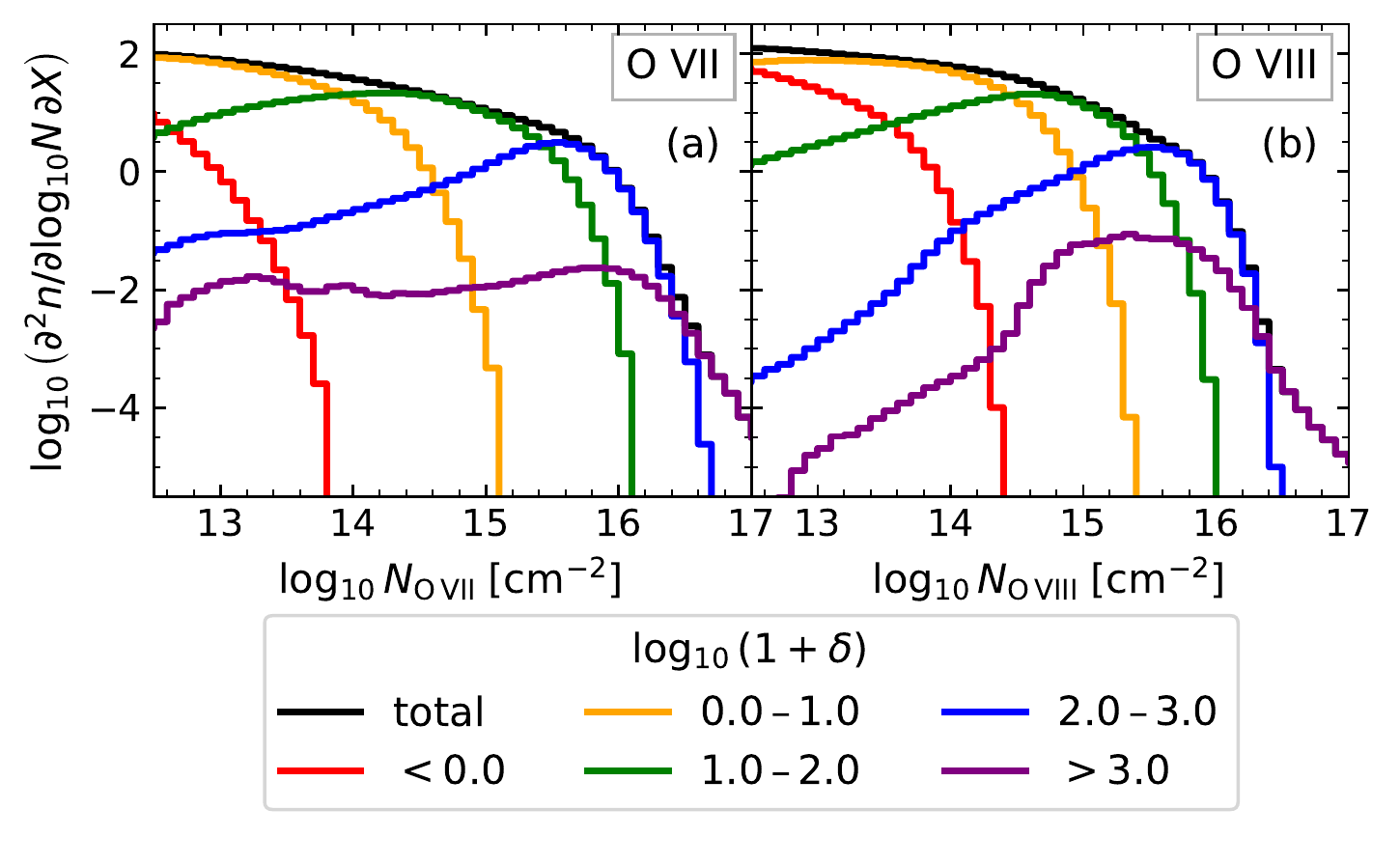}
\caption{The CDDFs for \ion{O}{vii} (a) and \ion{O}{viii} (b) absorption systems with ion-weighted overdensities in the ranges indicated in the legend. These distributions are for $6.25 \us \mathrm{cMpc}$ sightlines in the reference (\code{Ref-L100N1504}) simulation at redshift~$0$. This shows that the break in the distribution occurs at overdensities $\sim 10^2$ for both ions, and therefore likely derives from the transition from absorption tracing sheets and filaments to absorption tracing haloes.} 
%The bins are not exactly at integer log overdensity values, as the initial binning was done at $0.1 \us \mathrm{dex}$ round values in (cgs) mass density. }
\label{fig:cddfs_o78_by_rho}
\end{figure}
% spurious low values (rho < 10^-33 g/cm^-3 or log delta + 1 ~ -2.6) do not contribute at all

\begin{figure}
\includegraphics[width=\columnwidth]{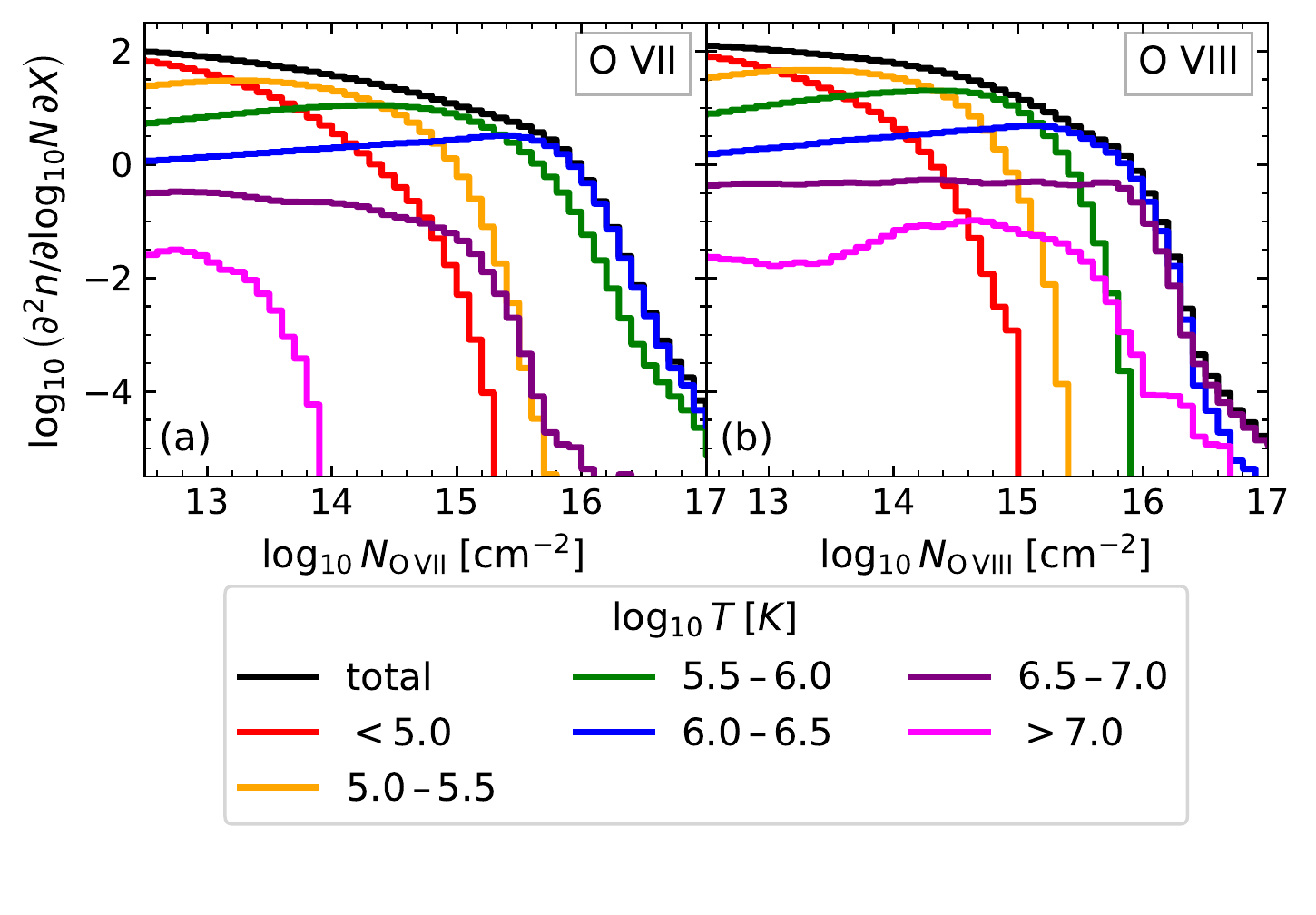}
\caption{The CDDFs for \ion{O}{vii} (a) and \ion{O}{viii} (b) decomposed into absorption systems with ion-weighted temperatures in the ranges indicated in the legend. This shows that WHIM gas dominates the distributions at all column densities likely to be observable in the foreseeable future, and that absorbers with column densities beyond the break are at temperatures associated with collisional ionization.}
\label{fig:cddfs_o78_by_T}
\end{figure}
% The black lines show the total column density distributions. These distributions are for $6.25 \us \mathrm{cMpc}$ sightlines in the reference (\code{Ref-L100N1504}) simulation at redshift~$0$. In both panels, the gray line indicates gas at WHIM temperatures.
% checked: spurious low values (T < 10^2.5 K) have no visible impact on this plot; in fact their contributions are identically zero

\begin{figure}
\includegraphics[width=\columnwidth]{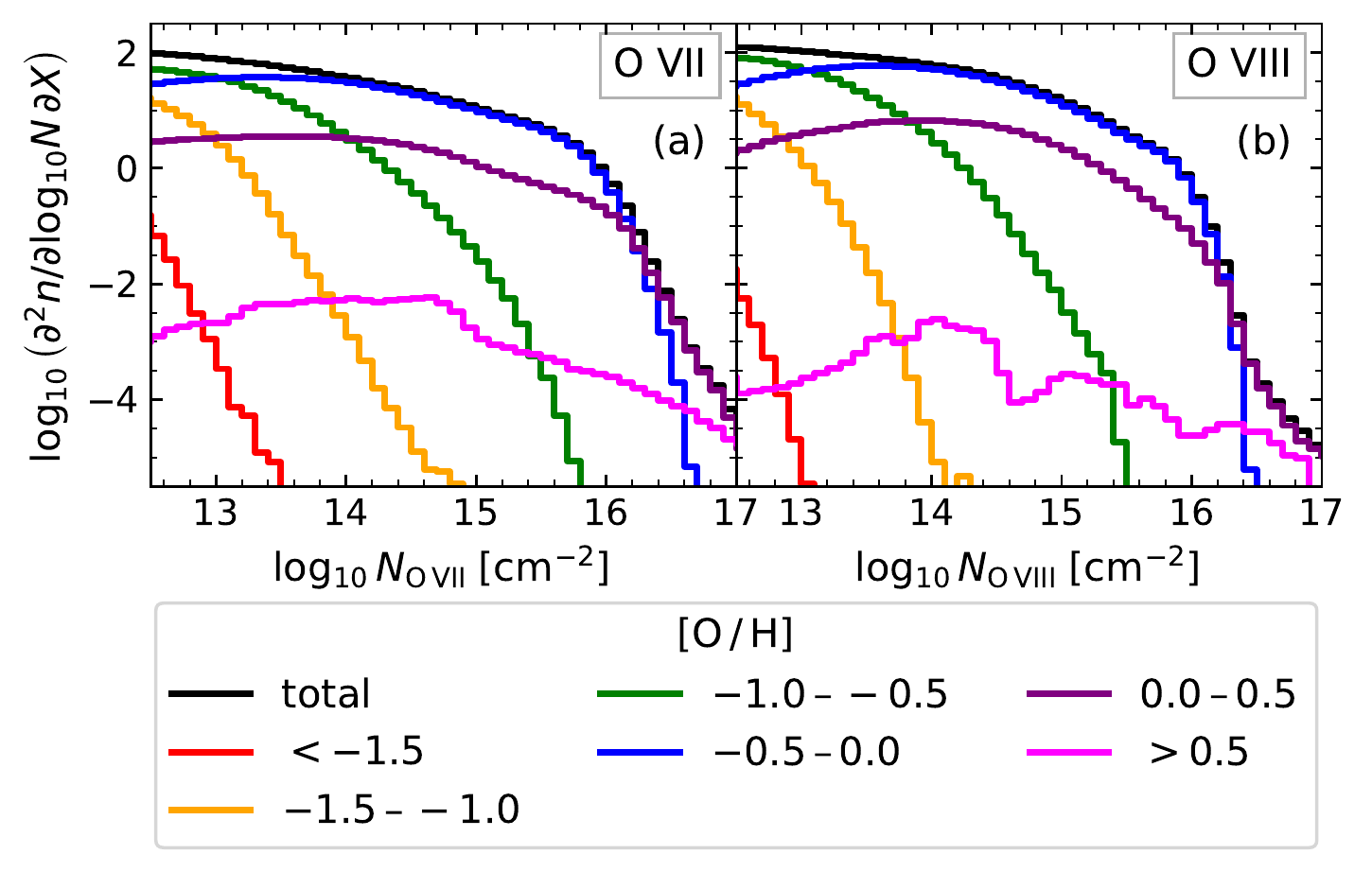}
\caption{The CDDFs for \ion{O}{vii} (a) and \ion{O}{viii} (b) decomposed into absorption systems with different ion-weighted SPH-smoothed oxygen abundances. The values indicated in the legend are ranges in the log oxygen mass fraction, relative to the solar oxygen mass fraction. This shows that most absorption systems we might detect in the foreseeable future should have oxygen-ion-weighted oxygen abundances around the solar value. (Note that mass- and volume-weighted metallicities may well be smaller.)}
\label{fig:cddfs_o78_by_fO-Sm}
\end{figure}

Figures~\ref{fig:cddfs_o78_by_rho}, \ref{fig:cddfs_o78_by_T}, and~\ref{fig:cddfs_o78_by_fO-Sm} show what type of gas dominates the CDDF at different column densities. Fig.~\ref{fig:cddfs_o78_by_rho} shows the contribution of absorption systems with different ion-weighted overdensities $\delta = \rho/\overline{\rho_b} - 1$. 
%This is with respect to the average total baryon density ($ \overline{\rho_b}= \Omega_b \rho_{\mathrm{crit}}$).  
The shape of the total distribution (black) is different to that shown in some other figures because it shows the number of absorption systems per unit \emph{log} column density (per unit absorption length), instead of per unit column density (per unit absorption length). 
%The other colours show the column density distributions of absorption systems with ion-weighted overdensities in the ranges shown in the legend.
 We show column densities up to $10^{17} \us \mathrm{cm}^{-2}$ here, unlike in most figures, to show what happens at the highest column densities. However, since the CDDFs are not converged above $10^{16.5} \us \mathrm{cm}^{-2}$, the values here should be taken as indicative.
%the column densities values here are just an indication of what happens at the extreme end of the distribution. The ion-weighted temperatures and densities should also be rough indicators for these columns, though non-convergence implies they may vary somewhat across the column in these systems.  

We see in Fig.~\ref{fig:cddfs_o78_by_rho} that higher densities tend to produce higher column densities.
%, ranging from underdense gas for the lowest column densities (only partially shown) to overdensities typical for haloes at the highest column densities.
At the break in the CDDFs, the ion-weighted overdensities are typically $\sim 10^2$
%$\sim 100$--$200$ (closer to $100$ for \ion{O}{vii} and closer to $200$ for \ion{O}{viii}). 
This points towards a cause for the break: gas at $\delta \gtrsim 10^2$ is typically within dark matter haloes, which have lower covering fractions than partially collapsed intergalactic structures (filaments, sheets, and voids) found at lower overdensities. The denser gas in haloes is also more likely to produce high column densities if it has the right temperature. This picture is broadly confirmed by visual inspection of the column density maps in Fig.~\ref{fig:coldensmaps}\footnote{Recall that this figure shows column densities at a much lower resolution than we use for the CDDFs, and measures column densities in $100 \us \mathrm{cMpc}$ columns.}, which show that the highest column densities ($\gtrsim 10^{16} \us \mathrm{cm}^{-2}$) mostly occur in nodes in the cosmic web. 
% so the covering fraction of this high-column-density gas in these maps will not reflect its occurrence as described by the $6.25 \us \mathrm{cMpc}$ CDDF. 
This result is also roughly consistent with the results of e.g., \citet{fang_bryan_canizares_2002}, who compared \ion{O}{vii} and \ion{O}{viii} CDDFs from their simulation to analytical models and found they could explain the CDDF at high column densities as coming from collapsed haloes.
%{\todo Michael Shull mentioned this interpretation of the \ion{O}{vi}(?) CDDF break in his Alabama WHIM 2018 talk. Track down reference? I think it was the Borgani simulations or something?}

Fig.~\ref{fig:cddfs_o78_by_T} shows the contribution of absorption systems with different ion-weighted temperatures to the \ion{O}{vii} and \ion{O}{viii} CDDFs. The black line shows the total CDDF.
%, and the gray line shows the contribution of the $10^5$--$10^7 \us \mathrm{K}$ gas. 
Gas with temperatures of $10^5$--$10^7 \us \txn{K}$ is canonically referred to as the warm-hot intergalactic medium (WHIM), where e.g.\ \citet{cen_ostriker_1999} found most of the `missing baryons' in their simulations. We see that absorbers tracing this gas dominate the CDDF at column densities likely to be observable with planned future missions. For \ion{O}{vii}, we see that absorbers around the break and above tend to trace gas somewhat above $\sim 10^6 \us \txn{K}$, while for \ion{O}{viii}, absorbers around the break tend to trace gas at $\sim 10^{6.5} \us \txn{K}$, with the temperature continuing to increase somewhat with column density. These temperature ranges are close to the temperatures where each ion attains its peak ion fraction in collisional ionization equilibrium (CIE). As we will show in more detail in Section~\ref{sec:pds}, the absorption systems with column densities above the break do indeed tend to be primarily collisionally ionized while below the break, lower temperatures dominate and photoionization is important.

Finally, in Fig.~\ref{fig:cddfs_o78_by_fO-Sm}, we turn to oxygen abundances. It shows that over a wide range in column densities, absorbers tend to trace gas with oxygen-ion-weighted metallicities at, or somewhat below, solar, while oxygen abundances become solar or larger at the largest column densities ($N \gtrsim 10^{16.3} \us \txn{cm}^{-2}$). Oxygen abundances above $\approx 3$ times the solar value do occur, but they are rare. It is important to note that what we show here is the mean oxygen abundance weighted by the contribution of each gas element to the oxygen ion column density. Mass-weighted and volume-weighted mean metallicities can be much lower.

\subsection{The effect of AGN feedback}
\label{sec:agn}
%{\todo Main question: how does AGN feedback (and how it is implemented) affect the o7/o8 cddf?}

\begin{figure}
\includegraphics[width=\columnwidth]{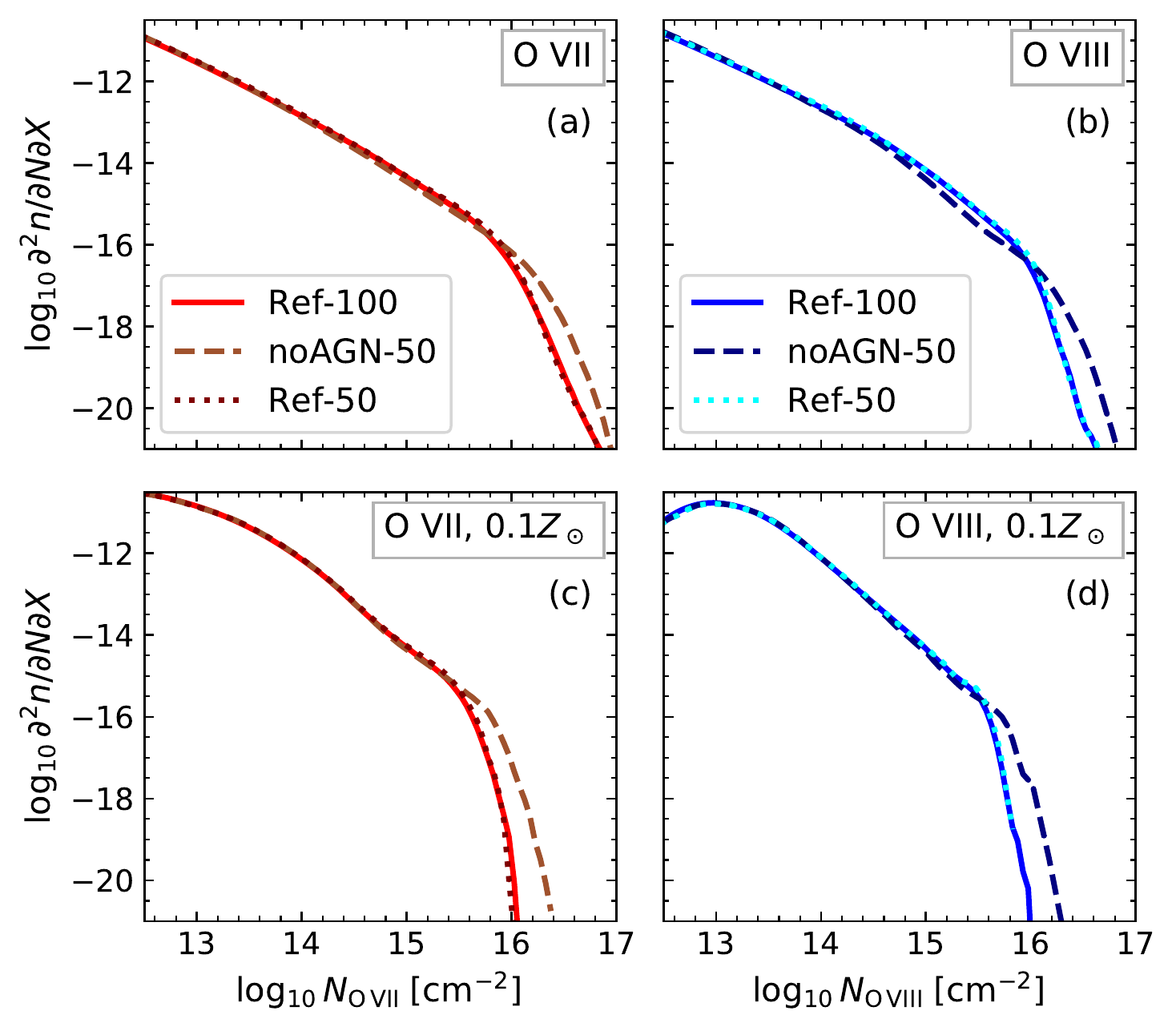}
\caption{The top panels (a, b) compare the reference and no AGN models, to show the effect of AGN feedback on the $z=0.1$ \ion{O}{vii} (left panels) and \ion{O}{viii} (right panels) CDDFs. The column density distribution in the no AGN simulation (\code{NoAGN-L050N0752}, `noAGN-50') is compared to the reference simulation (\code{Ref-L100N1504}, `Ref-100'). We also show the $50\us\txn{cMpc}$ reference simulation (\code{Ref-L050N0752}, `Ref-50') results to demonstrate that differences between the reference and no AGN simulation distributions are not due to the size of the simulation domain. To gauge how much of the differences are due to differences in gas oxygen content, we show CDDFs made assuming all gas has a metallicity of $0.1 \us Z_{\sun}$ in the bottom panels. This shows that beyond the break in the CDDF, a boost in metal enrichment due to AGN feedback partly offsets its other effects.}
\label{fig:noagn}
\end{figure}

To investigate the effect of AGN feedback on the CDDF, we compare the column density distributions we found in Section~\ref{sec:cddfs} to a recent {\eagle} simulation not described by \citet{eagle_paper} or \citet{eagle_calibration}: \code{NoAGN-L050N0752}. As the name suggests, this simulation does not include AGN feedback, while the rest of the (subgrid) physics is the same as in the reference model. This means it does not produce a realistic universe: AGN feedback is needed to quench star formation in high-mass galaxies and their progenitors in the {\eagle} model, and to regulate the gas fractions of haloes. Fig.~\ref{fig:noagn} also shows the CDDFs we get from the reference model in a $50 \us \txn{cMpc}$ box (\code{Ref-L050N0752}) to verify that any differences are not due to the difference in box size.

%This shows that above the CDDF break, the influence of AGN is mainly to enrich the IGM with metals, while below the break, changes in the temperature and/or density of the gas explain at least part of their effect.

The top panels of Fig.~\ref{fig:noagn} show that, when AGN feedback is disabled, the CDDF is barely affected at the lowest column densities, intermediate column densities are slightly less common, and the highest column densities occur more frequently. The decrease at intermediate column densities is larger for \ion{O}{viii} than for \ion{O}{vii}.
%The results for the two reference simulations generally differ very little, with some noise at higher column densities. 
At column densities $\gtrsim 10^{16.5} \us \txn{cm}^{-2}$, the CDDFs are not converged with pixel size in the column density maps (see Appendix~\ref{app:conv}). We show larger column densities in this plot mainly to show the difference with the no AGN simulation at high column densities more clearly, but relative differences in this range may not be reliable.  

Since massive galaxies in the no AGN simulations produce too many stars, but too weak outflows, we might expect that some of this difference, especially at higher column densities probing halo gas, might be due to differences in gas metallicity. To check this, we do a similar comparison in the bottom panels of Fig.~\ref{fig:noagn}, except we assume all gas has a metallicity of $0.1$ times the solar value when we calculate the number of ions in our columns.
% A one-to-one comparison of these column density distributions to those made using the tracked gas oxygen content from the simulation is difficult: we already saw in Fig.~\ref{fig:cddf_z0p1} that this metallicity prescription itself changes the CDDF, and we cannot simply compare the distributions at the same column density. We therefore take the break in the CDDFs calculated these two ways as a reference point for the comparison.

We then note that the differences below the break in the CDDF (almost) disappear when comparing constant metallicity results, while the differences above the CDDF break increase somewhat. 
%Due to the non-convergence of the reference model CDDF at its highest column densities, we do not entirely trust the largest column densities at fixed metallcity either, including the extreme relative differences between the models there. In the simulations without AGN, the break also seems to shift to larger column densities with both metallicity prescriptions: the part of the distribution after the break seems to shift to higher column densities.
We interpret the causes of these differences as follows. Within the haloes, we see larger column densities in the absence of AGN feedback. This could be due to higher densities, metallicities, or mass of the hot gas responsible for the absorption at these high column densities. The effect of AGN feedback at higher column densities increases, but only a bit, if we assume a fixed metallicity. This indicates that the main effect of AGN feedback is to decrease the density of the hot gas, while it also increases its metallicity. These effects partially cancel out, but the density effect dominates. The enhancement of metal ejection by AGN feedback also explains the increase in the CDDF below the break.  

This picture is supported by the work of \citet{davies_crain_etal_2019}. They showed that for haloes with masses $\gtrsim 10^{12} \us \txn{M}_{\sun}$ at $z=0$ in {\eagle}, the amount of AGN feedback injected into these halos is a good predictor of the halo gas fraction at fixed halo mass, and that this is due to AGN ejecting gas from the circumgalactic medium. In a follow-up paper, \citet{oppenheimer_davies_etal_2019} showed that this feedback also decreases ion column densities around galaxies, though they discuss \ion{H}{i}, \ion{C}{iv}, and \ion{O}{vi}.

\subsection{Physical properties of the absorbers}
\label{sec:pds}
%{\todo Main question: Which gas properties drive the column density distribution? Which gas is responsible for (most of) the absorption? }

\begin{figure}
\includegraphics[width=\columnwidth]{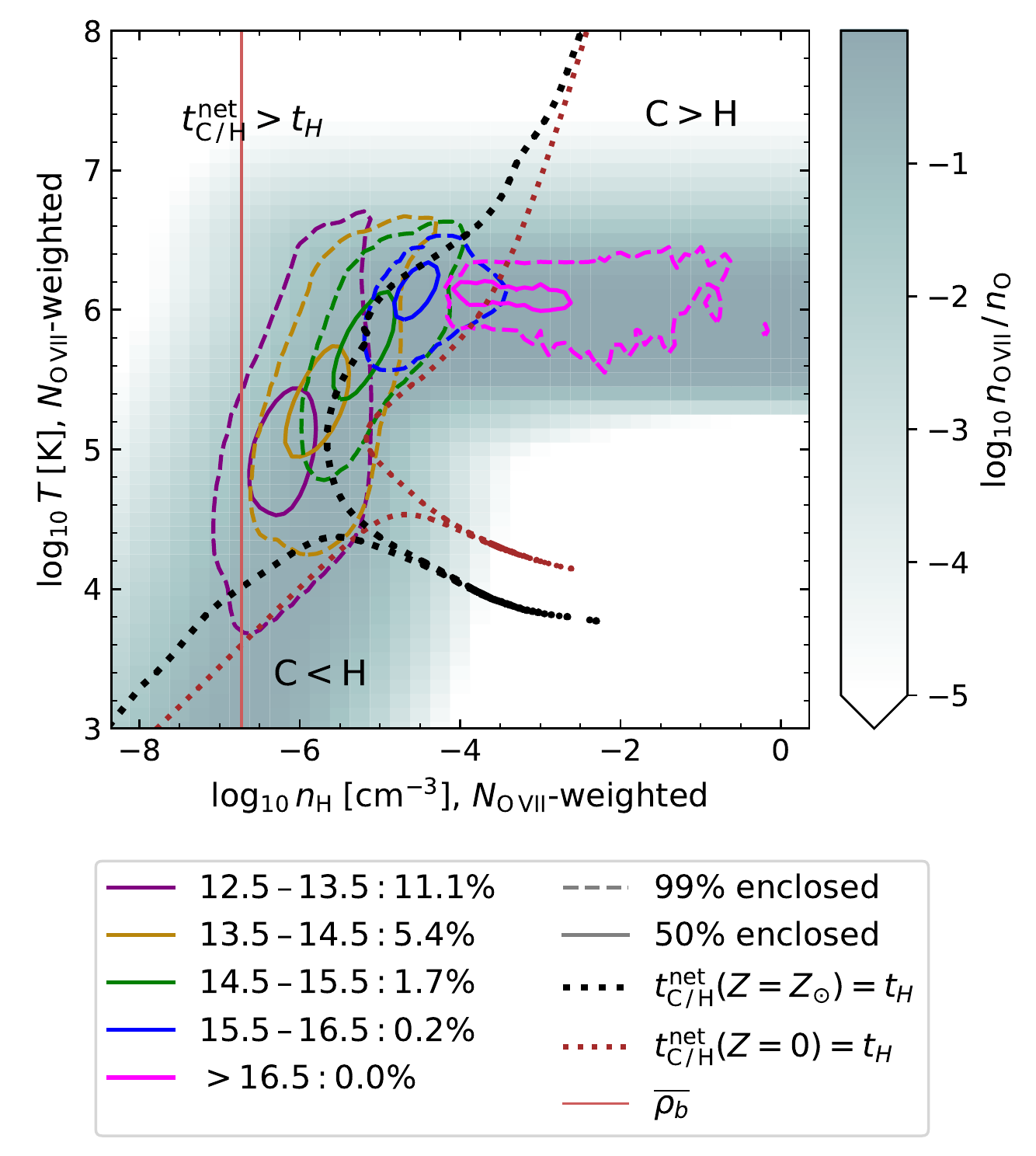}
\caption{Phase diagram of the \ion{O}{vii} absorption system density and temperature in different column density bins in  \code{Ref-L100N1504} at $z=0$, using $6.25 \us \txn{cMpc}$ long sightlines.  The rainbow-coloured contours indicate the distribution of ion-weighted average temperatures and densities for absorbers with column densities in different bins. The legend indicates the bins by their values of $\log_{10} N_{\mathrm{O VII}}\,/\, \mathrm{cm}^{-2}$; the percentages indicate what fraction of columns in the simulation (including those with zero column density) is in each bin.  Solid and dashed contours enclose $50$ and $99 \us \%$ of the columns in each bin, respectively. The greyscale in the background indicates which fraction of oxygen nuclei are \ion{O}{vii} ions. We indicate gas at fixed heating and cooling time scales using the dotted pairs of contours. They indicate constant net radiative cooling (upper contours) and heating (lower contours) time scales $t_{\mathrm{C}\,/\,\mathrm{H}}^{\mathrm{net}}$ equal to the Hubble time $t_H$ for gas with solar metallicity and with primordial abundances, using a \citet{HM01} background. The vertical brown line indicates the $z=0$ average baryon density.}
\label{fig:pds-no7w}
\end{figure}

\begin{figure}
\includegraphics[width=\columnwidth]{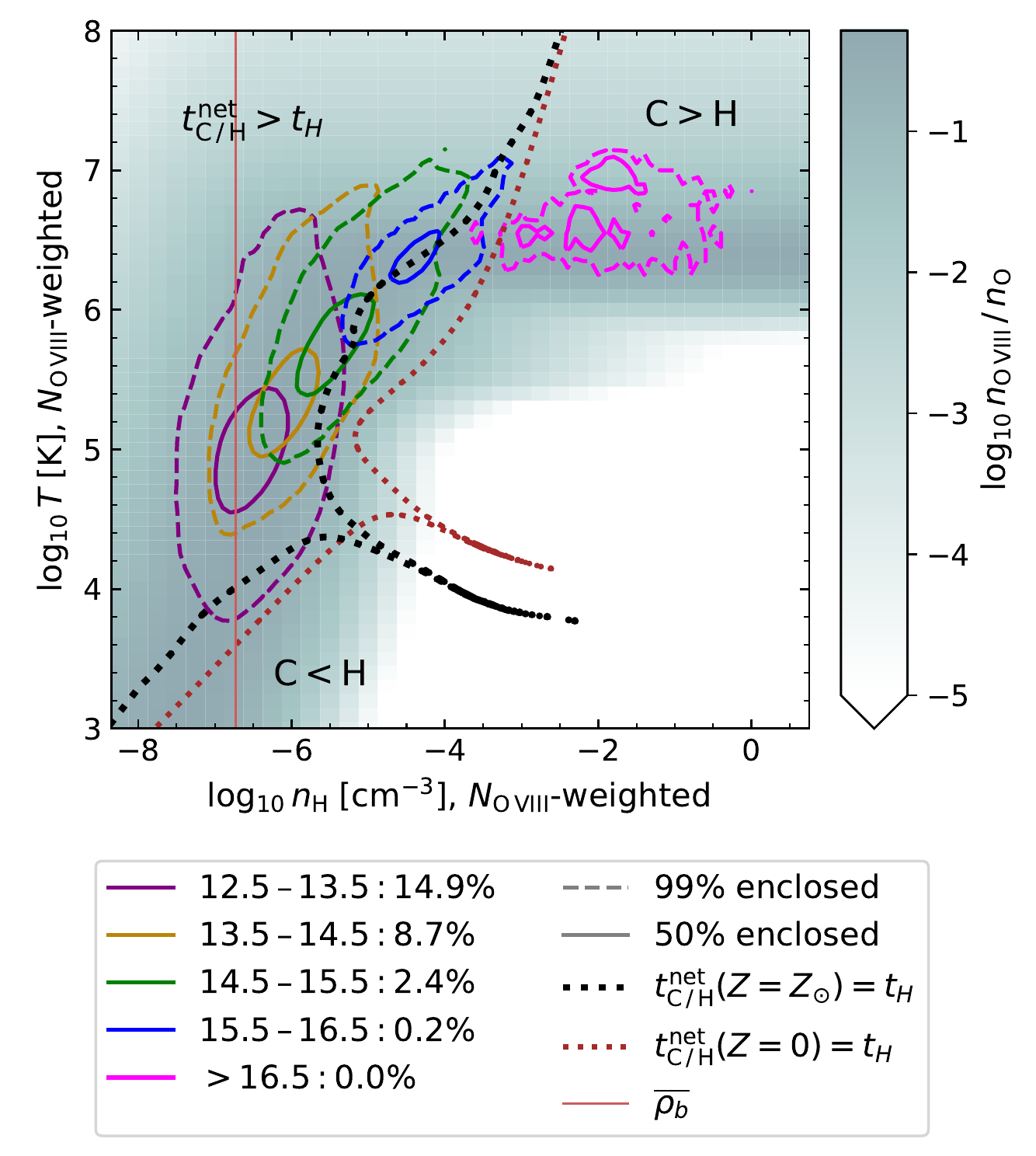}
\caption{As Fig.~\ref{fig:pds-no7w}, but for \ion{O}{viii}.}
\label{fig:pds-no8w}
\end{figure}

To investigate the physical properties of the absorption systems, we first look at the temperature and density of gas traced by the \ion{O}{vii} and \ion{O}{viii} ions. In Figs.~\ref{fig:pds-no7w} and~\ref{fig:pds-no8w}, contours in different colours show the distribution of absorption system temperatures and densities in different column density ranges for \ion{O}{vii} and \ion{O}{viii}, respectively. The greyscale shows the fraction of oxygen ions in the ionization state of interest, and the dotted contours show lines of constant net radiative cooling or heating time scale. 
The hydrogen number density displayed along the axis, used to calculate the ion balance as well as the cooling times, was calculated from the mass density assuming the primordial hydrogen mass fraction $0.752$, but the difference with a solar hydrogen mass fraction conversion is negligible. 

Comparison of the coloured contours and the greyscale shows that the absorbers reside in regions of the phase diagram where the fraction of oxygen ions in that state is high, as expected.
These ion fractions 
%(greyscale in figures~\ref{fig:pds-no7w} and~\ref{fig:pds-no8w}) are calculated assuming collisional ionization and photoionization equilibrium (CIE and PIE, respectively). In our ionization model, the UV/X-ray background \citep{HM01} is uniform (at each simulation output redshift). This means the impact of photoionization is density-dependent, but not explicitly position-dependent. Besides the ionization energy and cross-section, the fraction of oxygen atoms the background can ionize depends on the photon-to-element ratio. 
account for photoionization by the UV/X-ray background \citep{HM01}.
At high densities, the ion fraction becomes density-independent, as the influence of the ionizing radiation becomes negligible and the ion balance asymptotes to collisional ionization equilibrium (CIE). 
%(Since ionizing collisions and recombinations are both two-particle interactions, their rates have the same density-dependence.) At lower densities the ionizing radiation becomes important: the lower the density, the larger the fraction of any element that is ionized to higher levels than CIE alone would produce. Therefore, as density decreases, the peak temperature for each ion drops. 

%These cooling times were originally calculated using appropriate hydrogen mass fractions for the two abundance patterns, but the resulting hydrogen number densities have been scaled back to mass densities assuming primordial abundances. The contours The pink line indicates the average baryon density in the simulation, for reference.
%The density and temperature here are ion-weighted averages in each $6.25 \us \txn{cMpc}$ column with a non-zero column density.
%  $t_H = H(0)^{-1} = 14.4 \us \mathrm{Gyr}$
%  The `peaks' towards high density and low temperature are where radiative heating and cooling times are equal; gas to the right of the lines can cool within a Hubble time. The higher-temperature wings of these contours indicate cooling time scales, the lower-temperature wings indicate heating time scales. This shows which \ion{O}{vii} absorbers are likely to trace collisionally ionized and photoionized gas, as which absorption systems are likely to trace radiatively cooling gas.

As expected, the column density increases with the density of the absorbing gas, with the lowest column densities often coming from (almost) underdense gas.
%To make figures~\ref{fig:pds-no7w} and~\ref{fig:pds-no8w}, we retrieved the ion-mass-weighted temperature and density in each column we also used to measure column densities. In the figures, we show the distribution of columns (representing absorption systems) in those temperatures and densities for different ranges of column density for the measured columns. 
The percentages in the legend indicate the fraction of sightlines (including those with temperatures and densities outside the plotted range) in that column density range. The highest column densities are rare, accounting for less than $0.1\%$ of sightlines.  Furthermore, as the ion fraction colouring shows, most of the sightlines have their absorption coming mainly from photoionized gas: the gas at lower densities where the ion balance is density-dependent. However, at the highest column densities, especially where $N_{\mathrm{O VII}} \gtrsim 10^{15.5} \us \mathrm{cm}^{-2}$, the absorption comes mainly from collisionally ionized gas.  

For ionization models of  observed absorbers, it is important to know what are reasonable assumptions when trying to establish what sort of gas an absorber traces. For both \ion{O}{vii} and \ion{O}{viii}, photo-ionized gas dominates the absorption at lower column densities. For \ion{O}{vii}, gas at the CDDF break ($N_{\mathrm{O\,VII}} \sim 10^{16} \us \mathrm{cm}^{-2}$) is clearly dominated by gas in the CIE peak temperature range (Fig.~\ref{fig:cddfs_o78_by_T}). For \ion{O}{viii}, WHIM gas is also clearly dominant there, but CIE temperature gas is only just becoming dominant. 
%At the highest column densities, we see that gas at slightly higher temperatures is present; Fig.~\ref{fig:cddfs_o78_by_T} shows it is even dominant at some of the very highest column densities. At $10^7 \us \mathrm{K}$, the CIE \ion{O}{viii} fraction is still $\sim 5\%$ of the maximum value, so this higher-temperature contribution is not very alarming. 
At the CDDF break ($\sim 10^{16} \us \txn{cm}^{-2}$), CIE gas is important, but the absorption systems here have sufficiently low densities that photoionization still influences the ion fractions. For a given temperature, the ion fractions at $n_{\mathrm{H}} \sim 10^{-5}\us \txn{cm}^{-3}$ (the value used by \citet{nicastro_etal_2018} to model the absorbers they found) can differ from the CIE values by factors $\gtrsim 2$ within the CIE temperature range (i.e., at temperatures where, in CIE, the ion fraction is at least half the maximum value). Therefore, in modelling these absorption systems, a temperature consistent with CIE does not necessarily imply that CIE ion fractions will be accurate. 
%Also, note that extracting the absorption system density based on line width and the Hubble flow may not work, since, as we showed in Section~\ref{sec:EW}, line widths do not seem to correlate with column density, at least in the higher column density range, while absorber densities do.

\begin{figure}
\includegraphics[width=\columnwidth]{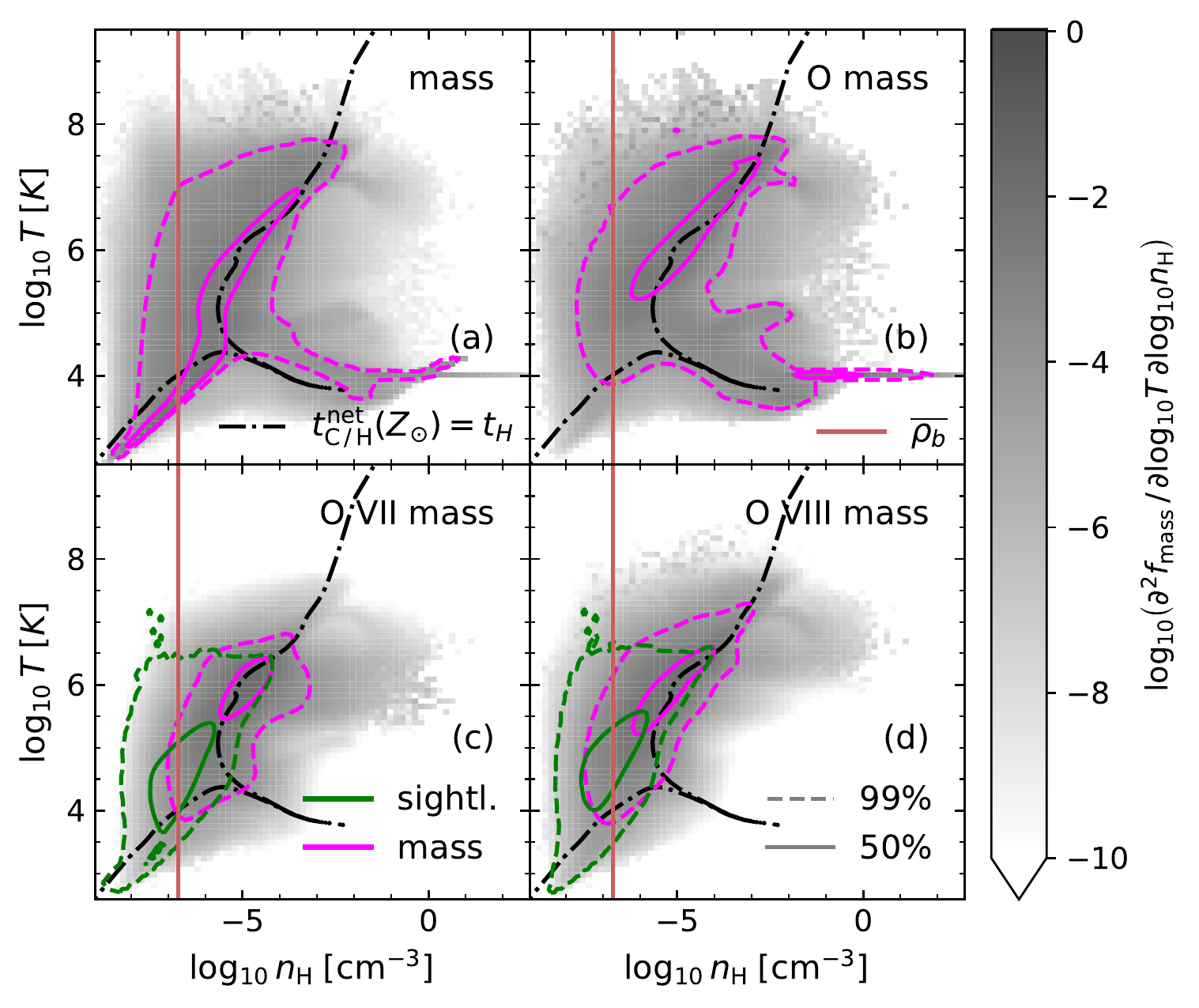}
\caption{Phase diagrams for the total gas mass (a), oxygen mass (b), and \ion{O}{vii} (c) and \ion{O}{viii} (d) masses in \code{Ref-L100N1504}  at $z=0$.  The solid and dashed fuchsia contours enclose $50$ and $99 \us \%$ of the total mass. The green contours show the distribution of ion-weighted temperatures and densities that were computed for  $6.25 \us \txn{cMpc}$ long columns. This is similar to the distributions shown in figures~\ref{fig:pds-no7w} and~\ref{fig:pds-no8w}, but for all non-zero ion column densities. The black, dot-dashed contours indicate a constant radiative cooling or heating time scale equal to the  Hubble time for gas with solar abundances. The vertical brown lines indicate the $z=0$ average baryon density.}
%{\todo The red line indicated the temperature and pressure floors used in the {\eagle} simulations.  check temperature floor ($8000 \us \txn{K}$ for $ 10^{-5} \us \txn{cm}^{-2} < n_{\mathrm{H}} \leq 10^{-1}\us \txn{cm}^{-2}$, $P_{\txn{EOS}} =\propto \rho^{\gamma_{\txn{EOS}} = 4/3}$ at $n_{\mathrm{H}} > 10^{-1}\us \txn{cm}^{-2}$)}
% We histogrammed the temperature and hydrogen number density of simulation gas particles by their mass (a), oxygen mass (b), \ion{O}{vii} mass (c), and \ion{O}{viii} mass (d).
%The `peaks' towards high density and low temperature are where radiative heating and cooling times are equal. The higher-temperature wings of these contours indicate cooling time scales, the lower-temperature wings indicate heating time scales. At $T \gtrsim 10^9 K$, the time scales are a very crude extrapolation. Vertical pink lines indicate the cosmic average baryon density for reference. This indicates how the total mass distribution and the metal enrichment in the simulation, influenced by which gas can cool, shapes the densities and temperatures at which we can find absorption systems, and their abundances.
\label{fig:pds-mass}
\end{figure}

To quantify what fractions of gas are affected by this, we divide the temperature-density plane into four regions, defined by a temperature and a density cut\footnote{Note that we use the intrinsic temperature and density of gas in EAGLE (not the ion-weighted values in projection), along with its ion content, for these calculations. This means we make cuts in the temperature-density plane of Fig.~\ref{fig:pds-mass}.}. 
For the temperature cut, we use the lowest temperature at which the ion fraction in CIE reaches half its maximum value: $T \approx 10^{5.5}$, $10^{6.2} \us \txn{K}$ for \ion{O}{vii} and \ion{O}{viii}, respectively. 
For the density cut, we focus on the same temperature ranges, where, in CIE, the ion fraction is at least half the maximum value. We consider the gas to be in CIE when it has a temperature above the cut, and a density above a minimum value. This minimum is the lowest density where the ion fractions in the CIE temperature range defined above, differ from those at the highest density we have data for ($10^3 \us \txn{cm}^{-3}$) by less than a factor $1.5$. This means 
$n_{\txn{H}} \approx 10^{-4.75}$, $10^{-4.25} \us \txn{cm}^{-3}$ 
for \ion{O}{vii} and \ion{O}{viii}, respectively. For \ion{O}{vii} and \ion{O}{viii}, 
$\approx 32$ and $ 8 \us \%$
of the ion mass is in CIE, respectively.
Gas at lower densities, but with a temperature above the cut, is the gas that may be mistaken for gas in CIE based on temperature diagnostics, but where ionization modelling based on this assumption may cause errors. 
This is 
$\approx 45$ and $ 19 \us \%$ 
of the \ion{O}{vii} and \ion{O}{viii} mass, respectively.
Gas below the density and the temperature cuts is not necessarily in pure PIE: the ion fractions are still influenced by collisional processes at the higher temperatures in this regime. This gas accounts for 
$\approx 23$ and $ 72 \us \%$ 
of the \ion{O}{vii} and \ion{O}{viii} mass, respectively. 
A small fraction of the \ion{O}{vii} and \ion{O}{viii} mass, 
$\leq 0.5 \us \%$, 
is at densities above the CIE threshold, but below the temperature cut. For these density thresholds and ions, there is more gas at CIE temperatures that is not actually in CIE than there is CIE gas. However, note that this depends on choices we made. If we choose our density thresholds based on a maximum factor~$2$ difference between the ion fraction and full CIE ion fraction, both ions have more gas in CIE than in the error-prone regime.    

%https://www.ast.cam.ac.uk/~pettini/STARS/Lecture06.pdf
We look into the effect of temperature further by comparing the $b$ parameter range of Fig.~\ref{fig:cog} with the thermal broadening for different temperature ranges we see in figures~\ref{fig:pds-no7w} and~\ref{fig:pds-no8w}. Note that those $b$ parameters come from comparing total EW and column density, not line fitting, and do not account for instrumental broadening.

The thermal contribution to the line widths 
$b_{\mathrm{th}} = \sqrt{{2 k T}{m}^{-1}}$ 
is equal to 
$16$--$ 57\us \mathrm{km}/\mathrm{s}$ for gas with temperatures 
$T = 10^{5.4}$--$10^{6.5} \us K$. 
This range dominates the \ion{O}{vii} CDDF at the column densities 
$\gtrsim 10^{15} \us \txn{cm}^{-2}$,
 where Fig.~\ref{fig:cog} shows we can estimate the line width from the column density and equivalent width. For \ion{O}{viii}, we find  $b_{\txn{th}} = 36$--$ 81\us \mathrm{km}/\mathrm{s}$ for temperatures around  the \ion{O}{viii} collisional ionization peak 
($T = 10^{6.1}$--$10^{6.8} \us K$). 
For the gas at $T = 10^{7.0} \us \txn{K}$, reached by some of the highest-column-density \ion{O}{viii}, we find 
 $b_{\th} = 102 \us \mathrm{km}/\mathrm{s}$. 
For both ions, the lowest EWs we find are consistent with thermal broadening, but the typical $b$ parameters for these absorption systems are larger than predicted by their ion-weighted temperatures. This strengthens our conclusion from visual inspection of virtual spectra that velocity structure is important for line widths, especially when the spectral resolution is too low to resolve components in absorption systems.  This also means that measured line widths for these ions may not provide meaningful constraints on absorber temperatures.

We show contours indicating where the radiative net cooling (or heating) time equals the Hubble time for gas with primordial and solar abundances in brown and black, respectively. We calculated these time scales using the tables of \citet{wiersma_schaye_smith_2009}, which were also used for radiative cooling in the simulations. These include the effects of the evolving \citet{HM01} UV/X-ray background.
% and the cosmic microwave background at each redshift. These contours are similar to those shown in fig.~4 of that paper. The cooling or heating time scale here is the internal energy divided by the cooling rate. 
%The lower-temperature `wings' of these contours are net heating time scales, with net heating within a Hubble time at lower temperatures ($\mathrm{C} < \mathrm{H}$). The higher-temperature parts indicate net cooling time scales, with net cooling within a Hubble time at higher densities ($\mathrm{C} > \mathrm{H}$). The `peaks' to the right at $\sim 10^4 \us \txn{K}$ are where radiative cooling and heating balance each other. 
%The cooling times for solar and primordial gas compositions roughly bracket those for gas of metallicities between those values. 
For each metallicity, the lower-temperature contour indicates where the net heating time scale is equal to the Hubble time , with net heating within a Hubble time at lower temperatures ($\mathrm{C} < \mathrm{H}$). The higher-temperature contours indicate net cooling time scales, with net cooling within a Hubble time at higher densities ($\mathrm{C} > \mathrm{H}$). The pairs of contours converge in `peaks' to the right at $\sim 10^4 \us \txn{K}$, where radiative cooling and heating balance each other.

%Therefore, for the higher column density gas, the solar metallicity contours are the better indicator of whether the gas is cooling. For the cool IGM gas, the lower metallicities are likely more appropriate; this gas typically has oxygen abundances $\lesssim 0.01$ times the solar value. Fig.~\ref{fig:pds-mass} shows that at solar metallicity, this gas would be heated to larger temperatures in a Hubble time than it actually has. 
As column densities increase, more of the absorbing gas can cool within a Hubble time. For gas at column densities $\gtrsim 10^{16}\us \txn{cm}^{-2}$, which we noted previously is dominated by halo-density gas, this means that the absorbing gas may be a part of cooling flows, where hot gas cools radiatively as it flows from the CGM to the central galaxy. As we saw in Fig.~\ref{fig:cddfs_o78_by_fO-Sm}, gas with such column densities typically has roughly solar or larger oxygen abundances. 
% $\gtrsim 10^{14.0} \us \txn{cm}^{-2}$ 
However, some of it may also be gas that has recently been shock-heated in outflows.
%This is true for a larger fraction of \ion{O}{vii} absorption systems than \ion{O}{viii} absorption systems. Gas at even higher column densities ($\gtrsim 10^{16.5} \us \txn{cm}^{-2}$) is typically cooling quite rapidly, within $\sim 1.4 \us \txn{Gyr}$. 

%As Fig.~\ref{fig:pds-no7w} shows, in the column density range where absorbers are not optically thin (see Fig.~\ref{fig:cog}), temperatures of \ion{O}{vii} absorption systems have little to no dependence on column density, so at least the thermal part of the $b$ parameters, and its scatter, would not be expected to have a column density dependence in this regime. For \ion{O}{viii}, Fig.~\ref{fig:pds-no8w} does show some temperature-dependence. However, Fig.~\ref{fig:cog}b does not seem to show that this causes a dependence of the $b$ parameter on the column density. That is consistent with our earlier conclusion from Section~\ref{sec:break} that thermal broadening does not seem to dominate the absorption system widths we find. 

We explore the effect of the gas phase distribution on the absorption system properties with Fig.~\ref{fig:pds-mass}. The panels show the distribution of gas for mass, oxygen mass, and \ion{O}{vii} and \ion{O}{viii} mass.  The horizontal lines for high densities at $10^4 \us \mathrm{K}$ in the upper panels are the result of setting all star-forming gas to that temperature.
The green contours (sightline distribution, for $6.25 \us \mathrm{cMpc}$ long sightlines, as in figures~\ref{fig:pds-no7w} and~\ref{fig:pds-no8w}) and fuchsia contours (mass distribution) seem to be similar, as expected, but the sightline distribution prefers lower-density gas. This is due to its higher volume-filling fraction, hence higher covering fraction, relative to its mass fraction. We also show the radiative heating/cooling time contours at solar abundances as in figures~\ref{fig:pds-no7w} and~\ref{fig:pds-no8w}. This demonstrates how the cooling and heating time scales shape gas conditions: most gas is either roughly at the temperature set by radiative heating over a Hubble time, or is too hot or diffuse to cool within about a Hubble time. 

%This is less of an issue for the WHIM gas at similar densities. Gas at high temperatures in {\eagle} was heated by either accretion onto massive structures or directly or indirectly by stellar or AGN feedback. This requires this gas to have at least been close to sources of metal enrichment (galaxies). Some of the absorption systems do trace this gas.

Comparing the ion mass distributions to the mass and oxygen distributions clearly shows the effect of ionization state fractions and metallicities. The low-density, low-temperature gas associated with the cool IGM contains few ions because it contains little oxygen. 
%However, some of the sightlines in our distribution (brown contours) do trace this gas. We also saw this gas in in figures~\ref{fig:pds-no7w} and~\ref{fig:pds-no8w}, as the diagonal 'tail' at low densities and temperatures which roughly traces the primordial gas $t_H$ radiative heating contour. The `peak' in this regime for columns with column densities $\leq 10^{12} \us \txn{cm}^{-2}$ in figures~\ref{fig:pds-no7w} and~\ref{fig:pds-no8w} does not appear unless we include gas with column densities $\lesssim 10^7 \us \txn{cm}^{-2}$. It probably dominates absorption in some sightlines simply because a relatively high volume filling fraction will mean some sightlines do not intersect much else. 
Comparing the ion mass distribution to the total oxygen mass distribution shows that the low-temperature, high-density gas (associated with haloes and galaxies) is not traced by these ions because neither collisional nor photoionization ionizes much of the oxygen to \ion{O}{vii} or \ion{O}{viii} in this regime. 

\subsection{Correlations between O~{\sc viii}, O~{\sc vii}, and Ne~{\sc viii}, O~{\sc vi}, and H~{\sc i} absorption}
\label{sec:ioncorr}

\begin{figure*}
\includegraphics[width=\textwidth]{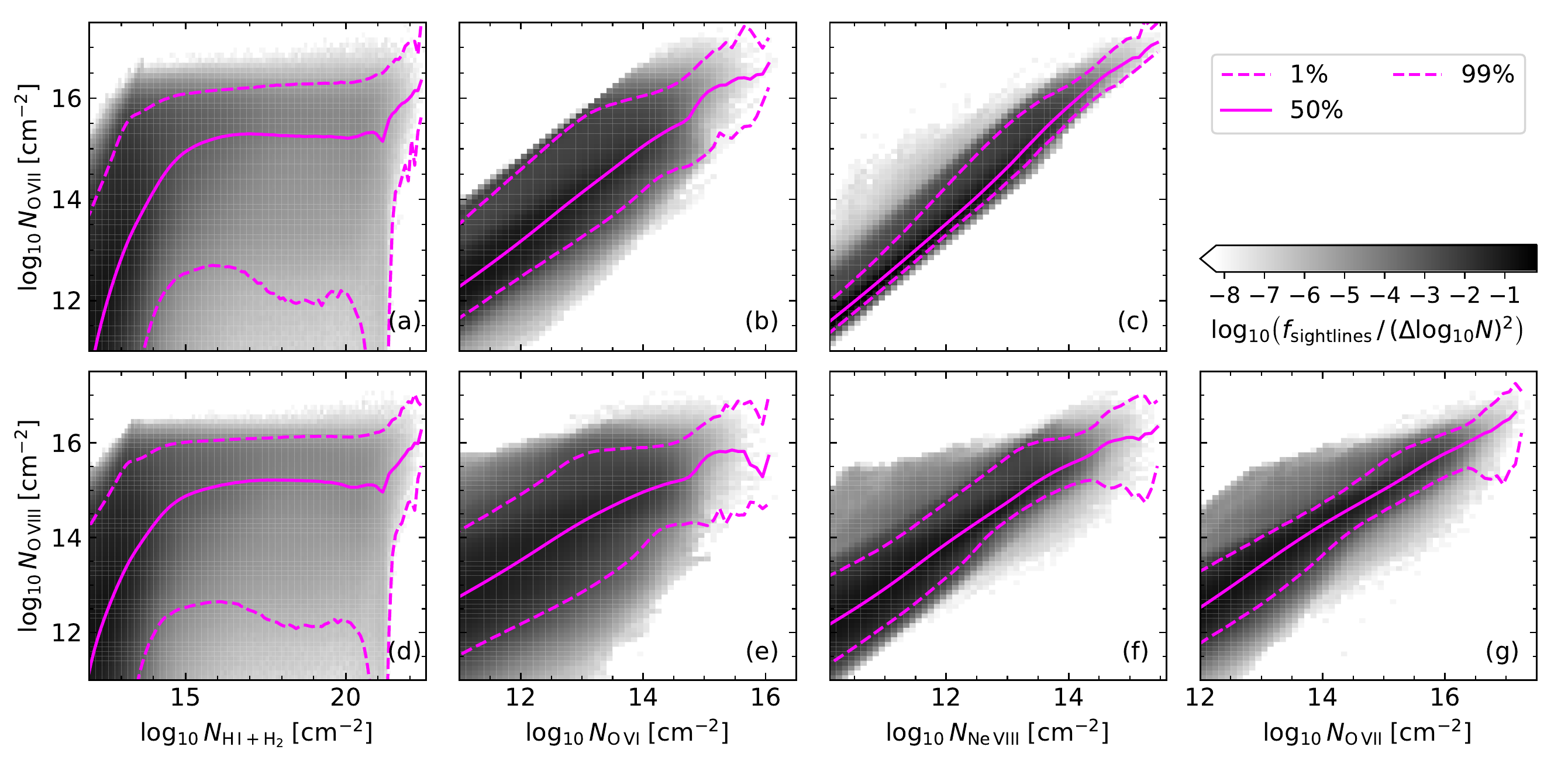}
\caption{The range of counterpart column densities for \ion{O}{vii} (a, b, c) and \ion{O}{viii} (d, e, f, g), given the column densities of neutral hydrogen (\ion{H}{i} and $\txn{H}_2$; a, d), \ion{O}{vi} (b, e), \ion{Ne}{viii} (c, f) and \ion{O}{vii} (g). We show the correlations for $6.25 \us \txn{cMpc}$ sightlines in the reference simulation (\code{Ref-L100N1504}) at $z=0.1$. The greyscale shows the distribution of sightlines in the column density of each pair of ions, while the lines show percentiles in the distribution of the ion on the y-axis at fixed column density on the x-axis. Large \ion{H}{i} column densities ($N_{\mathrm{H\,I} + \mathrm{H}_2} \gtrsim 10^{15} \us \txn{cm}^{-2}$) predict a $\sim 50 \us \%$ chance of finding a strong ($N \gtrsim 10^{15} \us \txn{cm}^{-2}$) \ion{O}{vii} or \ion{O}{viii} counterpart. Strong \ion{O}{vi} or \ion{Ne}{viii} absorption is a good predictor for strong \ion{O}{vii} and \ion{O}{viii} absorption.}
\label{fig:ioncorr_pred}
\end{figure*}

\begin{figure}
\includegraphics[width=\columnwidth]{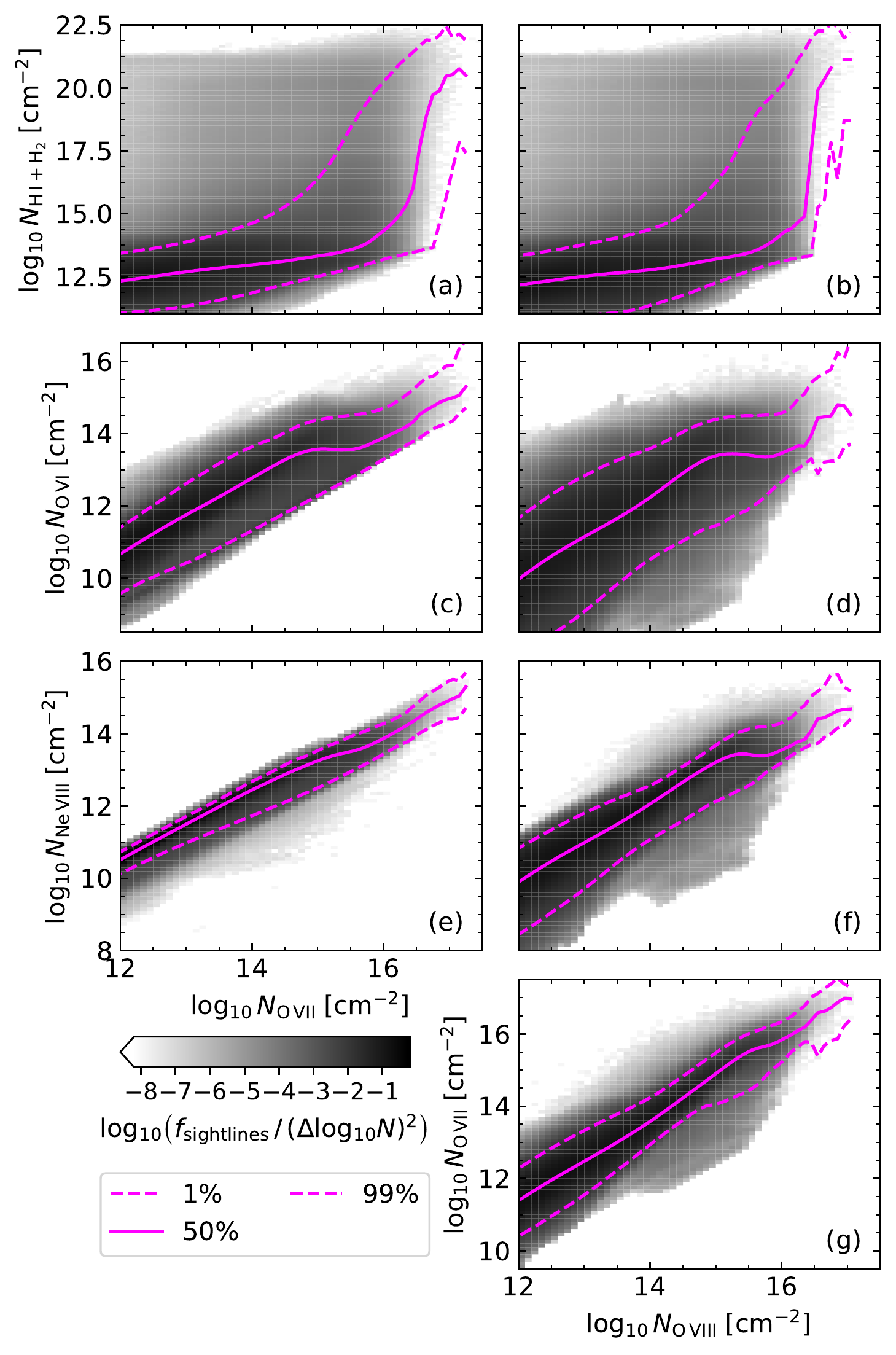}
\caption{As Fig.~\ref{fig:ioncorr_pred}, but with the axes reversed. For a given \ion{O}{vii} or \ion{O}{viii} column density, this indicates whether a counterpart agrees with what we find in {\eagle}.}
\label{fig:ioncorr_diag}
\end{figure}

We now look at how \ion{O}{vii} and \ion{O}{viii} absorption correlate, and also how this absorption correlates with neutral hydrogen, \ion{O}{vi} and \ion{Ne}{viii}. This is shown in Figs.~\ref{fig:ioncorr_pred} and~\ref{fig:ioncorr_diag}. 
The neutral hydrogen we show is \ion{H}{i} and $\txn{H}_2$ together, where we count the total number of hydrogen atoms. We use the prescription of \citet{rahamti_pawlik_etal_2013} to obtain the hydrogen neutral fraction, since the ionization balance tables we use for the other ions do not account for gas self-shielding against ionizing radiation, which is important for \ion{H}{I} fractions in cool, dense gas. For all but the highest column densities, the neutral fraction will be dominated by \ion{H}{I}. For neutral hydrogen, we see that over a very large range of column densities ($N \sim 10^{15}$--$10^{21} \us \txn{cm}^{-2}$), the trend between hydrogen column density and \ion{O}{vii} and \ion{O}{viii} column density is flat, with a lot of scatter. However, we also see that in this range $\sim 50 \us \%$ of the hydrogen absorbers have \ion{O}{vii} or \ion{O}{viii} counterparts with column densities $\gtrsim 10^{15} \us \txn{cm}^{-2}$, which is potentially observable, especially when searching at a specific redshift.   

In fact, \citet{kovacs_orsolya_etal_2019} have already reported extragalactic \ion{O}{vii} absorption by stacking quasar spectra using the known redshifts of 17 Ly$\alpha$ absorbers. They did make an extra selection, using only absorbers associated with massive galaxies (stellar mass $\gtrsim 10^{10} \us \txn{M}_{\sun}$). The average equivalent width of the Ly$\alpha$ absorbers was $174.4 \us$m{\AA} (corresponding to a column density $\sim 10^{14} \us \txn{cm}^{-2}$ for $b \approx 20 \us \txn{km}\, \txn{s}^{-1}$), and they found an \ion{O}{vii} column density of $1.4 \pm 0.4 \times 10^{15} \us \txn{cm}^{-2}$. Depending on the line width, this equivalent width may correspond to a column density close to or in the flat part of Fig.~\ref{fig:ioncorr_pred}a, and the \ion{O}{vii} column density they measure (an average for their stacked spectra) is similar to the median we  predict for that column density range. 

The other ions correlate better with \ion{O}{vii} and \ion{O}{viii} along $6.25 \us \mathrm{cMpc}$ sightlines.
Fig.~\ref{fig:ioncorr_pred} also shows how good \ion{O}{vi} and \ion{Ne}{viii} are at predicting strong absorption in our two X-ray lines, and what sort of \ion{O}{viii} counterparts to expect for an \ion{O}{vii} line. 
%The \ion{O}{vi} column densities along a sightline have a tighter correlation with the `adjacent' ion \ion{O}{vii} than with \ion{O}{viii}.
%This is especially clear at the highest column densities, where high \ion{O}{vi} column density sightlines favour lower \ion{O}{viii} column densities than \ion{O}{vii} column density alone would predict.  
The tightest correlation is between \ion{Ne}{viii} and \ion{O}{vii}, with a maximum $1$--$99\us \%$ scatter of $1.1 \us \txn{dex}$ at $N_{\txn{Ne\, VIII}} > 10^{11} \us \txn{cm}^{-2}$. (The maximum scatter at fixed \ion{O}{vii} column density, as shown in Fig.~\ref{fig:ioncorr_diag}, is the same.) The  tighter correlation between \ion{Ne}{viii} and \ion{O}{vii} than between \ion{O}{vi} and \ion{O}{vii} is likely because \ion{Ne}{viii} is a higher-energy ion than \ion{O}{vi}, with an ion fraction that peaks in almost the same band in the density-temperature plane as \ion{O}{vii}, but with a narrower peak.  We also notice from both figures in this section that \ion{O}{vii} is better correlated with both the UV ions than \ion{O}{viii}.  
%For this more energetic ion, high UV column densities do predict high X-ray column densities, but the correlation with \ion{O}{viii} column density flattens, and the scatter increases, at high UV column densities.
% {\todo similar production mechanisms as well? Both are mainly produced in type-II supernovae, according to wikipedia. Basically tackling the metallicity end of the correlation origin, since different elements to not need to be in the same place. (Will be correlated, but should introduce extra scatter.)} 

We can see from Fig.~\ref{fig:ioncorr_pred} that 
%while even a large survey is unlikely to find the highest \ion{O}{vii} and \ion{O}{viii} column densities we find in {\eagle}, 
a survey targeted at sightlines with observed high \ion{O}{vi} or \ion{Ne}{viii} column densities has a greater chance of detecting \ion{O}{vii} and \ion{O}{viii} systems than a blind survey. As an example, we showed in Section~\ref{sec:EW} that 
equivalent widths of $\sim 1 \us$m{\AA} require 
column densities of $\sim 10^{14.5} \us \mathrm{cm}^{-2}$ for \ion{O}{vii} and 
$\sim 10^{15} \us \mathrm{cm}^{-2}$ for \ion{O}{viii}. 
For those threshold column densities, we predict a $50 \us \% $ chance of detecting an \ion{O}{vii} (\ion{O}{viii}) counterpart to an \ion{O}{vi} absorber at $N_{\mathrm{O\,VI}} \approx 10^{13.4} \us \txn{cm}^{-2}$  ($10^{14.1}$). 
For \ion{Ne}{viii}, the corresponding column density is  
$N_{\mathrm{Ne\,VIII}} \approx 10^{12.9} \us \txn{cm}^{-2}$ ($10^{13.3}$). 
These column densities are close to the breaks in the CDDFs $f(N, z)$ (see equation~\ref{eq:cddf}) of the FUV ions, but not beyond them. We find 
$f_{\mathrm{O\,VI}}\left( 10^{13.4}, z=0.1 \right) = 3 \times 10^{-13} \us \txn{cm}^{2}$ and  
$f_{\mathrm{Ne\,VIII}}\left( 10^{12.9}, z=0.1 \right) = 1.3 \times 10^{-12} \us \txn{cm}^{2}$ at the \ion{O}{vii} threshold 
($1.7 \times 10^{-14}$, $3 \times 10^{-13} \us \txn{cm}^{2}$, 
respectively, for \ion{O}{viii}). 

Fig.~\ref{fig:ioncorr_diag} is  useful for checking whether an observed absorption system matches our expectations: given an \ion{O}{vii} or \ion{O}{viii} column density, it shows what counterparts to expect. For neutral hydrogen, the scatter is very large at \ion{O}{vii} or \ion{O}{viii} column densities $\gtrsim 10^{15} \us \txn{cm}^{-2}$, but for the other ions, the correlations are tighter. 

\subsection{Correlation between absorber properties}
\label{sec:ioncomp}

Coincident absorption from more than one ion can give powerful constraints on the temperature and/or density of the gas, particularly if the ions belong to the same element. However, this only works if the same gas is causing the different absorption lines.
Since gas clouds do not have constant densities and temperatures, it is possible that different absorption lines arise in different parts of the same clouds. 
 Since in planned blind X-ray absorption surveys with Athena, Arcus, and Lynx, \ion{O}{vii} and \ion{O}{viii} are the main targets \citep{brenneman_smith_etal_2016, Athena_2017_11, smith_abraham_etal_2016_arcus, lynx_2018_08}, it is useful to investigate how closely the absorption system properties for these ions match. To do this, we compare temperature and density weighted by \ion{O}{vii} and \ion{O}{viii} along the same $6.25 \us \mathrm{cMpc}$ sightlines. We also compare to \ion{O}{vi}, which has a strong doublet in the FUV range that may provide further constraints without element abundance ratio uncertainties, and to \ion{Ne}{viii}, which we found in Section~\ref{sec:ioncorr} correlates very well with \ion{O}{vii} and \ion{O}{viii}. We do not compare to neutral hydrogen gas properties, since we expect neutral hydrogen to reside in cooler, denser gas than the other ions we considered in Section~\ref{sec:ioncorr}. In addition, any temperature or density constraints from hydrogen to oxygen ion column density ratios would depend on the (generally  unknown) metallicity of the gas.

\begin{figure*}
\includegraphics[width=\textwidth]{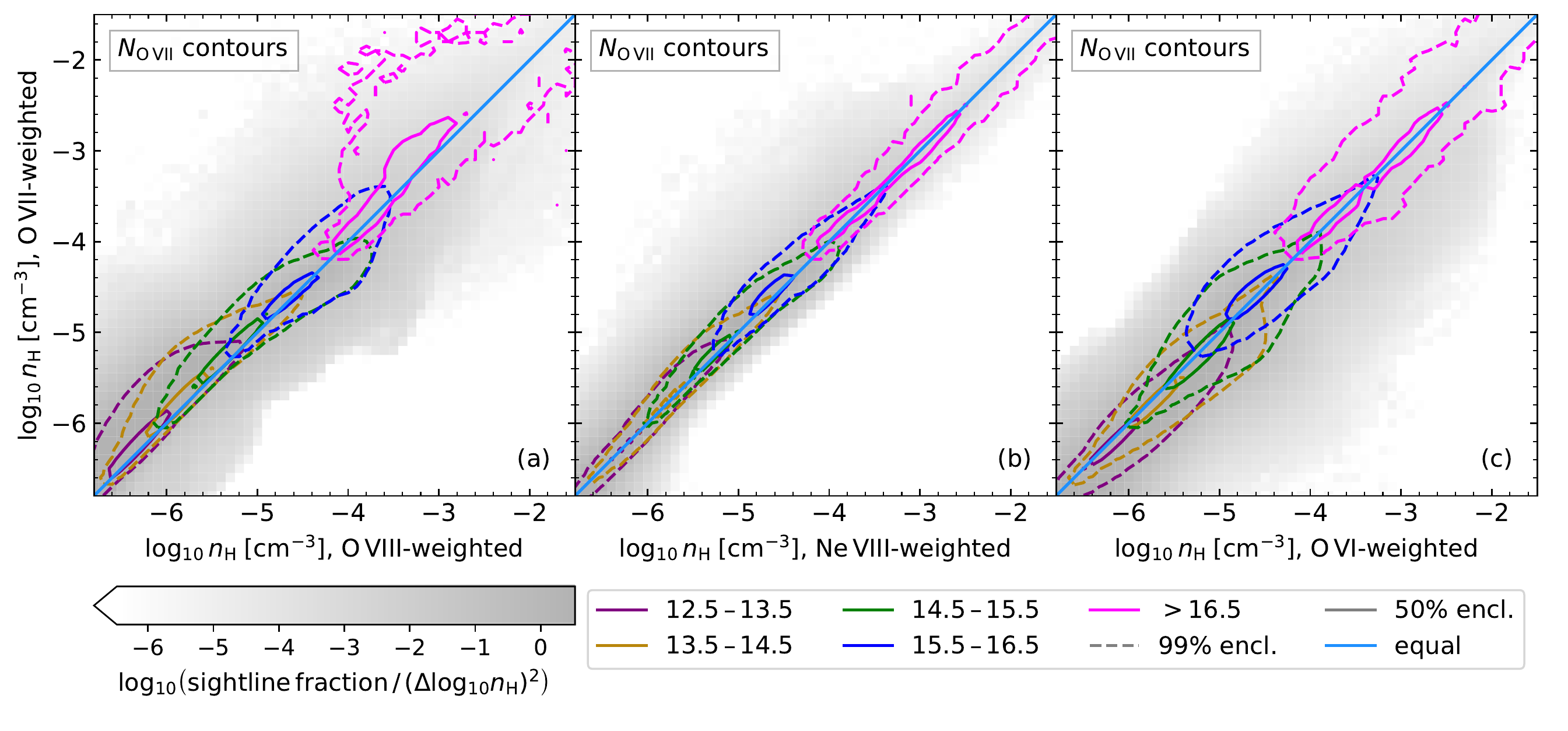}
\caption{The \ion{O}{vii}-weighted density (vertical) of absorption systems plotted against the \ion{O}{viii}-weighted density (a),  the \ion{Ne}{viii}-weighted density (b), and the \ion{O}{vi}-weighted density (c) along the same sightline. These sightlines are our $6.25 \us \mathrm{cMpc}$ long columns, and come from $32000^2$ pixel projections of the reference (\code{Ref-L100N1504}) simulation at redshift $0$.  The greyscale shows the total sightline distribution, while the contours show the density distributions for different \ion{O}{vii} column density bins, matching those in Fig.~\ref{fig:pds-no7w}. These ranges are given in units of $\log_{10} \mathrm{cm}^{-2}$. The light blue lines indicate where the densities are equal. This indicates that these ions tend to trace gas with similar densities along the same sightline, though there are small systematic differences at low column densities.}
\label{fig:rhodiff}
\end{figure*}

First, we explore density differences in Fig.~\ref{fig:rhodiff}. This shows the distribution of \ion{O}{vi}, \ion{Ne}{viii}, \ion{O}{vii} and \ion{O}{viii}-weighted densities\footnote{The axes show density as hydrogen number density, but the densities used for the plot are mass densities, converted to hydrogen number densities assuming a primordial hydrogen mass fraction of $0.752$.} in the $6.25 \us \txn{cMpc}$ columns we use to measure column densities. Contours show the distribution of sightlines in different \ion{O}{vii} column density ranges. The light blue lines indicate where the densities are equal. We can see that at low column densities, where the gas is photoionized ($n_{\txn{H}} \lesssim 10^{-5} \us \txn{cm}^{-3}$), \ion{O}{vii} tends to trace slightly higher-density gas than \ion{O}{viii}. This makes sense, since lower-density gas in the same radiation field is more highly ionized. The differences here are largely ($\geq 98 \%$ of sightlines) below $0.6\us \txn{dex}$ in all column density ranges and for all pairs of ions we show. The median differences are typically small, $\lesssim 0.1 \us \txn{dex}$, except at $N_{\mathrm{O \, VII}} > 10^{16.5} \us \txn{cm}^{-2}$ for \ion{O}{vii} and \ion{O}{viii}, where the median difference is still $< 0.2 \us\txn{dex}$.
%At the very highest column densities, it seems to take high-density gas with a large fraction of the ion in question to produce a large column density. That highest-density gas may not have the right temperature for a different ion. 
The differences between other combinations of these ions are similar. 

%If we consider $1 \us \txn{dex}$ bins in \ion{O}{vi} column density instead of the \ion{O}{vii} bins we show in Fig.~\ref{fig:rhodiff}b, we see a small bias towards \ion{O}{vi} tracing higher densities at all column densities. When comparing  Fig.~\ref{fig:rhodiff}a to a version using bins of \ion{O}{viii} column density, we see similar behaviour at lower column densities, while sightlines with the highest \ion{O}{viii} column densities show a bimodal traced density distribution: like at the highest \ion{O}{vii} column densities, some sightlines trace roughly the same density in both ions, while some trace higher densities with the ion constrained to have a large column density.

\begin{figure*}
\includegraphics[width=\textwidth]{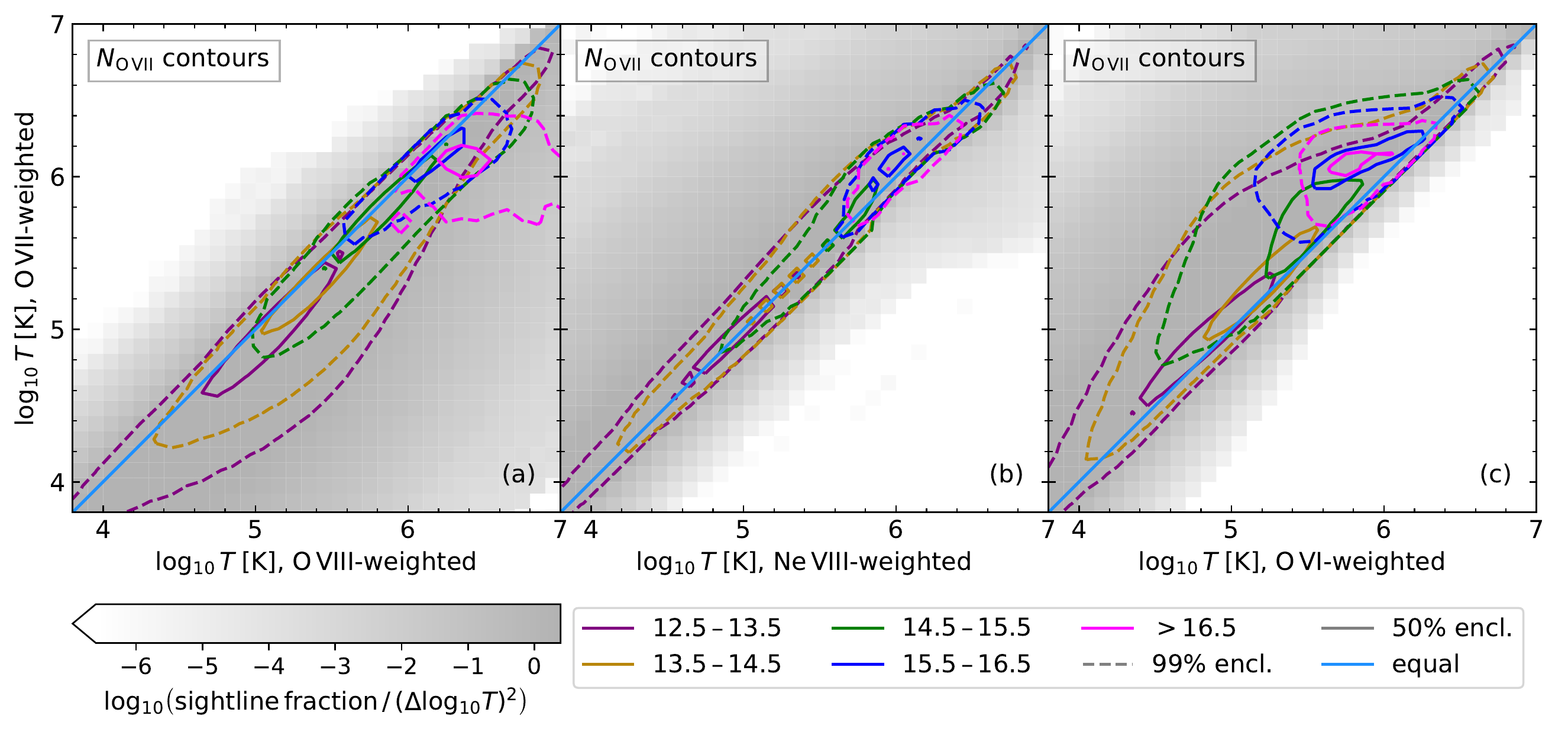}
\caption{As Fig.~\ref{fig:rhodiff}, but for temperature. Between \ion{O}{vii}, and \ion{O}{vi} and \ion{O}{viii}, there are systematic temperature differences at all column densities, but especially the larger ones. The \ion{O}{vii} and \ion{Ne}{viii} temperatures agree within $0.3 \us \txn{dex}$ for $ \gtrsim 98 \%$ of absorbers at all column densities shown.}
\label{fig:Tdiff}
\end{figure*}

Next, we look into temperature, shown in Fig.~\ref{fig:Tdiff}, similar to Fig.~\ref{fig:rhodiff}, but for ion-weighted temperatures. Between \ion{O}{vii} and \ion{O}{vi} or \ion{O}{viii}, differences are typically larger than for density, and clearly systematic: the higher-energy \ion{O}{viii} ions prefer hotter gas than \ion{O}{vii} at all (column) densities, while \ion{O}{vii} traces hotter gas than \ion{O}{vi}. The median differences at lower column densities ($N_{\mathrm{O \, VII}} \lesssim 10^{15.5} \us \txn{cm}^{-2}$) are $\lesssim 0.2 \us \txn{dex}$, but at higher column densities, the median temperatures differ by up to $\approx 0.3 \, \mathrm{dex}$ for \ion{O}{vii} and \ion{O}{vi} or \ion{O}{viii}. This occurs at column densities tracing collisionally ionized gas, where the ions both tend to trace their own CIE peak temperature gas. For \ion{Ne}{viii} and \ion{O}{vii}, we see a hint of a similar CIE peak preference in the narrowing of the range of absorber temperatures as the \ion{O}{vii} column density increases, but because these two ions have similar peak temperatures, their median differences remain $\lesssim 0.1 \us \txn{dex}$. 

We also compared temperatures and densities traced by \ion{O}{vi} and \ion{O}{viii} along the same $6.25 \us \txn{cMpc}$ sightlines. We found similar trends in the differences to those we found between \ion{O}{vi} and \ion{O}{vii} and between \ion{O}{vii} and \ion{O}{viii}, except that both the scatter and systematic differences were larger.

%If we consider $1 \us \txn{dex}$ bins in \ion{O}{vi} column density instead of the \ion{O}{vii} bins we show in Fig.~\ref{fig:Tdiff}b, we see similar behaviour at lower column densities. At higher \ion{O}{vi} column densities, as the ions become collisionally ionized, we see a smaller spread in \ion{O}{vi}-weighted temperature and a larger spread in \ion{O}{vii}-weighted temperature than at large \ion{O}{vii} column densities. This is not unexpected, since selecting large column densities for a particular ion mainly constrains the gas that ion traces to be at CIE temperatures. We see a similar, but smaller effect when comparing  Fig.~\ref{fig:Tdiff}a to a version using bins of \ion{O}{viii} column density; this might be a result of \ion{O}{viii} tracing a larger temperature range than \ion{O}{vii} at high column densities (figures~\ref{fig:pds-no7w} and~\ref{fig:pds-no8w}). 

%To investigate the impact of these differences on modelling, we find the ion balances for each of these absorption systems, assuming a primordial hydrogen mass fraction ($0.752$) to convert density to hydrogen number density. We use the \ion{O}{vii}-weighted and \ion{O}{viii}-weighted quantities to calculate the ion balance for each ion, then compare the results. This only uses the absorption system temperatures and densities, not their oxygen mass fractions. The results are shown in Fig.~\ref{fig:N_vs_ibreldiff}. Here, we show the distribution of sightlines in column density and relative ion fraction difference. 

%For \ion{O}{vii} and \ion{O}{viii}, we also computed the ion balance expected for an absorber at the temperature and density traced by each ion. %We compared these ion balances for each species, as calculated from the gas properties traced by each ion. We found that these reconstructed ion balances were most commonly close to zero, but could differ substantially: the region enclosing $99\%$ of sightlines reached at least a $0.5 \us \txn{dex}$ difference at all column densities, and a difference above $1 \us \txn{dex}$ for column densities $> 10^{16} \us \txn{cm}^{-2}$. The differences were larger for \ion{O}{vii} fractions, and became systematic at the largest column densities, where we also saw systematic temperature differences. 
The differences in gas temperature and density probed by different ions in the same structures may be large enough to be a potential issue when modelling the absorption as coming from the same gas. 
%Overall, most sightlines showed differences in ion balance smaller than $1 \us \mathrm{dex}$, while at most column densities $> 10^{12} \us \mathrm{cm}^{-2}$ in both ions, the ion balances from and for both ions differ by less than $0.5 \us \mathrm{dex}$ for at least half the sightlines. The most common differences are close to zero. For the \ion{O}{vii} ion balance, the differences seem to be larger more often,  and are more clearly systematic: using \ion{O}{vii}-weighted quantities almost always gives \ion{O}{vii} fractions roughly equal to or larger than those given by \ion{O}{vii}-weighted quantities. This is what we would naively expect, but \ion{O}{viii} seems to show more symmetric scatter, with the most common differences corresponding to slightly higher \ion{O}{viii} fractions from \ion{O}{vii}-weighted quantities at lower column densities and slightly higher \ion{O}{viii} fractions from \ion{O}{viii}-weighted quantities at higher column densities. The transitions occurs at a column density of $\sim 10^{14} \us \mathrm{cm}^{-2}$ in either ion. These differences seem to be large enough, at all column densities, to be a potential problem for modelling systems where there is more than one line from one absorption system. These differences seem to be large enough, at all column densities, to be a potential problem for modelling systems where there is more than one line from one absorption system.
In future work, we intend to test this by applying ionization models to column densities measured from virtual absorption spectra. This will also enable us to test whether enforcing different lines to have identical redshifts would help reduce the systematic errors given the energy resolution expected for upcoming missions.

\section{Discussion}

\label{sec:discussion}

\subsection{Detection prospects}
\label{sec:missions}

We examine the prospects for detecting \ion{O}{vii} and \ion{O}{viii} line absorption with the planned mission 
Athena \citep{Athena_2017_11, Athena_2018_07}, and the proposed missions
Arcus\footnote{This was one of three (later two) proposed NASA MIDEX missions in the most recent (February 2019) funding round, but it was not chosen.} \citep{smith_abraham_etal_2016_arcus, brenneman_smith_etal_2016}, 
and Lynx \citep{lynx_2018_08}. 
Detectability depends on the background source as well as on the absorber. Here, we consider quasars/blazars like those discussed in the mission science cases. The Arcus and Athena science cases include detecting the WHIM and measuring the CDDF for WHIM absorbers \citep{brenneman_smith_etal_2016, Athena_2017_11}. 
For Lynx \citep{lynx_2018_08}, which would launch later if approved, the focus is on characterising this gas. %Brenneman: Arcus

\begin{table}
\caption{Expected number of absorption systems per unit redshift above a given observer-frame EW threshold $\txn{d}N(>\txn{EW}_{\txn{obs}})\,/\,\txn{d}z$ for \ion{O}{vii} and \ion{O}{viii} at different redshifts. Here, we use $6.25 \us \txn{cMpc}$ slice CDDFs and $100 \us \txn{cMpc}$ column density equivalent width relations (with scatter) at each redshift. The results come from the reference (\code{Ref-L100N1504}) {\eagle} simulation.}
\label{tab:EWvals}
\begin{tabular}{r r@{.}l r@{.}l r@{.}l r@{.}l r@{.}l r@{.}l}
\hline
$\mathrm{EW}_{\mathrm{obs}}$ 
 & \multicolumn{2}{c}{\ion{O}{vii}} 	& \multicolumn{2}{c}{}	& \multicolumn{2}{c}{}	
 & \multicolumn{2}{c}{\ion{O}{viii}}	& \multicolumn{2}{c}{} 	& \multicolumn{2}{c}{}	\\
m{\AA} 
& \multicolumn{2}{c}{$z=0.1$}	& \multicolumn{2}{c}{$z=0.2$}	& \multicolumn{2}{c}{$z=1.0$} 	
& \multicolumn{2}{c}{$z=0.1$}	& \multicolumn{2}{c}{$z=0.2$}	& \multicolumn{2}{c}{$z=1.0$} 	\\
\hline
$6.8$	& 1&98	& 2&43	& 4&33	& 0&533	& 0&762	& 3&23	\\
$5.2$	& 2&95	& 3&53	& 5&59	& 0&987	& 1&33	& 4&49	\\
$4.0$	& 4&06	& 4&77	& 6&92	& 1&60	& 2&09	& 5&94	\\
$3.0$	& 5&57	& 6&37	& 8&61	& 2&49	& 3&23	& 7&74	\\
$1.0$	& 14&0	& 15&1	& 17&2	& 9&87	& 11&8	& 17&5	\\
\hline
\end{tabular}
\end{table}
%, and using the full (`all') and ion-selected (`ion') sample of sightlines to obtain the column density-EW relation for the longer sightlines.
% & all 	& ion	& all
% old version, using z=0.1 CDDF for all, and with a bug in the read-out function
%$6.8$	& 1&95	& 2&60	& 9&05	& 0&605	& 0&875	& 3&62	\\
%$5.2$	& 3&11	& 4&04	& 11&7	& 1&02	& 1&41	& 5&20	\\
%$4.0$	& 4&12	& 5&27	& 14&5	& 1&80	& 2&11	& 8&19	\\
%$3.0$	& 5&86	& 6&83	& 19&1	& 2&61	& 3&46	& 11&2	\\
%$1.0$	& 14&1	& 17&1	& 40&6	& 10&6	& 12&3	& 36&3	\\
% with the bug removed, but old method:
%       O VII                           O VIII 
%       z=0.0   0.1	    0.2     1.0     0.0     0.1     0.2     1.0         
%$6.8$	1&42	& 1&95	& 2&53	& 8&44	& 0&351	& 0&531	& 0&750	& 3&47	\\
%$5.2$	2&21	& 2&91	& 3&69	& 11&3	& 0&697	& 0&985	& 1&31	& 5&18	\\
%$4.0$   3&15	& 4&03	& 4&99	& 14&3	& 1&17	& 1&59	& 2&06	& 7&52	\\
%$3.0$	4&32	& 5&51	& 6&77	& 18&5	& 1&88	& 2&49	& 3&18	& 10&9	\\
%$1.0$  11&2    &13&9	&16&7	& 40&4	& 7&70	& 9&86	& 12&2	& 34&3	\\
% 6.25 cMpc sightlines, but snapshot 27 N-EW relation
%$6.8$	& 1&98	& 2&34	& 4&24	& 0&533	& 0&773	& 3&26	\\
%$5.2$	& 2&95	& 3&41	& 5&55	& 0&987	& 1&34	& 4&46	\\
%$4.0$	& 4&06	& 4&63	& 6&88	& 1&60	& 2&10	& 5&92	\\
%$3.0$	& 5&57	& 6&28	& 8&59	& 2&49	& 3&21	& 7&76	\\
%$1.0$	& 14&0	& 15&2	& 17&1	& 9&87	& 11&8	& 17&5	\\

The Arcus X-ray Grating Spectrometer (XGS) \citep{smith_abraham_etal_2016_arcus, brenneman_smith_etal_2016} should be able to find absorbers with equivalent widths $\geq 4 \us${m\AA} at $5 \sigma$ against blazars of a brightness threshold met by at least 40 known examples, at redshifts up to $0.2$ with exposure times $\leq 500 \us \mathrm{ks}$. There are also plans to measure absorption against gamma-ray bursts. Fig.~\ref{fig:ew_litcomp} shows that this threshold is roughly at the knee of the equivalent width (EW) distribution. Fig.~\ref{fig:cog} shows that Arcus would mainly be probing absorption systems that do not lie on the linear part of the curve of growth with this EW regime. \citet{smith_abraham_etal_2016_arcus} expect to find around 40 \ion{O}{vii} absorbers at $\txn{EW} > 4 \us${m\AA}, using about 20 blazar background sources probing redshift paths that add up to $\Delta z \approx 8$. 

We can make a similar prediction for the number of detections in such a survey. The results are shown in Table~\ref{tab:EWvals}. At each redshift, we make predictions using EW distributions obtained from $6.25 \us \txn{cMpc}$ slice CDDFs, using the (rest-frame) column density EW relation for a full sample of $100 \us \txn{cMpc}$ sightlines at each redshift. We obtain the column density equivalent width relation at $z=0.2$ and $z=1$ in the same way as described in Sections~\ref{sec:ew_methods} and~\ref{sec:EW} for $z=0.1$, except that we select the higher-redshift sightlines by \ion{O}{vii} and \ion{O}{viii} column density alone, ignoring \ion{O}{vi}. This should not significantly affect our results, since the ion-selected subsamples did not yield substantially different results from the full sample at $z=0.1$. 

In this calculation, changes in the predicted number of absorbers with redshift are due to four effects. First, the CDDFs evolve, as we saw in Fig.~\ref{fig:cddf_zev}. The evolution of the CDDFs between redshifts~$0.1$ and~$0.2$ is small, but differences with $z=1.0$ are larger. 
Secondly, the relation between column density and equivalent width evolves. This evolution is weak, affecting the values in Table~\ref{tab:EWvals} by $\lesssim 4 \us \%$.
Thirdly, these CDDFs concern the absorption system distribution with respect to $\diff X$, not $\diff z$. Between redshifts $0.0$ and $0.2$, $\diff X / \diff z$ increases from $1$ to $1.30$.
%, with a value of $1.15$ at redshift~$0.1$. 
Finally, the instrumental EW threshold applies to the observed EW. This means we can probe lower-EW systems at higher redshift, assuming e.g.\ quasar fluxes and backgrounds/foregrounds are equal. 
%This table shows that using the $6.25 \us \txn{cMpc}$ column CDDF instead of the $100 \us \txn{cMpc}$ CDDF makes a large difference in the predicted detection rates for \ion{O}{viii}. We verified that using the column density-EW relation from all sightlines (as shown in Table~\ref{tab:EWvals}) or only the ones selected for the ion in question makes a difference of a most a few percent in the tabulated values, which is well below the differences over the redshift range.

For $\Delta z = 8$, we expect to see 32 (38)~\ion{O}{vii} absorption systems and 13 (17)~\ion{O}{viii} absorption systems if redshift~$\approx0.1$ ($0.2$) contributes most to the survey path length. These values agree roughly with the $\approx 40$ \ion{O}{vii} absorbers expected by \citet{smith_abraham_etal_2016_arcus}. They expect 10--15 \ion{O}{viii} counterparts to \ion{O}{vii} lines along these sightlines, which is compatible with the total number of \ion{O}{viii} absorbers we expect. From the correlations between \ion{O}{vii} and \ion{O}{viii} absorption systems, (Fig.~\ref{fig:ioncorr_diag}g), and the column-density-equivalent-width relation we find (Fig.~\ref{fig:cog}), we expect that some of the detectable \ion{O}{viii} absorbers do not have observable \ion{O}{vii} counterparts: for absorbers with 
$N_{\mathrm{O\,VIII}} \geq 10^{15.5} \us \txn{cm}^{-2}$,
$31 \us \%$ ($3.7 \us \%$) have \ion{O}{vii} counterparts with 
$N_{\mathrm{O\,VII}} \leq 10^{15.5} \us \txn{cm}^{-2}$ ($\leq 10^{15}$).

%Using the column density-EW relation from all sightlines or only the ones selected for the ion in question makes a difference of a most a few percent, as shown in Table~\ref{tab:EWvals}, well below the differences over the redshift range. 
As shown in appendix~\ref{app:dXvar}, we do not see major effects of the large-scale structure on variations in measured CDDFs in the {\eagle} reference simulation (\code{Ref-L100N1504}). We expect the variations in measured EW distributions are Poisson as well. That would imply that the expected survey-to-survey scatter in the number of absorption systems detected is similar to the maximum effect of varying the background source redshift distribution (as long as $z \lesssim 0.2$ still holds). 
%Given a more detailed survey description, we could calculate the expected number of absorption systems more accurately. 

If the sensitivity can be increased to a minimum EW of $3\us${m\AA}, we would expect to see about $1.3$--$1.6$ times as many absorption systems at redshifts~$0.1$--$0.2$. If, for deep surveys, a sensitivity of $1\us${m\AA} can be achieved, we would expect to find  
double to triple the number of \ion{O}{vii} absorption systems and double to quadruple the number \ion{O}{viii} absorption systems per unit redshift as at $3 \us$m{\AA}.
 
%{\todo At the Alabama WHIM 2018 conference, Andy Ptak stated that a minimum equivalent width of $3 \us${m\AA} was supported by instrument simulations, with $1 \us${m\AA} for deep surveys. Of course, the detection limit will depend on the S/N of the observations, and the limiting equivalent width will therefore depend on the survey depth.}

As for resolving line widths or line shapes, Fig.~\ref{fig:mockspectra} shows the expected effects of line broadening and Poisson noise on observations. This figure suggests that constraining the line widths of single absorption components will be difficult, especially since what would appear to be single absorption components often are not. Comparing our approximate line widths from Fig.~\ref{fig:cog} to the spectral resolution expected for the instrument supports this.

The Arcus XGS resolving power was planned to be $\lambda/\Delta \lambda = 2000, 2500$ below and above $21.6 \us${\AA}, respectively \citep{smith_abraham_etal_2016_arcus}. For our broadest \ion{O}{vii} lines ($\approx 220 \us \mathrm{km}/\mathrm{s}$, we have $\lambda/\Delta \lambda \approx 1.4 \times 10^3$ assuming a single line. For \ion{O}{viii}, $\lambda/\Delta \lambda \approx 1.0 \times 10^3$ for the broadest lines. The $b$ parameters for \ion{O}{viii} in Fig.~\ref{fig:cog} were calculated by fitting the relation between column density and equivalent width by the two doublet lines, each with a width $b$. 
%This means that the actual doublet width is also broadened by the line separation. This separation alone gives $\lambda/\Delta \lambda \sim 3.5 \times 10^3$, and 
Fitting the \ion{O}{viii} curve of growth using a single line instead (using the sum of oscillator strengths and oscillator-strength-weighted average wavelength) only adds about $10 \us \mathrm{km}/\mathrm{s}$ to the best-fit $b$ parameter. 
%(for all the fitting variations)
This would suggest we may be able to constrain some of the line widths with the Arcus XGS, though a good measurement of the line widths may be possible for only the broadest lines.
Bear in mind that the $b$ parameters we use are a rough measure, determined by the curve of growth rather than the actual structure of the absorption system, and that unresolved turbulence may cause lines to be broader than predicted from the velocity structure resolved in the simulation alone. Although resolving individual components will be difficult, Arcus will be able to decompose many systems into multiple components. 
%These COG fit $b$ parameters suggest that it would be unlikely we can resolve these lines well enough to actually measure single-component $b$ parameters with Arcus. Fig.~\ref{fig:mockspectra} suggests this may indeed be difficult, especially since what would appear to be single absorption components often are not. 
%(Note the shaded bands, which show the effect of resolution and indicate noise only by the thickness of the band.)

On Athena, the instrument of interest is the X-IFU \citep{Athena_2018_07}. The science requirements for Athena as a whole, including this instrument, are described by \citet{Athena_2017_11}. Weak line $5 \sigma$ detections should be possible from $EW = 0.18 \us \mathrm{eV}$ against `bright sources'. The plan is to measure WHIM absorption against 100 BLLacs and 100 gamma-ray bursts, to study the WHIM at $z<1$. 

%These requirements are for measurements at $1 \us \mathrm{keV}$, while the rest-frame ion lines are at $0.5740 \us \mathrm{eV}$ (\ion{O}{vii}) and  $0.6536 \us \mathrm{eV}$ (\ion{O}{viii}). These are the closest values \citet{Athena_2017_11} give, so we will assume they give a good idea of the requirements at the ion absorption line energies we are interested in. 
The minimum EW translates to $6.8$ and $5.2 \us${m\AA}  (rest-frame) for \ion{O}{vii} and \ion{O}{viii} respectively, for which we show the expected number of absorption systems per unit redshift in Table~\ref{tab:EWvals}. 
These limits are for $50 \us \mathrm{ks}$ observations against a point source with a 0.5~mCrab  flux  in 
the 2--10~keV  energy  band  ($F_{2\mathrm{-}10 \us \mathrm{keV}} = 1 \times 10^{-11}\us \mathrm{erg}\,\mathrm{cm}^{-2} \mathrm{s}^{-1}$ for $\Gamma= 1.8$). 
%{\todo A quick search shows this seems to be a reasonable quasar flux: \url{https://arxiv.org/pdf/0808.0184.pdf} for a measurement including the brightest blazar \citep{nicastro_etal_2016}.} 
These are larger EWs than expected for Arcus, though still above the knee of the EW distribution. Note that the observing time used here is ten times smaller than that of the Arcus specification, so some differences are expected. Rather than attempt to correct for this difference, we explore what Athena would see if the X-IFU is used as described. 

The exact number of absorption systems expected from the survey will depend on the redshift distribution of the quasars and gamma-ray bursts used to probe the WHIM. We make predictions for Athena in the same way as for Arcus. Over the redshift range~$0.1$--$1$, Table~\ref{tab:EWvals} shows large changes in predicted absorber densities in redshift space. The expected number of absorption systems at the Athena sensitivity thresholds is less than expected in the Arcus survey, given the current factor~$10$ difference in planned exposure times, but if $\Delta z = 50$--$100$ is achieved, we would expect the Athena survey to find a larger number of \ion{O}{vii} absorption systems. Since the aim is to observe 100~BLLacs and 100~gamma-ray burst afterglows, such a survey size seems reasonable.

For the Lynx~XGS \citep{lynx_2018_08}, a goal is to use \ion{O}{vii} and \ion{O}{viii} absorption to characterise the hot CGM of galaxies with halo mass $\gtrsim 10^{12} \us \txn{M}_{\sun}$ and the hot gas in filaments with overdensities $\gtrsim 30$. For the filaments, this means looking for column densities $\gtrsim 10^{15} \us \txn{cm}^{-2}$ and rest-frame EWs $\sim 1 \us$m{\AA}. Indeed, \citet{lynx_2018_08} expects to be able to detect lines at $\txn{EW} \gtrsim 1 \us$m{\AA} outside halo virial radii, against bright AGN 
($F_{0.5\mathrm{-}2 \us \mathrm{keV}} \sim 1 \times 10^{-11}\us \mathrm{erg}\,\mathrm{cm}^{-2} \mathrm{s}^{-1}$)
with an average exposure time of $\approx 60 \us\txn{ks}$. Table~\ref{tab:EWvals} shows the predicted number of detected \ion{O}{vii} and \ion{O}{viii} absorbers in blind surveys with this sensitivity.

In the future we plan to improve on the rough estimates provided in this section by creating many virtual Arcus and X-IFU spectra, processing them through the instrument models, and analysing them as if they were real data.

\subsection{Caveats}
\label{sec:caveats}

We have studied the impact of resolution and some technical choices in appendices~\ref{app:conv} and~\ref{app:projchoice}. There are some other approximations and uncertainties we will discuss here. 

First, we assume the ion fractions are set by the temperature and hydrogen number density of the gas, assuming collisional and photoionization equilibrium with a uniform (but evolving) \citet{HM01} UV/X-ray background. 
%{\todo Uncertainties on the part of the spectrum that matters for these high ions is not super well known -- Michael Shull at the WHIM conference, I think. }     
\citet{cen_fang_2006} found that there were differences in the equivalent width distributions derived with and without equilibrium assumptions. These differences are comparable to differences between different simulations for \ion{O}{vii}, shown in Fig.~\ref{fig:ew_litcomp} and smaller than those differences for \ion{O}{viii}. We therefore consider ionization equilibrium a reasonable model simplification. \citet{yoshikawa_sasaki_2006} also discuss non-equilibrium effects, and plot at what temperatures and densities they expect non-equilibrium effects to be strongest. They found that WHIM detectability was not impacted much by non-equilibrium effects, but that line ratios were impacted more significantly, further complicating measurements of absorber temperatures.  \citet{oppenheimer_etal_2016} found using much 
higher-resolution simulations that non-equilibrium ionization was generally unimportant for \ion{O}{vi}. 

Another issue relating to ionization and equilibrium is the impact of local ionization sources. For these high-energy ions, we would expect AGN to be the most important. \citet{oppenheimer_schaye_2013} pointed out that if, as expected, AGN vary on time scales similar to 
or shorter than the recombination time scale in the IGM, then a large fraction of high-ionization metal absorbers may be more highly ionized than they would be if they were only ever illuminated by the metagalactic background radiation. This could then even affect gas near galaxies that currently do not show detectable AGN activity. 
\citet{segers_oppenheimer_etal_2017} found that in the inner parts of the halo, a variable AGN could have large effects on \ion{O}{vi} column densities, with smaller effects out to $2 \us R_{\txn{vir}}$. These effects persisted between periods where the AGN was `on'. They did not study the effect of this on \ion{O}{vii} and \ion{O}{viii}, so the precise impact here is unknown. 
The effect could be important in regions where the temperature is too low for collisional ionization to be effective. Here, AGN radiation could add a population of cooler, high column density systems. However, it might also affect the population of strong absorbers in our current sample, if the X-ray flux is high enough to ionize their hot, dense gas even further.  
%Figures~\ref{fig:coldensmaps} and~\ref{fig:cddfs_o78_by_rho} show this should mainly affect the gas at the highest column densities. As figures~\ref{fig:pds-no7w} and~\ref{fig:pds-no8w} show, this is mostly collisionally ionized gas in a \citet{HM01} UV/X-ray background, but this gas may become photoionized in a stronger X-ray radiation field. 

%The key to this fossil AGN effect is having ion recombination times much longer than the time scale on which AGN `flicker'. \citet{segers_oppenheimer_etal_2017} discuss that these are time scales are highly uncertain, but likely are somewhere between $\sim10^{-1}$ and $\sim 30 \us \txn{Myr}$. \citet{yoshikawa_sasaki_2006} show that with the same \citep{HM01} ionizing background, recombination time scales for gas at overdensities~$\sim 100$--$1000$ (column densities $\sim 10^{15.5}$--$10^{16.5} \us \txn{cm}^{-2}$) in the \ion{O}{vii} and \ion{O}{viii} CIE peak ranges are $\gtrsim 1 \us \txn{Gyr}$, so this effect may be important for the CDDF at columns densities above the break.

A further difficulty lies in defining absorbers. We have somewhat avoided this issue when discussing column densities, by limiting our work to a reasonable proxy for absorption \emph{systems}. However, in observations, observers usually fit line profiles to their spectra and gather information on each line. Although we have synthetic spectra, we have not attempted to identify and characterise individual absorption components in these, since the finding and identification of these components will generally depend on the observing instrument and exposure time, and is time-consuming. From visual inspection of Athena and Arcus mock spectra, it seems that absorption systems would usually look more or less like single lines at the spectral resolution of Arcus, and would nearly always look this way at the Athena spectral resolution. This justifies limiting our discussion to entire absorption systems, though we do plan to revisit this question in future work.

\section{Conclusions}
\label{sec:conclusion}

We have used the EAGLE cosmological, hydrodynamical simulations to predict the rate of incidence and physical conditions of intergalactic \ion{O}{vii} ($\lambda = 21.6019\us${\AA}, $E=574 \us \txn{eV}$) and \ion{O}{viii} ($\lambda = 18.9671, 18.9725 \us${\AA}, $E=654 \us \txn{eV}$) absorbing gas. In the largest simulation, $40 \us \%$ of gas-phase oxygen is in the form of these two ions, making them the key tracers of cosmic metals.  
We have extracted column density distribution functions (CDDFs) for \ion{O}{vii}  and \ion{O}{viii} by measuring the number of ions in a grid of long, thin columns within slices through the simulation box. 
Assuming a uniform metallicity instead of the predicted metallicities results in much steeper CDDFs (Fig.~\ref{fig:cddf_z0p1}) because higher column densities are predicted to correspond to higher metallicities. 
The most notable feature in the CDDFs, which evolve only mildly between $z=1$ and $z=0$, looks like a power-law break, and occurs at column densities $\sim 10^{16} \us \mathrm{cm}^{-2}$ (Fig.~\ref{fig:cddf_zev}). This break occurs where absorption systems reach overdensities $\sim 10^2$ (Fig.~\ref{fig:cddfs_o78_by_rho}), which indicates it is likely caused by a transition from absorption by sheet and filament gas to absorption by denser halo gas. 
%This break density roughly matches the density where typical absorbers start to become dense enough to cool radiatively (figures~\ref{fig:pds-no7w} and~\ref{fig:pds-no8w}). 
%We find that there is some evolution in the column density distributions between redshifts zero and one, but it is not much, especially around $N \sim 10^{16} \us \mathrm{cm}^{-2}$ where absorbers are most likely to be detected in blind surveys (Fig.~\ref{fig:cddf_zev}). 

AGN feedback substantially modifies the CDDF at high column densities. It impacts both the position of the break in the CDDF and its slope after the break, indicating an impact on haloes. Without AGN feedback, there are slightly fewer absorption systems at column densities of $\sim 10^{15}$--$10^{16} \us \txn{cm}^{-2}$ and significantly more at column densities $\gtrsim 10^{16} \us \txn{cm}^{-2}$ (Fig.~\ref{fig:noagn}). Before the break, this is because AGN feedback increases the metallicity of the gas. Beyond the break, AGN feedback still increases the metallicity, but column densities decrease due to the decrease in the hot gas densities. %This suggests that less metals escape haloes without AGN, and are possibly not reaching the outer (lower-density) parts of the haloes. Within the haloes, metallicity differences do not explain (all) the differences, and the AGN impact on the CGM temperature and/or density must play a part. 

We used a large set of synthetic absorption spectra to determine the relation between equivalent width (EW) and projected column density. The best-fit $b$-parameters to the curve of growth are $90$ and $158 \us \txn{km}\, \txn{s}^{-1}$ for \ion{O}{vii} and \ion{O}{viii}, respectively, but the scatter in $b(N)$ is large (Fig.~\ref{fig:cog}). 
We used the CDDF to predict the EW distribution in the {\eagle} reference simulation at redshift~$0.1$ (Fig.~\ref{fig:ew_litcomp}), accounting for the scatter by using our sample's EW distribution at fixed column density in the conversion. We predict an \ion{O}{vii} EW distribution consistent with the recent observations of \citet{nicastro_etal_2018}. %Our predictions are a bit more pessimistic at lower equivalent widths, while our \ion{O}{viii} predictions are somewhat more pessimistic than those of \citet{branchini_ursino_etal_2009} and much more pessimistic than those of \citet{cen_fang_2006}.

By measuring the ion-weighted gas temperatures and densities along the same columns, we investigated the physical properties of our absorption systems. 
Overall, for $32 \us \%$ ($8 \us \%$) of \ion{O}{vii} (\ion{O}{viii}) the ion fractions are within a factor 1.5 of the values corresponding to collisional ionization equilibrium, while for most of the remaining gas both collisional ionization and photoionization by the metagalactic background radiation are important for the ion fractions.
Although \ion{O}{vii} and \ion{O}{viii} often trace gas that is (partly) photoionized, \ion{O}{vii} and \ion{O}{viii} absorption systems with column densities $\gtrsim 10^{16} \us \txn{cm}^{-2}$ mostly trace gas at temperatures close to those where their ion fractions peak in collisional ionization equilibrium: $\sim 10^6$ and $\sim 10^{6.5} \us \txn{K}$, respectively. 
At column densities $\sim 10^{15}$--$10^{16} \us \mathrm{cm}^{-2}$, much of the gas is at temperatures consistent with collisional ionization equilibrium, but the ionization fractions are not the same as the purely collisional values (figures~\ref{fig:pds-no7w} and~\ref{fig:pds-no8w}). There are systematic differences between the temperatures traced by \ion{O}{vii} and \ion{O}{viii} for the same absorbing structures (Fig.~\ref{fig:Tdiff}). Along with density differences, this may lead to difficulties with using column density ratios to estimate gas temperatures. Instrumental line broadening and unresolved absorption system structure mean that temperatures will also typically not be well-constrained by line widths. 

On the other hand, the column densities of \ion{O}{vii} and \ion{O}{viii} are correlated with each other, and with those of ions with detectable absorption lines in the far UV (Figs.~\ref{fig:ioncorr_pred}, \ref{fig:ioncorr_diag}). The presence of neutral hydrogen absorption with column density $\gtrsim 10^{15} \us \txn{cm}^{-2}$ implies  a $\sim 50 \us \%$ chance of finding an \ion{O}{vii} or \ion{O}{viii} absorber with column density $\gtrsim 10^{15} \us \txn{cm}^{-2}$, but the probability does not increase further for even stronger \ion{H}{i} absorption (until ISM densities are reached). High \ion{O}{vi} or \ion{Ne}{viii} column densities are good predictors of the presence of detectable \ion{O}{vii} and \ion{O}{viii} column densities (Fig.~\ref{fig:ioncorr_pred}). This suggests that pre-selection based on these UV lines is an efficient (though necessarily biased) search strategy for the X-ray lines.

%\section{Conclusions}
%{ \todo
%The last numbered section should briefly summarise what has been done, and describe
%the final conclusions which the authors draw from their work.
%}
\section*{Acknowledgements}
%The Acknowledgements section is not numbered. Here you can thank helpful
%colleagues, acknowledge funding agencies, telescopes and facilities used etc.
We thank the referee for their helpful comments.
We would like to thank Dylan Nelson and Enzo Branchini for sharing their data, John Helly for programs we used to access {\eagle} data (\textsc{read{\_}eagle}), and Volker Springel for the original version of the code we use to project particles onto a grid (\textsc{HsmlAndProject}). We would also like to thank Ali Rahmati for help testing the code we use to make projections and setting up {\specwizard}. We also made use of WebPlotDigitizer \citep{webplotdigitizer} and the \textsc{numpy} \citep{numpy}, \textsc{scipy} \citep{scipy}, \textsc{h5py} \citep{h5py}, and \textsc{matplotlib} \citep{matplotlib} \textsc{python} libraries, as well as the \textsc{ipython} \citep{ipython} command-line interface. 
This work used the DiRAC@Durham facility managed by the Institute for
Computational Cosmology on behalf of the STFC DiRAC HPC Facility
(\url{www.dirac.ac.uk}). The equipment was funded by BEIS capital funding
via STFC capital grants ST/K00042X/1, ST/P002293/1, ST/R002371/1 and
ST/S002502/1, Durham University and STFC operations grant
ST/R000832/1. DiRAC is part of the National e-Infrastructure. 
RAC is a Royal Society University Research Fellow. 
FN acknowledges funding from the INAF PRIN-SKA 2017 program 1.05.01.88.04.
%%%%%%%%%%%%%%%%%%%%%%%%%%%%%%%%%%%%%%%%%%%%%%%%%%



\bibliographystyle{mnras}
\bibliography{bibliography} % if your bibtex file is called example.bib

\begin{thebibliography}{}
\makeatletter
\relax
\def\mn@urlcharsother{\let\do\@makeother \do\$\do\&\do\#\do\^\do\_\do\%\do\~}
\def\mn@doi{\begingroup\mn@urlcharsother \@ifnextchar [ {\mn@doi@}
  {\mn@doi@[]}}
\def\mn@doi@[#1]#2{\def\@tempa{#1}\ifx\@tempa\@empty \href
  {http://dx.doi.org/#2} {doi:#2}\else \href {http://dx.doi.org/#2} {#1}\fi
  \endgroup}
\def\mn@eprint#1#2{\mn@eprint@#1:#2::\@nil}
\def\mn@eprint@arXiv#1{\href {http://arxiv.org/abs/#1} {{\tt arXiv:#1}}}
\def\mn@eprint@dblp#1{\href {http://dblp.uni-trier.de/rec/bibtex/#1.xml}
  {dblp:#1}}
\def\mn@eprint@#1:#2:#3:#4\@nil{\def\@tempa {#1}\def\@tempb {#2}\def\@tempc
  {#3}\ifx \@tempc \@empty \let \@tempc \@tempb \let \@tempb \@tempa \fi \ifx
  \@tempb \@empty \def\@tempb {arXiv}\fi \@ifundefined
  {mn@eprint@\@tempb}{\@tempb:\@tempc}{\expandafter \expandafter \csname
  mn@eprint@\@tempb\endcsname \expandafter{\@tempc}}}

\bibitem[\protect\citeauthoryear{{Allende Prieto}, {Lambert}  \&
  {Asplund}}{{Allende Prieto}
  et~al.}{2001}]{allendeprieto_lambert_asplund_2001}
{Allende Prieto} C.,  {Lambert} D.~L.,   {Asplund} M.,  2001, \mn@doi [\apj]
  {10.1086/322874}, \href
  {https://ui.adsabs.harvard.edu/#abs/2001ApJ...556L..63A} {556, L63}

\bibitem[\protect\citeauthoryear{{Bahcall} \& {Peebles}}{{Bahcall} \&
  {Peebles}}{1969}]{bahcall_peebles_1969}
{Bahcall} J.~N.,  {Peebles} P.~J.~E.,  1969, \mn@doi [\apjl] {10.1086/180337},
  \href {http://adsabs.harvard.edu/abs/1969ApJ...156L...7B} {156, L7}

\bibitem[\protect\citeauthoryear{{Barret} et~al.,}{{Barret}
  et~al.}{2016}]{athena_ifu_2016}
{Barret} D.,  et~al., 2016, in Society of Photo-Optical Instrumentation
  Engineers (SPIE) Conference Series. p. 99052F (\mn@eprint {arXiv}
  {1608.08105}), \mn@doi{10.1117/12.2232432}

\bibitem[\protect\citeauthoryear{{Barret} et~al.,}{{Barret}
  et~al.}{2018}]{Athena_2018_07}
{Barret} D.,  et~al., 2018, in Society of Photo-Optical Instrumentation
  Engineers (SPIE) Conference Series. p. 106991G (\mn@eprint {arXiv}
  {1807.06092}), \mn@doi{10.1117/12.2312409}

\bibitem[\protect\citeauthoryear{{Bertone}, {Schaye}, {Dalla Vecchia}, {Booth},
  {Theuns}  \& {Wiersma}}{{Bertone} et~al.}{2010a}]{bertone_schaye_etal_2010}
{Bertone} S.,  {Schaye} J.,  {Dalla Vecchia} C.,  {Booth} C.~M.,  {Theuns} T.,
   {Wiersma} R.~P.~C.,  2010a, \mn@doi [\mnras]
  {10.1111/j.1365-2966.2010.16932.x}, \href
  {http://adsabs.harvard.edu/abs/2010MNRAS.407..544B} {407, 544}

\bibitem[\protect\citeauthoryear{{Bertone}, {Schaye}, {Booth}, {Dalla Vecchia},
  {Theuns}  \& {Wiersma}}{{Bertone}
  et~al.}{2010b}]{bertone_schaye_etal_2010_uv}
{Bertone} S.,  {Schaye} J.,  {Booth} C.~M.,  {Dalla Vecchia} C.,  {Theuns} T.,
   {Wiersma} R.~P.~C.,  2010b, \mn@doi [\mnras]
  {10.1111/j.1365-2966.2010.17188.x}, \href
  {http://adsabs.harvard.edu/abs/2010MNRAS.408.1120B} {408, 1120}

\bibitem[\protect\citeauthoryear{{Bonamente}, {Nevalainen}, {Tilton},
  {Liivam{\"a}gi}, {Tempel}, {Hein{\"a}m{\"a}ki}  \& {Fang}}{{Bonamente}
  et~al.}{2016}]{bonamente_nevalainen_etal_2016}
{Bonamente} M.,  {Nevalainen} J.,  {Tilton} E.,  {Liivam{\"a}gi} J.,  {Tempel}
  E.,  {Hein{\"a}m{\"a}ki} P.,   {Fang} T.,  2016, \mn@doi [\mnras]
  {10.1093/mnras/stw285}, \href
  {http://adsabs.harvard.edu/abs/2016MNRAS.457.4236B} {457, 4236}

\bibitem[\protect\citeauthoryear{{Booth} \& {Schaye}}{{Booth} \&
  {Schaye}}{2009}]{booth_schaye_2009}
{Booth} C.~M.,  {Schaye} J.,  2009, \mn@doi [\mnras]
  {10.1111/j.1365-2966.2009.15043.x}, \href
  {http://adsabs.harvard.edu/abs/2009MNRAS.398...53B} {398, 53}

\bibitem[\protect\citeauthoryear{{Borgani} et~al.,}{{Borgani}
  et~al.}{2004}]{borgani_murante_etal_2004}
{Borgani} S.,  et~al., 2004, \mn@doi [\mnras]
  {10.1111/j.1365-2966.2004.07431.x}, \href
  {https://ui.adsabs.harvard.edu/#abs/2004MNRAS.348.1078B} {348, 1078}

\bibitem[\protect\citeauthoryear{{Branchini} et~al.,}{{Branchini}
  et~al.}{2009}]{branchini_ursino_etal_2009}
{Branchini} E.,  et~al., 2009, \mn@doi [\apj] {10.1088/0004-637X/697/1/328},
  \href {http://adsabs.harvard.edu/abs/2009ApJ...697..328B} {697, 328}

\bibitem[\protect\citeauthoryear{{Bregman}}{{Bregman}}{2007}]{bregman_2007_review}
{Bregman} J.~N.,  2007, \mn@doi [Annual Review of Astronomy and Astrophysics]
  {10.1146/annurev.astro.45.051806.110619}, \href
  {https://ui.adsabs.harvard.edu/#abs/2007ARA&A..45..221B} {45, 221}

\bibitem[\protect\citeauthoryear{{Brenneman} et~al.,}{{Brenneman}
  et~al.}{2016}]{brenneman_smith_etal_2016}
{Brenneman} L.~W.,  et~al., 2016, in Space Telescopes and Instrumentation 2016:
  Ultraviolet to Gamma Ray. p. 99054P, \mn@doi{10.1117/12.2231193}

\bibitem[\protect\citeauthoryear{{Burchett} et~al.,}{{Burchett}
  et~al.}{2019}]{burchett_tripp_etal_2018}
{Burchett} J.~N.,  et~al., 2019, \mn@doi [\apj] {10.3847/2041-8213/ab1f7f},
  \href {https://ui.adsabs.harvard.edu/abs/2019ApJ...877L..20B} {877, L20}

\bibitem[\protect\citeauthoryear{{Cen}}{{Cen}}{2012}]{cen_2012}
{Cen} R.,  2012, \mn@doi [\apj] {10.1088/0004-637X/753/1/17}, \href
  {http://adsabs.harvard.edu/abs/2012ApJ...753...17C} {753, 17}

\bibitem[\protect\citeauthoryear{{Cen} \& {Fang}}{{Cen} \&
  {Fang}}{2006}]{cen_fang_2006}
{Cen} R.,  {Fang} T.,  2006, \mn@doi [\apj] {10.1086/506506}, \href
  {http://adsabs.harvard.edu/abs/2006ApJ...650..573C} {650, 573}

\bibitem[\protect\citeauthoryear{{Cen} \& {Ostriker}}{{Cen} \&
  {Ostriker}}{1999}]{cen_ostriker_1999}
{Cen} R.,  {Ostriker} J.~P.,  1999, \mn@doi [\apj] {10.1086/306949}, \href
  {http://adsabs.harvard.edu/abs/1999ApJ...514....1C} {514, 1}

\bibitem[\protect\citeauthoryear{{Cen} \& {Ostriker}}{{Cen} \&
  {Ostriker}}{2006}]{cen_ostriker_2006}
{Cen} R.,  {Ostriker} J.~P.,  2006, \mn@doi [\apj] {10.1086/506505}, \href
  {http://adsabs.harvard.edu/abs/2006ApJ...650..560C} {650, 560}

\bibitem[\protect\citeauthoryear{Collette}{Collette}{2013}]{h5py}
Collette A.,  2013, Python and HDF5.
O'Reilly, \url {http://www.h5py.org/}

\bibitem[\protect\citeauthoryear{{Crain} et~al.,}{{Crain}
  et~al.}{2015}]{eagle_calibration}
{Crain} R.~A.,  et~al., 2015, \mn@doi [\mnras] {10.1093/mnras/stv725}, \href
  {http://adsabs.harvard.edu/abs/2015MNRAS.450.1937C} {450, 1937}

\bibitem[\protect\citeauthoryear{Cyburt, Fields, Olive  \& Yeh}{Cyburt
  et~al.}{2016}]{cyburt_fields_etal_2016}
Cyburt R.~H.,  Fields B.~D.,  Olive K.~A.,   Yeh T.-H.,  2016, \mn@doi [Rev.
  Mod. Phys.] {10.1103/RevModPhys.88.015004}, 88, 015004

\bibitem[\protect\citeauthoryear{{Dalla Vecchia} \& {Schaye}}{{Dalla Vecchia}
  \& {Schaye}}{2012}]{dalla-vecchia_schaye_2012}
{Dalla Vecchia} C.,  {Schaye} J.,  2012, \mn@doi [\mnras]
  {10.1111/j.1365-2966.2012.21704.x}, \href
  {http://adsabs.harvard.edu/abs/2012MNRAS.426..140D} {426, 140}

\bibitem[\protect\citeauthoryear{{Davies}, {Crain}, {McCarthy}, {Oppenheimer},
  {Schaye}, {Schaller}  \& {McAlpine}}{{Davies}
  et~al.}{2019}]{davies_crain_etal_2019}
{Davies} J.~J.,  {Crain} R.~A.,  {McCarthy} I.~G.,  {Oppenheimer} B.~D.,
  {Schaye} J.,  {Schaller} M.,   {McAlpine} S.,  2019, \mn@doi [\mnras]
  {10.1093/mnras/stz635}, \href
  {http://adsabs.harvard.edu/abs/2019MNRAS.tmp..621D} {}

\bibitem[\protect\citeauthoryear{{Dubois} et~al.,}{{Dubois}
  et~al.}{2014}]{dubois_pichon_etal_2014}
{Dubois} Y.,  et~al., 2014, \mn@doi [\mnras] {10.1093/mnras/stu1227}, \href
  {http://adsabs.harvard.edu/abs/2014MNRAS.444.1453D} {444, 1453}

\bibitem[\protect\citeauthoryear{{Fang}, {Bryan}  \& {Canizares}}{{Fang}
  et~al.}{2002}]{fang_bryan_canizares_2002}
{Fang} T.,  {Bryan} G.~L.,   {Canizares} C.~R.,  2002, \mn@doi [\apj]
  {10.1086/324400}, \href {http://adsabs.harvard.edu/abs/2002ApJ...564..604F}
  {564, 604}

\bibitem[\protect\citeauthoryear{{Ferland}, {Korista}, {Verner}, {Ferguson},
  {Kingdon}  \& {Verner}}{{Ferland} et~al.}{1998}]{cloudy}
{Ferland} G.~J.,  {Korista} K.~T.,  {Verner} D.~A.,  {Ferguson} J.~W.,
  {Kingdon} J.~B.,   {Verner} E.~M.,  1998, \mn@doi [Publications of the
  Astronomical Society of the Pacific] {10.1086/316190}, \href
  {https://ui.adsabs.harvard.edu/#abs/1998PASP..110..761F} {110, 761}

\bibitem[\protect\citeauthoryear{{Furlanetto}, {Phillips}  \&
  {Kamionkowski}}{{Furlanetto}
  et~al.}{2005}]{furlanetto_phillips_kamionkowski_2005}
{Furlanetto} S.~R.,  {Phillips} L.~A.,   {Kamionkowski} M.,  2005, \mn@doi
  [\mnras] {10.1111/j.1365-2966.2005.08885.x}, \href
  {http://adsabs.harvard.edu/abs/2005MNRAS.359..295F} {359, 295}

\bibitem[\protect\citeauthoryear{{Haardt} \& {Madau}}{{Haardt} \&
  {Madau}}{2001}]{HM01}
{Haardt} F.,  {Madau} P.,  2001, in {Neumann} D.~M.,  {Tran} J.~T.~V.,  eds,
  Clusters of Galaxies and the High Redshift Universe Observed in X-rays.
  (\mn@eprint {} {astro-ph/0106018})

\bibitem[\protect\citeauthoryear{Hunter}{Hunter}{2007}]{matplotlib}
Hunter J.~D.,  2007, \mn@doi [Computing in Science \& Engineering]
  {10.1109/MCSE.2007.55}, 9, 90

\bibitem[\protect\citeauthoryear{{Johnson}, {Chen}  \& {Mulchaey}}{{Johnson}
  et~al.}{2013}]{johnson_chen_mulchaey_2013}
{Johnson} S.~D.,  {Chen} H.-W.,   {Mulchaey} J.~S.,  2013, \mn@doi [\mnras]
  {10.1093/mnras/stt1137}, \href
  {http://adsabs.harvard.edu/abs/2013MNRAS.434.1765J} {434, 1765}

\bibitem[\protect\citeauthoryear{Jones, Oliphant, Peterson  et~al.}{Jones
  et~al.}{2001}]{scipy}
Jones E.,  Oliphant T.,  Peterson P.,   et~al., 2001, {SciPy}: Open source
  scientific tools for {Python}, \url {http://www.scipy.org/}

\bibitem[\protect\citeauthoryear{{Kaastra}}{{Kaastra}}{2018}]{kaastra_2018_pc}
{Kaastra} J.,  2018, private communication

\bibitem[\protect\citeauthoryear{{Kaastra}, {Werner}, {Herder}, {Paerels}, {de
  Plaa}, {Rasmussen}  \& {de Vries}}{{Kaastra}
  et~al.}{2006}]{kaastra_werner_etal_2006}
{Kaastra} J.~S.,  {Werner} N.,  {Herder} J.~W.~A.~d.,  {Paerels} F.~B.~S.,  {de
  Plaa} J.,  {Rasmussen} A.~P.,   {de Vries} C.~P.,  2006, \mn@doi [\apj]
  {10.1086/507835}, \href {http://adsabs.harvard.edu/abs/2006ApJ...652..189K}
  {652, 189}

\bibitem[\protect\citeauthoryear{{Kov{\'a}cs}, {Bogd{\'a}n}, {Smith}, {Kraft}
  \& {Forman}}{{Kov{\'a}cs} et~al.}{2019}]{kovacs_orsolya_etal_2019}
{Kov{\'a}cs} O.~E.,  {Bogd{\'a}n} {\'A}.,  {Smith} R. a.~K.,  {Kraft} R.~P.,
  {Forman} W.~R.,  2019, \mn@doi [\apj] {10.3847/1538-4357/aaef78}, \href
  {https://ui.adsabs.harvard.edu/\#abs/2019ApJ...872...83K} {872, 83}

\bibitem[\protect\citeauthoryear{{Lide}}{{Lide}}{2003}]{crc_handbook}
{Lide} D.~R.,  ed. 2003, CRC Handbook of Chemistry and Physics, 84 edn.
CRC Press LLC, {Boca Raton}

\bibitem[\protect\citeauthoryear{{Lumb}, {den Herder}  \& the Athena
  Science~Team}{{Lumb} et~al.}{2017}]{Athena_2017_11}
{Lumb} D.,  {den Herder} J.-W.,   the Athena Science~Team 2017,
  Issue/Revision~2.01, Athena Science Requirements Document.
European Space Agency, European Space Research and Technology Centre,
  Keplerlaan 1, 2201~AZ Noordwijk, The Netherlands

\bibitem[\protect\citeauthoryear{{McAlpine} et~al.,}{{McAlpine}
  et~al.}{2016}]{mcalpine_helly_etal_2016}
{McAlpine} S.,  et~al., 2016, \mn@doi [Astronomy and Computing]
  {10.1016/j.ascom.2016.02.004}, \href
  {http://adsabs.harvard.edu/abs/2016A%26C....15...72M} {15, 72}

\bibitem[\protect\citeauthoryear{{McCarthy}, {Schaye}, {Bird}  \& {Le
  Brun}}{{McCarthy} et~al.}{2017}]{mccarthy_schaye_etal_2017}
{McCarthy} I.~G.,  {Schaye} J.,  {Bird} S.,   {Le Brun} A.~M.~C.,  2017,
  \mn@doi [\mnras] {10.1093/mnras/stw2792}, \href
  {http://adsabs.harvard.edu/abs/2017MNRAS.465.2936M} {465, 2936}

\bibitem[\protect\citeauthoryear{{Meiring}, {Tripp}, {Werk}, {Howk}, {Jenkins},
  {Prochaska}, {Lehner}  \& {Sembach}}{{Meiring}
  et~al.}{2013}]{meiring_tripp_etal_2013}
{Meiring} J.~D.,  {Tripp} T.~M.,  {Werk} J.~K.,  {Howk} J.~C.,  {Jenkins}
  E.~B.,  {Prochaska} J.~X.,  {Lehner} N.,   {Sembach} K.~R.,  2013, \mn@doi
  [\apj] {10.1088/0004-637X/767/1/49}, \href
  {http://adsabs.harvard.edu/abs/2013ApJ...767...49M} {767, 49}

\bibitem[\protect\citeauthoryear{{Monaghan} \& {Lattanzio}}{{Monaghan} \&
  {Lattanzio}}{1985}]{monaghan_lattanzio_1985}
{Monaghan} J.~J.,  {Lattanzio} J.~C.,  1985, {A refined particle method for
  astrophysical problems}

\bibitem[\protect\citeauthoryear{{Mroczkowski} et~al.,}{{Mroczkowski}
  et~al.}{2019}]{mroczkowski_nagai_etal_2018}
{Mroczkowski} T.,  et~al., 2019, \mn@doi [\ssr] {10.1007/s11214-019-0581-2},
  \href {https://ui.adsabs.harvard.edu/\#abs/2019SSRv..215...17M} {215, 17}

\bibitem[\protect\citeauthoryear{{Nelson} et~al.,}{{Nelson}
  et~al.}{2018}]{Nelson_etal_2018_hiO}
{Nelson} D.,  et~al., 2018, \mn@doi [\mnras] {10.1093/mnras/sty656}, \href
  {http://adsabs.harvard.edu/abs/2018MNRAS.477..450N} {477, 450}

\bibitem[\protect\citeauthoryear{{Nicastro}}{{Nicastro}}{2018}]{nicastro_2018}
{Nicastro} F.,  2018, preprint, \href
  {http://adsabs.harvard.edu/abs/2018arXiv181103498N} {} (\mn@eprint {arXiv}
  {1811.03498})

\bibitem[\protect\citeauthoryear{{Nicastro} et~al.,}{{Nicastro}
  et~al.}{2005}]{nicastro_mathur_etal_2005}
{Nicastro} F.,  et~al., 2005, \mn@doi [\apj] {10.1086/431270}, \href
  {http://adsabs.harvard.edu/abs/2005ApJ...629..700N} {629, 700}

\bibitem[\protect\citeauthoryear{{Nicastro}, {Krongold}, {Mathur}  \&
  {Elvis}}{{Nicastro} et~al.}{2017}]{nicastro_krongold_etal_2017}
{Nicastro} F.,  {Krongold} Y.,  {Mathur} S.,   {Elvis} M.,  2017, \mn@doi
  [Astronomische Nachrichten] {10.1002/asna.201713343}, \href
  {http://adsabs.harvard.edu/abs/2017AN....338..281N} {338, 281}

\bibitem[\protect\citeauthoryear{{Nicastro} et~al.,}{{Nicastro}
  et~al.}{2018}]{nicastro_etal_2018}
{Nicastro} F.,  et~al., 2018, \mn@doi [\nat] {10.1038/s41586-018-0204-1}, 558,
  406

\bibitem[\protect\citeauthoryear{{Oliphant}}{{Oliphant}}{2006}]{numpy}
{Oliphant} T.~E.,  2006, Guide to NumPy.
Trelgol Publishing, USA

\bibitem[\protect\citeauthoryear{{Oppenheimer} \& {Schaye}}{{Oppenheimer} \&
  {Schaye}}{2013}]{oppenheimer_schaye_2013}
{Oppenheimer} B.~D.,  {Schaye} J.,  2013, \mn@doi [\mnras]
  {10.1093/mnras/stt1150}, \href
  {http://adsabs.harvard.edu/abs/2013MNRAS.434.1063O} {434, 1063}

\bibitem[\protect\citeauthoryear{{Oppenheimer} et~al.,}{{Oppenheimer}
  et~al.}{2016}]{oppenheimer_etal_2016}
{Oppenheimer} B.~D.,  et~al., 2016, \mn@doi [\mnras] {10.1093/mnras/stw1066},
  \href {http://adsabs.harvard.edu/abs/2016MNRAS.460.2157O} {460, 2157}

\bibitem[\protect\citeauthoryear{{Oppenheimer}, {Segers}, {Schaye}, {Richings}
  \& {Crain}}{{Oppenheimer} et~al.}{2018}]{oppenheimer_2018_fossilAGN_cos}
{Oppenheimer} B.~D.,  {Segers} M.,  {Schaye} J.,  {Richings} A.~J.,   {Crain}
  R.~A.,  2018, \mn@doi [\mnras] {10.1093/mnras/stx2967}, \href
  {https://ui.adsabs.harvard.edu/#abs/2018MNRAS.474.4740O} {474, 4740}

\bibitem[\protect\citeauthoryear{{Oppenheimer} et~al.,}{{Oppenheimer}
  et~al.}{2019}]{oppenheimer_davies_etal_2019}
{Oppenheimer} B.~D.,  et~al., 2019, arXiv e-prints, \href
  {https://ui.adsabs.harvard.edu/abs/2019arXiv190405904O} {p. arXiv:1904.05904}

\bibitem[\protect\citeauthoryear{P\'erez \& Granger}{P\'erez \&
  Granger}{2007}]{ipython}
P\'erez F.,  Granger B.~E.,  2007, \mn@doi [Computing in Science \&
  Engineering] {10.1109/MCSE.2007.53}, 9, 21

\bibitem[\protect\citeauthoryear{{Perna} \& {Loeb}}{{Perna} \&
  {Loeb}}{1998}]{perna_loeb_1998}
{Perna} R.,  {Loeb} A.,  1998, \mn@doi [\apjl] {10.1086/311544}, \href
  {http://adsabs.harvard.edu/abs/1998ApJ...503L.135P} {503, L135}

\bibitem[\protect\citeauthoryear{{Pierre}, {Bryan}  \& {Gastaud}}{{Pierre}
  et~al.}{2000}]{pierre_bryan_gastaud_2000}
{Pierre} M.,  {Bryan} G.,   {Gastaud} R.,  2000, \aap, \href
  {http://adsabs.harvard.edu/abs/2000A%26A...356..403P} {356, 403}

\bibitem[\protect\citeauthoryear{{Pillepich} et~al.,}{{Pillepich}
  et~al.}{2018}]{pillepich_springel_etal_2018}
{Pillepich} A.,  et~al., 2018, \mn@doi [\mnras] {10.1093/mnras/stx2656}, \href
  {http://adsabs.harvard.edu/abs/2018MNRAS.473.4077P} {473, 4077}

\bibitem[\protect\citeauthoryear{{Planck Collaboration} et~al.,}{{Planck
  Collaboration} et~al.}{2014}]{planck_2013}
{Planck Collaboration} et~al., 2014, \mn@doi [\aap]
  {10.1051/0004-6361/201321529}, \href
  {http://adsabs.harvard.edu/abs/2014A%26A...571A...1P} {571, A1}

\bibitem[\protect\citeauthoryear{{Rahmati}, {Pawlik}, {Rai{\v c}evi{\'c}}  \&
  {Schaye}}{{Rahmati} et~al.}{2013}]{rahamti_pawlik_etal_2013}
{Rahmati} A.,  {Pawlik} A.~H.,  {Rai{\v c}evi{\'c}} M.,   {Schaye} J.,  2013,
  \mn@doi [\mnras] {10.1093/mnras/stt066}, \href
  {http://adsabs.harvard.edu/abs/2013MNRAS.430.2427R} {430, 2427}

\bibitem[\protect\citeauthoryear{{Rahmati}, {Schaye}, {Crain}, {Oppenheimer},
  {Schaller}  \& {Theuns}}{{Rahmati} et~al.}{2016}]{rahmati_etal_2016}
{Rahmati} A.,  {Schaye} J.,  {Crain} R.~A.,  {Oppenheimer} B.~D.,  {Schaller}
  M.,   {Theuns} T.,  2016, \mn@doi [\mnras] {10.1093/mnras/stw453}, \href
  {http://adsabs.harvard.edu/abs/2016MNRAS.459..310R} {459, 310}

\bibitem[\protect\citeauthoryear{Rohatgi}{Rohatgi}{2018}]{webplotdigitizer}
Rohatgi A.,  2018, WebPlotDigitizer, \url
  {https://automeris.io/WebPlotDigitizer}

\bibitem[\protect\citeauthoryear{{Schaller}, {Dalla Vecchia}, {Schaye},
  {Bower}, {Theuns}, {Crain}, {Furlong}  \& {McCarthy}}{{Schaller}
  et~al.}{2015}]{anarchy_effect}
{Schaller} M.,  {Dalla Vecchia} C.,  {Schaye} J.,  {Bower} R.~G.,  {Theuns} T.,
   {Crain} R.~A.,  {Furlong} M.,   {McCarthy} I.~G.,  2015, \mn@doi [\mnras]
  {10.1093/mnras/stv2169}, \href
  {https://ui.adsabs.harvard.edu/#abs/2015MNRAS.454.2277S} {454, 2277}

\bibitem[\protect\citeauthoryear{{Schaye}}{{Schaye}}{2004}]{schaye_2004}
{Schaye} J.,  2004, \mn@doi [\apj] {10.1086/421232}, \href
  {http://adsabs.harvard.edu/abs/2004ApJ...609..667S} {609, 667}

\bibitem[\protect\citeauthoryear{{Schaye} \& {Dalla Vecchia}}{{Schaye} \&
  {Dalla Vecchia}}{2008}]{schaye_dalla-vecchia_2008}
{Schaye} J.,  {Dalla Vecchia} C.,  2008, \mn@doi [\mnras]
  {10.1111/j.1365-2966.2007.12639.x}, \href
  {http://adsabs.harvard.edu/abs/2008MNRAS.383.1210S} {383, 1210}

\bibitem[\protect\citeauthoryear{{Schaye} et~al.,}{{Schaye}
  et~al.}{2010}]{schaye_dalla-vecchia_etal_2010_owls}
{Schaye} J.,  et~al., 2010, \mn@doi [\mnras]
  {10.1111/j.1365-2966.2009.16029.x}, \href
  {http://adsabs.harvard.edu/abs/2010MNRAS.402.1536S} {402, 1536}

\bibitem[\protect\citeauthoryear{{Schaye} et~al.,}{{Schaye}
  et~al.}{2015}]{eagle_paper}
{Schaye} J.,  et~al., 2015, \mn@doi [\mnras] {10.1093/mnras/stu2058}, \href
  {http://adsabs.harvard.edu/abs/2015MNRAS.446..521S} {446, 521}

\bibitem[\protect\citeauthoryear{{Segers}, {Oppenheimer}, {Schaye}  \&
  {Richings}}{{Segers} et~al.}{2017}]{segers_oppenheimer_etal_2017}
{Segers} M.~C.,  {Oppenheimer} B.~D.,  {Schaye} J.,   {Richings} A.~J.,  2017,
  \mn@doi [\mnras] {10.1093/mnras/stx1633}, \href
  {https://ui.adsabs.harvard.edu/#abs/2017MNRAS.471.1026S} {471, 1026}

\bibitem[\protect\citeauthoryear{{Shull}, {Smith}  \& {Danforth}}{{Shull}
  et~al.}{2012}]{shull_smith_danforth_2012}
{Shull} J.~M.,  {Smith} B.~D.,   {Danforth} C.~W.,  2012, \mn@doi [\apj]
  {10.1088/0004-637X/759/1/23}, \href
  {http://adsabs.harvard.edu/abs/2012ApJ...759...23S} {759, 23}

\bibitem[\protect\citeauthoryear{{Smith} et~al.,}{{Smith}
  et~al.}{2016}]{smith_abraham_etal_2016_arcus}
{Smith} R.~K.,  et~al., 2016, in Space Telescopes and Instrumentation 2016:
  Ultraviolet to Gamma Ray. p. 99054M, \mn@doi{10.1117/12.2231778}

\bibitem[\protect\citeauthoryear{{Springel}}{{Springel}}{2005}]{springel_2005}
{Springel} V.,  2005, \mn@doi [\mnras] {10.1111/j.1365-2966.2005.09655.x},
  \href {http://adsabs.harvard.edu/abs/2005MNRAS.364.1105S} {364, 1105}

\bibitem[\protect\citeauthoryear{{Tanimura} et~al.,}{{Tanimura}
  et~al.}{2019}]{tanimura_hinshaw_etal_2019}
{Tanimura} H.,  et~al., 2019, \mn@doi [\mnras] {10.1093/mnras/sty3118}, \href
  {http://adsabs.harvard.edu/abs/2019MNRAS.483..223T} {483, 223}

\bibitem[\protect\citeauthoryear{{Tepper-Garc{\'{\i}}a}, {Richter}, {Schaye},
  {Booth}, {Dalla Vecchia}, {Theuns}  \& {Wiersma}}{{Tepper-Garc{\'{\i}}a}
  et~al.}{2011}]{tepper-garcia_richter_etal_2011}
{Tepper-Garc{\'{\i}}a} T.,  {Richter} P.,  {Schaye} J.,  {Booth} C.~M.,  {Dalla
  Vecchia} C.,  {Theuns} T.,   {Wiersma} R.~P.~C.,  2011, \mn@doi [\mnras]
  {10.1111/j.1365-2966.2010.18123.x}, \href
  {http://adsabs.harvard.edu/abs/2011MNRAS.413..190T} {413, 190}

\bibitem[\protect\citeauthoryear{{Tepper-Garc{\'{\i}}a}, {Richter}, {Schaye},
  {Booth}, {Dalla Vecchia}  \& {Theuns}}{{Tepper-Garc{\'{\i}}a}
  et~al.}{2012}]{tepper-garcia_richter_etal_2012}
{Tepper-Garc{\'{\i}}a} T.,  {Richter} P.,  {Schaye} J.,  {Booth} C.~M.,  {Dalla
  Vecchia} C.,   {Theuns} T.,  2012, \mn@doi [\mnras]
  {10.1111/j.1365-2966.2012.21545.x}, \href
  {http://adsabs.harvard.edu/abs/2012MNRAS.425.1640T} {425, 1640}

\bibitem[\protect\citeauthoryear{{Tepper-Garc{\'{\i}}a}, {Richter}  \&
  {Schaye}}{{Tepper-Garc{\'{\i}}a}
  et~al.}{2013}]{tepper-garcia_richter_etal_2013}
{Tepper-Garc{\'{\i}}a} T.,  {Richter} P.,   {Schaye} J.,  2013, \mn@doi
  [\mnras] {10.1093/mnras/stt1712}, \href
  {http://adsabs.harvard.edu/abs/2013MNRAS.436.2063T} {436, 2063}

\bibitem[\protect\citeauthoryear{{The~Lynx~Team}}{{The~Lynx~Team}}{2018}]{lynx_2018_08}
{The~Lynx~Team} 2018, arXiv e-prints, \href
  {http://adsabs.harvard.edu/abs/2018arXiv180909642T} {}

\bibitem[\protect\citeauthoryear{{Tumlinson} et~al.,}{{Tumlinson}
  et~al.}{2011}]{tumlinson_etal_2011}
{Tumlinson} J.,  et~al., 2011, \mn@doi [Science] {10.1126/science.1209840},
  \href {http://adsabs.harvard.edu/abs/2011Sci...334..948T} {334, 948}

\bibitem[\protect\citeauthoryear{{Tumlinson}, {Peeples}  \& {Werk}}{{Tumlinson}
  et~al.}{2017}]{tumlinson_peeples_werk_2017_cgmreview}
{Tumlinson} J.,  {Peeples} M.~S.,   {Werk} J.~K.,  2017, \mn@doi [\araa]
  {10.1146/annurev-astro-091916-055240}, \href
  {http://adsabs.harvard.edu/abs/2017ARA%26A..55..389T} {55, 389}

\bibitem[\protect\citeauthoryear{{Verner}, {Verner}  \& {Ferland}}{{Verner}
  et~al.}{1996}]{verner_verner_ferland_1996}
{Verner} D.~A.,  {Verner} E.~M.,   {Ferland} G.~J.,  1996, \mn@doi [Atomic Data
  and Nuclear Data Tables] {10.1006/adnd.1996.0018}, \href
  {http://adsabs.harvard.edu/abs/1996ADNDT..64....1V} {64, 1}

\bibitem[\protect\citeauthoryear{Wendland}{Wendland}{1995}]{wendland_1995}
Wendland H.,  1995, \mn@doi [Advances in Computational Mathematics]
  {10.1007/BF02123482}, 4, 389

\bibitem[\protect\citeauthoryear{{Wiersma}, {Schaye}  \& {Smith}}{{Wiersma}
  et~al.}{2009a}]{wiersma_schaye_smith_2009}
{Wiersma} R.~P.~C.,  {Schaye} J.,   {Smith} B.~D.,  2009a, \mn@doi [\mnras]
  {10.1111/j.1365-2966.2008.14191.x}, \href
  {http://adsabs.harvard.edu/abs/2009MNRAS.393...99W} {393, 99}

\bibitem[\protect\citeauthoryear{{Wiersma}, {Schaye}, {Theuns}, {Dalla Vecchia}
   \& {Tornatore}}{{Wiersma} et~al.}{2009b}]{wiersema_etal_2009_insim}
{Wiersma} R.~P.~C.,  {Schaye} J.,  {Theuns} T.,  {Dalla Vecchia} C.,
  {Tornatore} L.,  2009b, \mn@doi [\mnras] {10.1111/j.1365-2966.2009.15331.x},
  \href {http://adsabs.harvard.edu/abs/2009MNRAS.399..574W} {399, 574}

\bibitem[\protect\citeauthoryear{{Yoshikawa} \& {Sasaki}}{{Yoshikawa} \&
  {Sasaki}}{2006}]{yoshikawa_sasaki_2006}
{Yoshikawa} K.,  {Sasaki} S.,  2006, \mn@doi [Publications of the Astronomical
  Society of Japan] {10.1093/pasj/58.4.641}, \href
  {https://ui.adsabs.harvard.edu/#abs/2006PASJ...58..641Y} {58, 641}

\bibitem[\protect\citeauthoryear{{Yoshikawa}, {Yamasaki}, {Suto}, {Ohashi},
  {Mitsuda}, {Tawara}  \& {Furuzawa}}{{Yoshikawa}
  et~al.}{2003}]{yoshikawa_yamasaki_etal_2003}
{Yoshikawa} K.,  {Yamasaki} N.~Y.,  {Suto} Y.,  {Ohashi} T.,  {Mitsuda} K.,
  {Tawara} Y.,   {Furuzawa} A.,  2003, \mn@doi [\pasj] {10.1093/pasj/55.5.879},
  \href {http://adsabs.harvard.edu/abs/2003PASJ...55..879Y} {55, 879}

\bibitem[\protect\citeauthoryear{{de Graaff}, {Cai}, {Heymans}  \&
  {Peacock}}{{de Graaff} et~al.}{2019}]{de-graaff_cai_etal_2017}
{de Graaff} A.,  {Cai} Y.-C.,  {Heymans} C.,   {Peacock} J.~A.,  2019, \mn@doi
  [\aap] {10.1051/0004-6361/201935159}, \href
  {https://ui.adsabs.harvard.edu/abs/2019A&A...624A..48D} {624, A48}

\makeatother
\end{thebibliography}

%\begin{thebibliography}{99}
%\end{thebibliography}

%%%%%%%%%%%%%%%%%%%%%%%%%%%%%%%%%%%%%%%%%%%%%%%%%%

%%%%%%%%%%%%%%%%% APPENDICES %%%%%%%%%%%%%%%%%%%%%

\appendix

\section{CDDF convergence}
\label{app:conv}

In this appendix, we test the convergence of the CDDFs with pixel size, slice thickness, box size, and simulation resolution, and give more details on how we tested the effect of the projection kernel shape.
The default projection we will compare other CDDFs to is that obtained from the reference simulation (\code{Ref-L100N1504}) at redshift~$0.1$, with 16 slices of $6.25 \us \txn{cMpc}$ each along the $z$-axis, and $32000^2$ pixels in the $x$- and $y$-directions.
%, using element abundances assigned to each SPH particle individually (and not SPH-smoothed abundances) to calculate the number of ions from the ion fraction and SPH particle mass. 
The assumed gas distribution for one SPH particle in this `standard' CDDF is given by the C2-kernel \citep{wendland_1995} used in the simulation. The temperature of star-forming gas is set to $10^4 \us \txn{K}$ for the ion fraction calculation. Any properties of CDDFs that are not otherwise mentioned here, are as described above.

\begin{figure}
\includegraphics[width=\columnwidth]{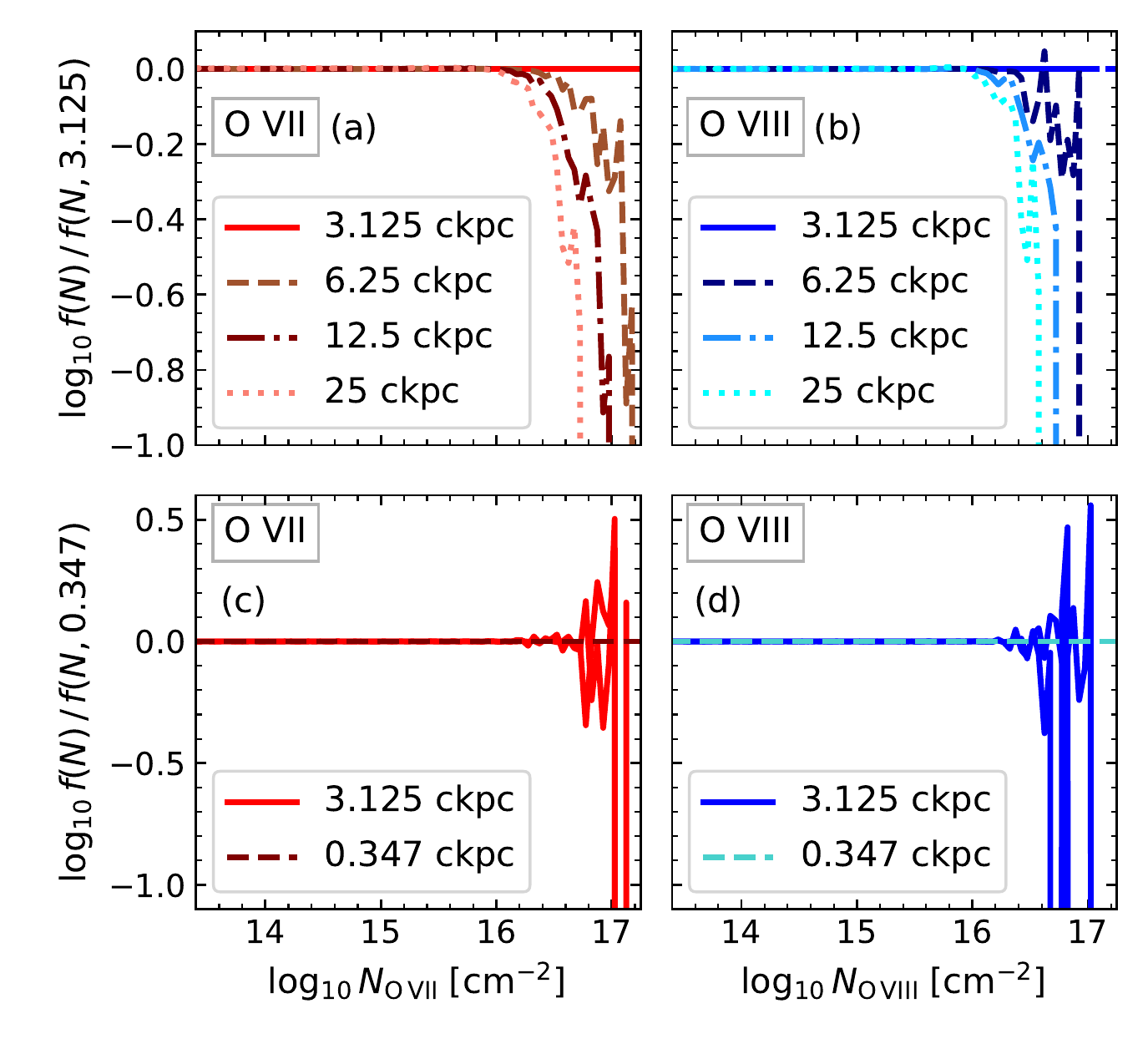}
\caption{Convergence of the column density distribution function (CDDF, $f(N)$) with pixel size. The left panels show \ion{O}{vii}, the right panels show \ion{O}{viii}. In all panels, the solid line shows the CDDF as used throughout the paper: the column density is calculated on a grid of $32000^2$ pixels, for each of~16 slices through the $100 \us \txn{cMpc}$ the box along the projection axis. These CDDFs are from the reference (\code{Ref-L100N1504}) simulation at $z=0.1$. In the top panels (a, b), this CDDF is compared to that obtained by degrading the resolution of the $3.125^2 \us \txn{ckpc}^2$ pixel grid to get grids of $6.25^2$ (dashed), $12.5^2$ (dot-dashed) and $25^2 \us \txn{ckpc}^2$ (dotted) pixels. The legends indicate the pixel sizes. In the bottom panels (c, d), the column density distributions extracted from two smaller regions ($10 \times 10 \us \txn{cMpc}^2$) at this standard resolution ($3.125^2 \us \txn{ckpc}^2$ per pixel) are compared with the CDDFs from the same region at nine time higher resolution. The CDDFs at different resolution are shown relative to the highest-resolution CDDF for the same region. }
\label{fig:pixconv}
\end{figure}

In Fig.~\ref{fig:pixconv}, we look at the effect of changing the pixel size. 
%These calculations are all done using the same snapshot of the same simulation (\code{Ref-L100N1504}, redshift~$0.1$), with columns along the same axis ($z$). Therefore, we are looking at the same absorption systems from the same angle in all cases, and the length or area of the columns is what is causing all the differences between the different CDDFs.  
Using lower resolution means underestimating the fraction of columns with large column densities, while assessments of lower column densities remain accurate. 
%We also note that the drop-off is relatively fast, going from well-converged to no systems at all in less that $1 \us \txn{dex}$ in column density. This drop-off seems to be fairly consistent between resolutions.
Our CDDFs appear to be converged at our standard pixel size for column densities $\lesssim 10^{16.5} \us \txn{cm}^{-2}$. At higher resolutions than our standard, we see problems arise at a similar column density. 
%Here, however, the effect looks more like noise. Comparing the degraded-resolution CDDF to our standard resolution, we see that the degraded-resolution CDDF is roughly indicative of the behaviour of the directly projected CDDF. (The fact that the match is not exact is due to how we handle SPH particles smaller than the pixel size in the projection.) We note that the agreement between column densities obtained with {\specwizard} and the projected column densities increases with decreasing pixel size up to our highest  ($2880 \us \txn{pixels}\, \txn{cMpc}^{-1}$) resolution. Here, the pixels are about the size of the minimum softening length of the gas particles, indicating that much higher resolutions should not be necessary.

\begin{figure}
\includegraphics[width=\columnwidth]{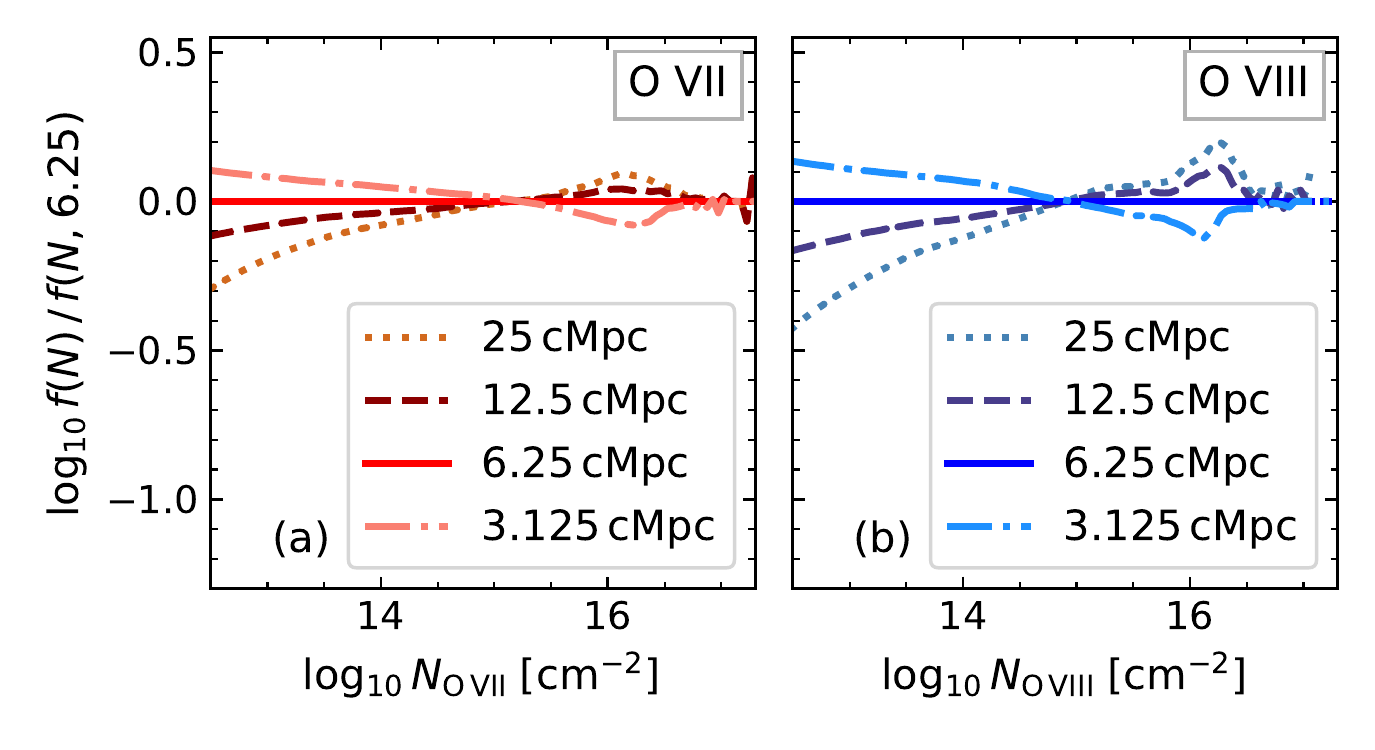}
\caption{Convergence of the column density distribution function (CDDF, $f(N)$) with slice thickness. The left panel shows \ion{O}{vii}, the right panel shows \ion{O}{viii}. In both panels, the solid line shows the CDDF as used throughout the paper: the column density is calculated on a grid of $32000^2$ pixels, for each of 16 slices the the $100 \us \txn{cMpc}$ box along the projection axis. For the other lines, the thickness of the box slices for which the column density is calculated is varied as indicated. These CDDFs are from the reference (\code{Ref-L100N1504}) simulation at $z=0.1$. The differences between our default $6.25 \us \txn{cMpc}$ slice CDDF and CDDFs using thicker or thinner slices are mostly $\lesssim 0.2 \us \txn{dex}$, and always $\lesssim 0.2 \us \txn{dex}$ in this column density range when comparing to $3.125$ and $12.5 \us \txn{cMpc}$ slices. These differences are small compared to the more than 8~orders of magnitude the CDDFs span in this column density range. }
\label{fig:sliceconv}
\end{figure}

In Fig.~\ref{fig:sliceconv}, we investigate the effect of using slices of different thickness to calculate the column densities.
The clearest effects are at low column densities. Thinner slices result in larger values of $f(N)$, which suggests that at low column densities, absorption often comes from multiple systems along the same line of sight. An optimal slice thickness is difficult to choose; we want to use a size which is small enough that the absorption in one column is (mostly) coming from a single system, but large enough that single systems are (mostly) not split over two columns along the line of sight. 
%At the higher column densities, there is also not a clear cut-off point; two absorption systems along these lines of sight are common enough, but these should typically already fall into different slices when using 4~or 8~slices. Mock spectra also show that 32~slices may risk cutting larger absorption systems in half, though this is in velocity space, so the comparison is not entirely straightforward due to redshift space distortions and temperature broadening.
We choose to use slices of $6.25 \us \txn{cMpc}$, which matches UV-ion CDDFs well below the CDDF break \citep{rahmati_etal_2016}, but bear in mind that this choice of, essentially, how to define absorption systems affects our results somewhat. Note, however, that the differences between CDDFs of different slice thickness are small compared to the range spanned by the CDDF. This standard slice thickness corresponds to $404 \us \txn{km}\, \txn{s}^{-1}$ in velocity space ($z=0.1$, rest-frame) when only accounting for the Hubble flow.

\begin{figure}
\includegraphics[width=\columnwidth]{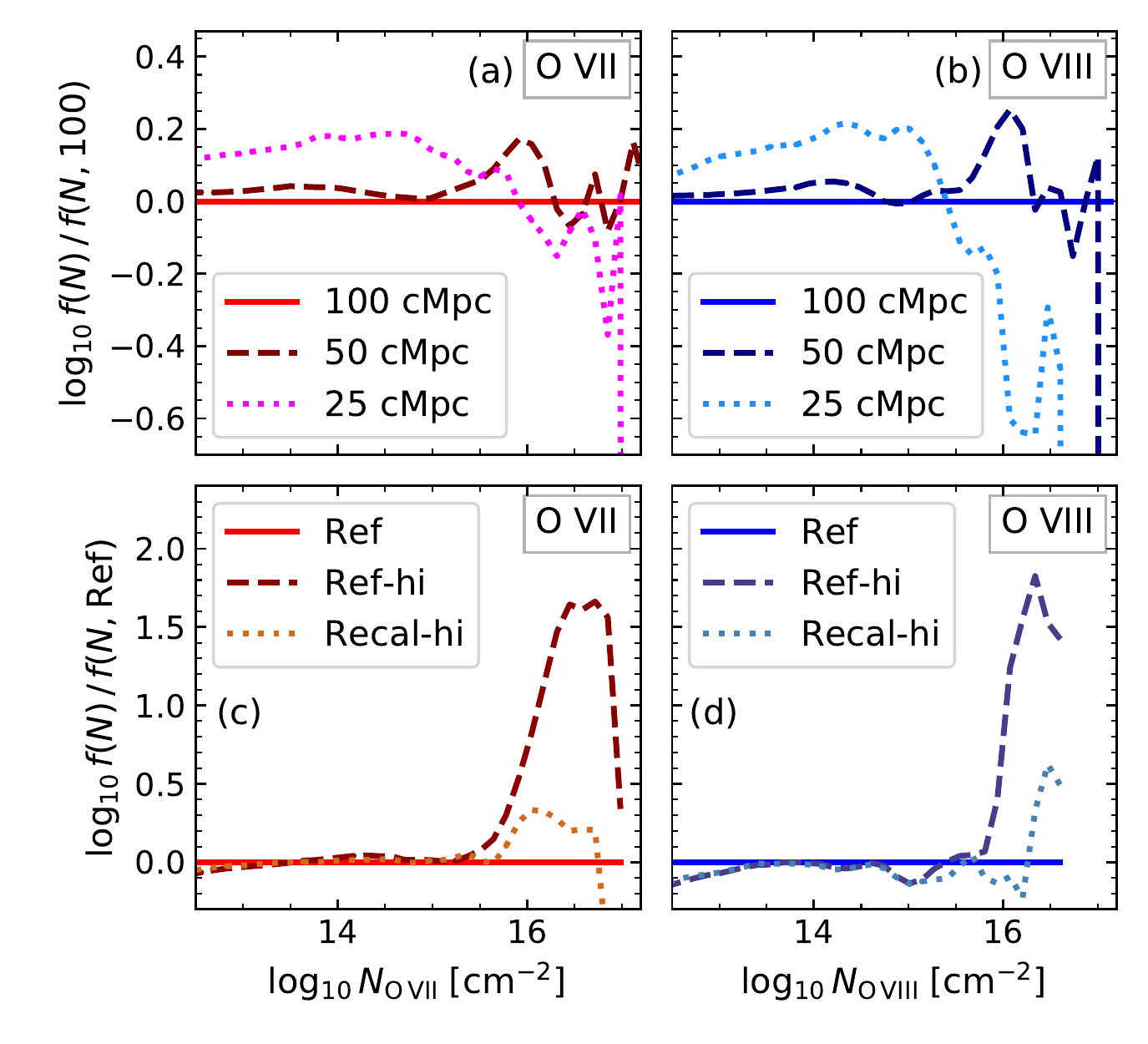}
\caption{Convergence of the column density distribution function with simulation box size (top panels) and resolution (bottom panels). The left panels show \ion{O}{vii}, the right panels show \ion{O}{viii}. The plots show the CDDF $f(N, z=0.1)$, as described in equation~\ref{eq:cddf}, relative to a reference CDDF. In the top panels, the CDDFs from \code{Ref-L025N0376} and  \code{Ref-L050N0752} are compared to the standard \code{Ref-L100N1504} CDDF. In the bottom panels,  \code{Ref-L025N0376} (`Ref') is compared to the \code{Ref-L025N0752} and \code{Recal-L025N0752} (`Ref-hi', `Recal-hi') CDDFs.}
\label{fig:boxresconv}
\end{figure}

Next, we investigate whether the CDDFs are converged in the simulations themselves. Specifically, we investigate the effect of simulation volume and resolution. In Fig.~\ref{fig:boxresconv}, we compare different simulations to each other. Here, differences may also come from differences in initial conditions (cosmic variance). In the top panels, we examine the effect of box size. In smaller boxes, the distributions are noisier because they are less well sampled, but there is also a physical difference: smaller boxes do not sample some of the large-scale modes of the initial density variations, so smaller boxes will contain fewer high-mass, large, hot systems. The $100 \us \txn{cMpc}$ box has no haloes with masses $M_{200c} > 10^{15} \us \txn{M}_{\sun}$, and only seven with $M_{200c} > 10^{14} \us \txn{M}_{\sun}$, so we do not expect complete convergence with box size. However, the figure indicates that, at column densities $\lesssim 10^{16.5}\us \txn{cm}^{-2}$, convergence with box size is sufficient. The $\lesssim 0.2 \us \txn{dex}$ difference between the CDDFs for the largest two boxes is much smaller than the range spanned by the CDDFs in the same column density range, and larger differences occur at column densities where the CDDF is not converged with projection resolution. 

In the bottom panels, we investigate the effect of simulation resolution (distinct from the projection resolution we investigated in Fig.~\ref{fig:pixconv}). We compare three $25 \us \txn{cMpc}$ boxes, since we do not have higher-resolution simulations of larger boxes. The comparison to \code{Ref-L025N0752} tests strong convergence and \code{Recal-L025N0752} tests weak convergence in the terminology of \citet{eagle_paper}. The \code{Ref-L025N0752} simulation was run using the exact same parameters as \code{Ref-L025N0376}, except for the resolution, and the addition of initial density perturbations on smaller scales in the higher-resolution box. However, the stellar and AGN feedback in the {\eagle} simulation is calibrated to reproduce the redshift $0.1$ galaxy stellar mass function, the relation between black hole mass and galaxy mass, and reasonable galaxy sizes. This calibration depends on the simulation resolution, so \cite{eagle_paper} also recalibrated the subgrid model for feedback to produce a similarly calibrated simulation at higher resolution: \code{Recal-L025N0752}. We consider this the most relevant comparison. 

%It is clear from the figure that there are some differences, particularly at high column denisities, indicating that the CDDFs we show here are reasonably weakly converged with simulation resolution. The strong convergence of \ion{O}{vii} and \ion{O}{viii} column density distributions is bad at high column densities.     
The weak convergence of the CDDFs is particularly good: for column densities $< 10^{16.3} \us \txn{cm}^{-2}$, differences are $\lesssim 0.3 \us \txn{dex}$. They get larger at higher column densities (particularly for \ion{O}{viii}), but this is in a regime where the CDDFs in the $25\us \txn{cMpc}$ box are poorly converged compared to the larger boxes (top panels of Fig.~\ref{fig:boxresconv}), and the behaviour of the CDDFs here may therefore not be representative of what the effect of higher simulation resolution might be in a larger box.

Finally, as we mentioned in section~\ref{sec:cddf_methods}, we found that the effect of the choice of SPH kernel is small. Here, we give some more details on the kernels we compared. Our default is the \citet{wendland_1995} C2 kernel, with shape $k$ given by
\begin{equation}
k_{\mathrm{C2}}(u) = (1 - u)^4  (1 + 4u),
\end{equation} 
where $u$ is the distance to the particle position normalized by the smoothing length of the particle. The kernel is zero at $u > 1$. Similarly, the {\textsc Gadget} kernel (used by \citet{springel_2005}, from \citet{monaghan_lattanzio_1985}) is described at $u \leq 1$ by
\begin{equation}
k_{\mathrm{Gadget}}(u) = 
\begin{cases}
2.5465 + 15.2789 \, (u - 1)\, u^2 & \text{if } u < 0.5\\
5.0930\, (1-u)^3             & \text{if } 0.5 \leq u \leq 1 .
\end{cases}
\end{equation}
% 2.546479089470 + 15.278874536822  (u - 1) u^2 & \text{if } u < 0.5\\
% 5.092958178941 (1-u)^3             & \text{if } 0.5 \leq u \leq 1 .
Both kernels are normalized to unity when integrated over the surface of a unit circle to ensure mass conservation in the projection. The C2 kernel is more centrally concentrated. When both kernels are normalized, their absolute difference is largest at the centre of the distribution: $k_{\mathrm{C2}}(0) = 2.23$ and $k_{\mathrm{Gadget}}(0) = 1.82$. The maximum relative difference is at the edge of the distribution, and diverges as $u \rightarrow 1$ due to the different slopes of the kernels near zero. 
The differences between the CDDFs we obtain using these kernels is $\lesssim 0.05\us\txn{dex}$ for $10^{11}\us \txn{cm}^{-2} < N_{\mathrm{O\, VII, VIII}} < 10^{16.5} \us\txn{cm}^{-2}$.

\section{Technical choices for the EW distribution}
\label{app:projchoice}

\begin{figure}
\includegraphics[width=\columnwidth]{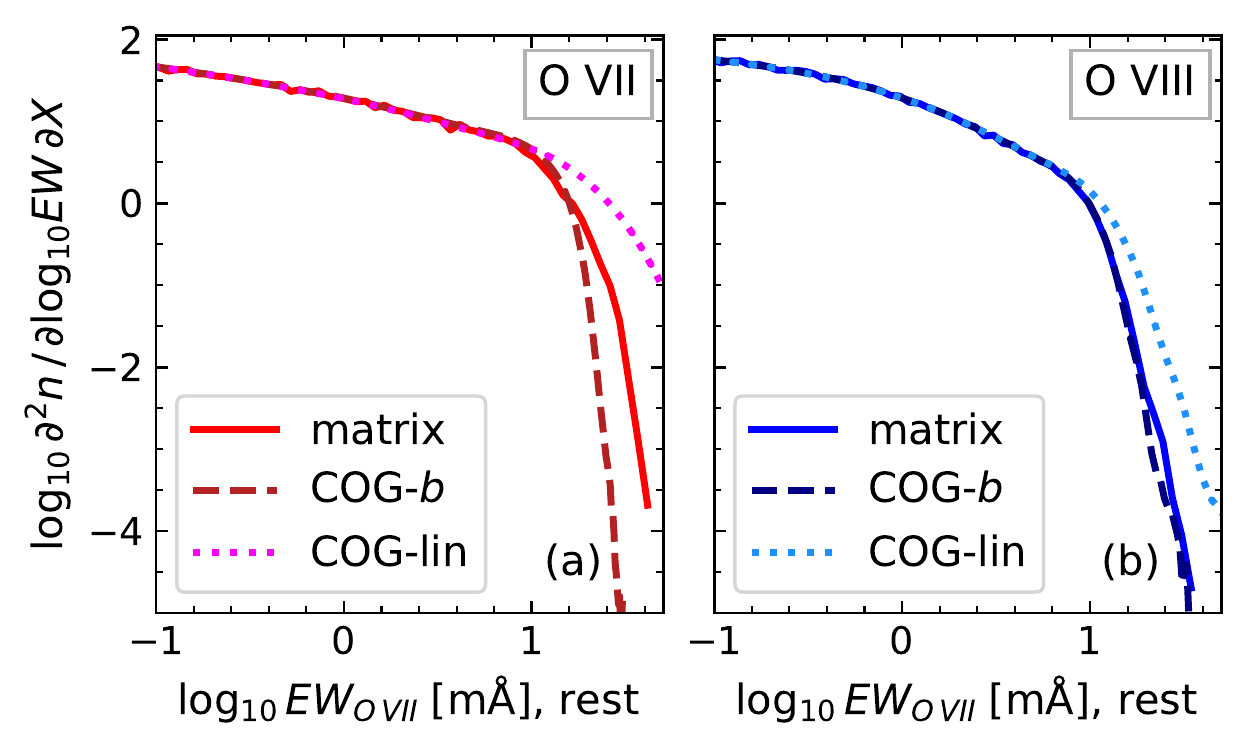}
\caption{A comparison of methods to obtain the rest-frame equivalent width distribution from the CDDF. The rest frame EW distributions for \ion{O}{vii}, $\lambda = 21.60\us${\AA}   (a) and the \ion{O}{viii}, $\lambda =18.97\us${\AA} doublet (b) in the \code{Ref-L100N1504} simulation at redshift~$0.1$ are shown. The lines labelled `COG-$b$' are for EWs calculated from the CDDF using a single curve of growth with the best-fit $b$ parameters to the relation between log EW and the column density obtained for a set of absorption spectra. The dotted line (`COG-lin') is for EWs calculated using the linear curve of growth for each ion. For the lines labelled `matrix', the conversion from CDDF to EW distribution was done using a matrix generated from the column density-EW relation in Fig.~\ref{fig:cog}, which accounts for scatter. We use the CDDF for $6.25 \us \txn{cMpc}$ slices.}
\label{fig:EWdist}
\end{figure} 
% The top panels (a, b) show the equivalent width distribution as a histogram, the bottom panels (c, d) show the cumulative distribution: the number of absorption systems with a given minimum equivalent width per absorption length $dX$. The distributions in the top panel are not entirely the equivalent width counterpart to the CDDF of equation~\ref{eq:cddf}, since the derivative of the number of absorber is with respect to $\log_{10} EW$, not the equivalent width $EW$ itself.

Fig.~\ref{fig:EWdist} shows the rest-frame EW distribution of \ion{O}{vii} and \ion{O}{viii} calculated from the CDDF using different methods.  In the first two, we impose a one-to-one relation between column density and EW. We use the linear curve of growth or the curve of growth with the best-fit $b$ parameter to the relation between log EW and column density obtained for a sample of absorption spectra. The third method, which is our fiducial model, accounts for scatter in the relation between column density and EW by using a matrix generated from the points in the column density-EW relation from Fig.~\ref{fig:cog}. We obtain this matrix by choosing column density and EW bins, then making a histogram of our sample of sightlines in column density and EW. After normalisation, we use this matrix to convert our column density histograms into EW histograms. 

%For the best-fit $b$ parameter and matrix methods, we try both the full sightline sample and the sample selected only on each separate ion. We also tried different methods to extrapolate the relation to lower column densities: best-fit and linear curve-of-growth conversions. However, this does not affect the equivalent width range shown in Fig.~\ref{fig:EWdist}. For the best-fit $b$ parameters, we used fits to the $\log EW$, and used the column densities obtained from specwizard. 
%Fits using the grid-projected column densities produced some values that were more sensitive to the sample choice and the choice between fitting to $EW$ or $\log_{10} EW$, presumably due to the larger scatter in this relation, particularly at lower column densities. This scatter will tend to favour unrealistic $b$ parameters in the optically thin regime, since we do not account for errors in column densities in our fits.  

The differences between the EWs calculated using the different methods are as expected. The matrix conversion is our preferred method, because it models the scatter. 
%The difference between the best-fit $b$ parameter curve-of-growth conversion and the matrix method is whether or not scatter is accounted for. This scatter is only important at high column densities and equivalent widths, so at lower column densities, the methods agree on the equivalent width distributions. The scatter becomes more important as the curve of growth starts deviating from linear, and the linear curve of growth conversion fails to be accurate starting at somewhat lower column densities than the best fit. 
Since the linear curve of growth is only valid for unsaturated absorption, it gives the maximum EW for any column density. This method overestimates the number of high EW systems ($\txn{EW} \gtrsim 10$m{\AA}).
The difference between the matrix and best-fit curve-of-growth methods is whether scatter is included. Including scatter increases the predicted number of absorption systems at the highest EWs because the column density distribution is steep and declining at the highest column densities. This means that scatter at fixed column density attributes larger EWs to more lower column density absorption systems than it attributes smaller EWs to larger column density absorption systems, in absolute terms. This is why the single best-fit $b$-parameter curve of growth conversion underestimates the number of high EW absorption systems. 
The discrepancy is larger for \ion{O}{vii} than for \ion{O}{viii}, because, as Fig.~\ref{fig:cog} shows, \ion{O}{vii} EWs have a larger scatter at fixed column density and have significant scatter over a larger range of column densities than \ion{O}{viii} EWs. This means that including scatter in the modelling of the EW distribution has a larger effect for \ion{O}{vii}.

\section{Mock survey CDDFs}
\label{app:dXvar}

\begin{figure}
\includegraphics[width=\columnwidth]{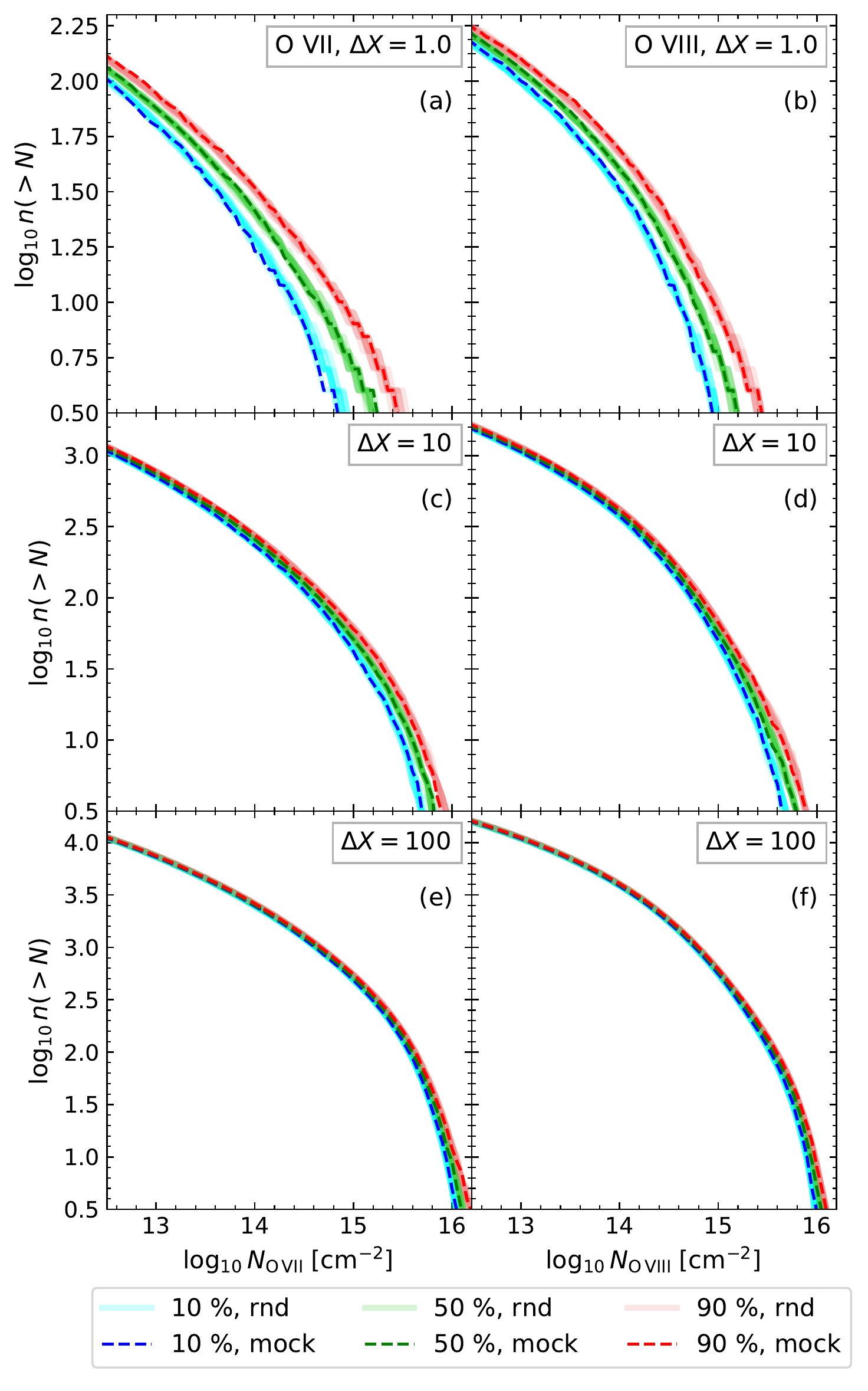}
\caption{The cumulative column density distributions for mock surveys of sizes $\Delta X = 1.01$ ($\Delta z = 0.878$; a, b), $\Delta X = 10.0$ ($ \Delta z = 8.66$; c, d), and $\Delta X = 100$ ($ \Delta z = 86.7$; e, f), of \ion{O}{vii} (a, c, e) and \ion{O}{viii} (b, d, f) in the reference (\code{Ref-L100N1504}) simulation at redshift $0.1$. The column densities are calculated for 16 $6.25 \us \txn{cMpc}$ slices, each with $3.125^2 \us \txn{ckpc}^2$ pixels. The dashed lines show mock surveys, where $(x, y)$ positions were selected at random, and then each slice along the $z$-axis at that position was included, mimicking the measurement of different absorption systems along full $100 \us \mathrm{cMpc}$ sightlines. This accounts for clustering along the line of sight on scales between $6.25$ and $50\us\txn{cMpc}$ (i.e.\ half the size of the periodic box). We obtained a sample of 100 such mock surveys. At each column density, show the $10^{\mathrm{th}}$ (blue), $50^{\mathrm{th}}$ (green, median), and $90^{\mathrm{th}}$ (red) percentiles of the distribution of absorber counts from the different mock surveys. To find the effect of large-scale structure on this distribution, we repeated this process for random surveys, show in lighter solid colors, where we select the same number of $6.25 \us \txn{cMpc}$ columns, but choose them randomly. For these random surveys, we show the percentiles for ten different sets of 100~random surveys. The good agreement between the solid and dashed curves indicates that large-scale structure does not seem to affect the statistical variation in CDDFs that surveys may measure on scales of $6.25$--$50\us\txn{cMpc}$.}
\label{fig:cddf_dXvar}
\end{figure}

%\begin{figure}
%\includegraphics[width=\columnwidth]{./figs/dXvar_rel.png}
%\caption{Same as Fig.~\ref{fig:cddf_dXvar}, but numbers of absorption systems are relative to the values predicted by a binomial distribution (dotted lines), and the values predicted by a Poisson distribution (dot-dashed lines) are also shown.}
%\label{fig:cddf_dXvar_rel}
%\end{figure}

We are interested in the variation in measured CDDFs for different survey sizes.  In particular, we investigate whether clustering along the line of sight on scales of $6.25$--$50\us\txn{cMpc}$ causes deviations from Poisson statistics, which we find is not the case. 
We measure survey size by the absorption distance $\Delta X$, as given by equation~\ref{eq:dX}. We measure this at redshift $0.1$, and use the Hubble flow across the box to calculate $\mathrm{d}z$. 
We conduct mock surveys of the $100 \us \mathrm{cMpc}$ box, of sizes $\Delta X = 0.5, 1.0., 10, 100, 1000$.  To create one mock survey, we choose $(x, y)$ positions at random, then take the column densities for all the $6.25 \us \txn{cMpc}$ columns along the $z$-axis at that position. This mimics searching for absorption systems along longer sightlines than we use to measure the column densities. Note that due to the size (and periodicity) of the box, and the extent over which we measure a single column density, this only probes the effect of large-scale structure correlations for separations between $6.25$ and $50 \us \txn{cMpc}$. 

Because we are interested in the statics of these surveys, we conduct 100~such mock surveys for \ion{O}{vii} and \ion{O}{viii} for each $\Delta X$. For comparison, we also create `random' surveys of the same size, where for each random $(x, y)$, we select a single $6.25 \us \txn{cMpc}$ slice at random. We make cumulative distributions for these surveys. Then at each threshold column density, we considered the distribution in absorption system counts among the different mock surveys. We show the $10^{\mathrm{th}}$ (blue), $50^{\mathrm{th}}$ (green, median), and $90^{\mathrm{th}}$ (red) percentiles of this distribution as the dashed lines in Fig.~\ref{fig:cddf_dXvar}.

%We compared these to the Poisson distribution predictions for these percentiles, given the number of absorption systems in above each threshold in the total CDDF and the size of the mock survey. At low column densities, we need to use a binomial distribution instead, since we define absorption systems to be a given portion of a sightline, meaning that the total number of absorption systems is fixed by our survey size. However, this is not at all an issue in the column density range we show here. 

In Fig.~\ref{fig:cddf_dXvar}, we do not show theoretical distributions, but the random surveys described above.
%but a different kind of mock surveys we call random surveys. This is because we expect some deviations from the theoretical percentiles simply due to using a finite set of mock surveys. For these random surveys we draw the column densities from the CDDF without considering any spatial information. We again take samples of 100 such surveys and compute the $10^{\mathrm{th}}$, $50^{\mathrm{th}}$, and $90^{\mathrm{th}}$ percentiles for the distribution between surveys of the number of absorption systems above each threshold. 
To probe what range in these percentiles is consistent with variation between survey samples, we use ten different samples of 100 random surveys. We plot these in Fig.~\ref{fig:cddf_dXvar} with colors matching the same percentiles for the mock surveys.

The figure shows that the mock surveys for the different ions and survey sizes are consistent with the random surveys: in other words, large-scale structure on the scales we can probe does not appear to affect the survey statistics for measuring CDDFs. 
%The results for the smallest and largest surveys are consistent with these.
We also checked the relative differences between these percentiles and Poisson and binomial distributions: these were centred on zero, and the mock survey distributions differed from the theoretical values by amounts consistent with the random surveys. Using the CDDF instead of the cumulative distribution gives the same results.  

Note that Fig.~\ref{fig:cddf_dXvar} can also be used to estimate a reasonable range of detected absorption systems above a given column density threshold for different survey sizes. 
\bsp	% typesetting comment
\label{lastpage}
\end{document}